\documentclass[aps,amsmath,amssymb, nofootinbib,  twocolumn, longbibliography,
superscriptaddress,prx]{revtex4-1}

\usepackage{amsmath,amsthm,amssymb,amsfonts}

\usepackage{braket}
\usepackage{xypic}
\usepackage{graphicx}

\usepackage[usenames,dvipsnames]{color}
\usepackage[colorlinks=true,citecolor=blue,linkcolor=magenta]{hyperref}

\usepackage{relsize,mdframed,lipsum}

\usepackage{dsfont}
\usepackage{changepage}
\usepackage{array}

\newcommand{\bit}{\begin{itemize}}
\newcommand{\eit}{\end{itemize}}

\newcommand{\half}{\frac{1}{2}}
\newcommand{\f}{\frac}
\renewcommand{\>}{\right\rangle}
\newcommand{\<}{\left\langle}
\newcommand{\ba}{\begin{align}}
\newcommand{\ea}{\end{align}}
\newcommand{\be}{\begin{equation}}
\newcommand{\ee}{\end{equation}}
\newcommand{\bi}{\begin{itemize}}
\newcommand{\ei}{\end{itemize}}
\newcommand{\lf}{\left(}
\newcommand{\ri}{\right)}
\newcommand{\dd}{\mathrm{d}}

\newcommand{\OO}{\mathcal{O}}
\newcommand{\mS}{\mathcal{S}}
\newcommand{\mU}{\mathcal{U}}
\newcommand{\rb}{\boldsymbol{r}}

\newcommand{\Tr}{\operatorname{Tr}}
\newcommand{\tr}{\operatorname{tr}}

\newcommand{\blue}{}

\newcommand{\avg}{\mathbb{E}}

\newcommand{\opnorm}[1]{\left\| {#1} \right\|_\infty}

\newcommand{\abs}[1]{\left| {#1} \right|}
\newcommand{\rd}{\mathrm{d}} 
\newcommand{\kup}{\ket{\uparrow}}
\newcommand{\kdown}{\ket{\downarrow}}

\newcommand{\bdown}{\bra{\downarrow}}

\newcommand{\otc}{{\mathcal{C}}} 

\begin{document}

\title{ Operator Spreading in Random Unitary Circuits}
\author{Adam Nahum}
\affiliation{Theoretical Physics, Oxford University, 1 Keble Road, Oxford OX1 3NP, United Kingdom}
\affiliation{Department of Physics, Massachusetts Institute of Technology, Cambridge, Massachusetts 02139, USA}
\author{Sagar Vijay}
\affiliation{Department of Physics, Massachusetts Institute of Technology, Cambridge, Massachusetts 02139, USA}
\author{Jeongwan Haah}
\affiliation{Station Q Quantum Architectures and Computation, Microsoft Research, One Microsoft Way, Redmond, Washington, USA}

\begin{abstract}

Random quantum circuits yield minimally structured models for chaotic quantum dynamics, able to capture
 for example universal properties of entanglement growth. We provide exact results, and coarse-grained 
 models, for the spreading of operators by quantum circuits made of Haar-random unitaries. 
 We study both 1+1D and higher dimensions, and argue that the coarse-grained pictures carry over to 
 operator spreading in generic many-body systems. In 1+1D, we demonstrate that the
  out-of-time-order correlator (OTOC) satisfies a biased diffusion equation, which gives exact results for 
  the spatial profile of the OTOC, and determines the butterfly speed $v_{B}$. We find that in 1+1D the `front' of the OTOC broadens diffusively, with a width scaling in time as $t^{1/2}$.  We address 
 fluctuations in the OTOC between different realizations of the random circuit, arguing that they are 
 negligible in comparison to the broadening of the front within a realization. Turning to higher
 dimensions, we show that the averaged OTOC can be understood exactly via a remarkable  correspondence 
 with a purely classical droplet growth problem. This implies that the width of 
 the front of the averaged OTOC scales as $t^{1/3}$ in 2+1D and as in 3+1D as $t^{\,0.240}$ (exponents
 of the Kardar-Parisi-Zhang universality class). We support our analytic argument with simulations in 2+1D. We point out that, in two or higher spatial dimensions, 
the shape of the spreading operator at late times is affected by underlying lattice symmetries,
and is in general not spherical.
 However when full spatial rotational symmetry is present in 2+1D, our mapping implies
 an exact asymptotic form for the OTOC, in terms of the Tracy-Widom distribution.
 
For an alternative perspective on the OTOC in 1+1D,
we map it to the partition function of an Ising-like statistical mechanics model.
As a result of special structure arising from unitarity,
this partition function reduces to a random walk calculation which can be performed exactly. 
We also use this mapping to give exact results for entanglement growth in 1+1D circuits.

\end{abstract}

\maketitle

\section{Introduction}


A key challenge for many-body physics is to identify universal properties of  quantum dynamics and the approach to thermalization. Particularly important are universal results that hold for generic  quantum systems. Examples of such universal properties include the existence of effective light cones for the propagation of quantum information \cite{lieb_robinson} and the existence of universal scaling forms for the growth and saturation of the von Neumann entanglement entropy in 1+1D \cite{CardyCalabrese2005EntanglementEvolution,
CalabreseCardyQuenchReview,KimHuse2013,ZhangKimHuse,LiuSuh2014,Asplund,CasiniLiuMezei2015,ho2015entanglement,NRVH2016, mezei2016entanglement,
SerbynPapicAbanin2013,HuseNandkishoreOganesyan2014,VoskAltman2013,GreinerEntanglement2016}.

By definition, generic systems lack the structures (for example large numbers of symmetries or conservation laws) that allow for exact results in typical solvable many-body systems. Surprisingly, insight into generic systems can come from studying dynamics with even \textit{less} structure than a generic Hamiltonian system, such as the dynamics generated by a random quantum circuit.
Random circuit dynamics provide a minimally structured  model with which real Hamiltonian dynamics can be compared \cite{hayden_preskill, sekino_susskind, brown2012scrambling, Lashkari2013, Hosur2016,OliveiraDahlstenPlenioPRL2007,Dahlsten,Znidaric,BHH2016}. Despite its simplicity, this model is able to capture universal scaling forms for entanglement growth both in 1+1D and in higher dimensions \cite{NRVH2016}. Random circuits are also toy models for information scrambling in black holes and other strongly coupled systems \cite{hayden_preskill, sekino_susskind, brown2012scrambling, Lashkari2013, Hosur2016,OliveiraDahlstenPlenioPRL2007,Dahlsten,Znidaric,BHH2016, Shenker_New}. 

In this paper we provide both exact results and coarse-grained descriptions for the spreading of quantum operators under random circuit dynamics, as measured by the `out-of-time-order correlator' (OTOC). The OTOC originally appeared in the study of quasi-classical approximations to superconductivity  \cite{larkin}, {\blue and is closely related to the commutator norm that appears in Lieb-Robinson bounds \cite{lieb_robinson},} but it has been studied recently as a means of quantifying the scrambling of quantum information \cite{kitaev, shenker_stanford_1, shenker_stanford_2, maldacena2016}.  {\blue It has been argued that} early-time exponential growth of the OTOC is a characteristic feature of chaotic quantum systems, and such growth has been obtained within the AdS/CFT correspondence and in a range of physical systems \cite{Aleiner, Stanford, Sachdev_CriticalFermiSurface, Altman, gu2016local, roberts2016lieb, chowdhury2017onset, Sachdev_DiffusiveMetals}. {\blue The OTOC has also been applied to characterize  slow dynamics in the presence of disorder and in the many-body localized phase \cite{swingle2017slow, ForthcomingGriffiths, huang2016out, chen2016out, chen2016universal, he2017characterizing, fan2017out, chapman2017classical}.  It has  been calculated in 2D conformal field theories \cite{roberts2015diagnosing} and integrable chains \cite{dora2016out}, and studied numerically in nonintegrable 1D systems \cite{bohrdt2016scrambling,luitz2017information,leviatan2017quantum}.} Following theoretical proposals \cite{Tarun, Swingle, Yao}, experiments {\blue addressing} the OTOC have been conducted \cite{AMRey,  li2016measuring, wei2016exploring}.

Random quantum circuits provide an ideal theoretical setting for the exact calculation of quantities such as the OTOC.  While the behaviour of the OTOC in a random circuit is interesting in its own right, we conjecture that the long-distance properties of the OTOC that we derive will also be applicable to deterministic dynamics. {\blue Therefore we believe that these results will provide a useful starting point for understanding the generic spatial structure of spreading operators.}

An operator $\mathcal{O}_0$ which is initially localized near the spatial origin (say, on a single site of a spin chain) will evolve under Heisenberg time evolution into a vastly more complicated operator $\mathcal{O}_0(t)=U^\dag(t) \mathcal{O}_0 U(t)$ that acts nontrivially on many sites. The `size' of $\mathcal{O}_0(t)$ is the size of the region in which $\mathcal{O}(t)$ fails to commute with a typical local operator $Y_x$ at position $x$. This may be measured by
\begin{align}\label{eq:introductionOfCommutator}
\otc(x,t) \equiv \half \Tr \rho [\mathcal{O}_0(t), Y_x]^\dag [\mathcal{O}_0(t), Y_x]
\end{align}
where the expectation value has been taken in an appropriate Gibbs state. (For our purposes this will be taken to be the infinite temperature Gibbs state $\rho_\infty$, which is the state to which random circuit dynamics equilibrate.) To make the connection with the out-of-time-order correlator (OTOC), we may expand out  the commutators in (\ref{eq:introductionOfCommutator}). For simplicity let us assume for the moment that the operators $\mathcal{O}_0$ and $Y_x$ are both Pauli-like operators squaring to the identity. We then have 
\begin{align}
\otc(x,t) = 1 - \Tr \rho_\infty \mathcal{O}_0(t) Y_x \mathcal{O}_0(t) Y_x.
\end{align}
The second term, in which the operators are not time ordered, is the OTOC.

\begin{figure}
    \includegraphics[width=0.79\linewidth]{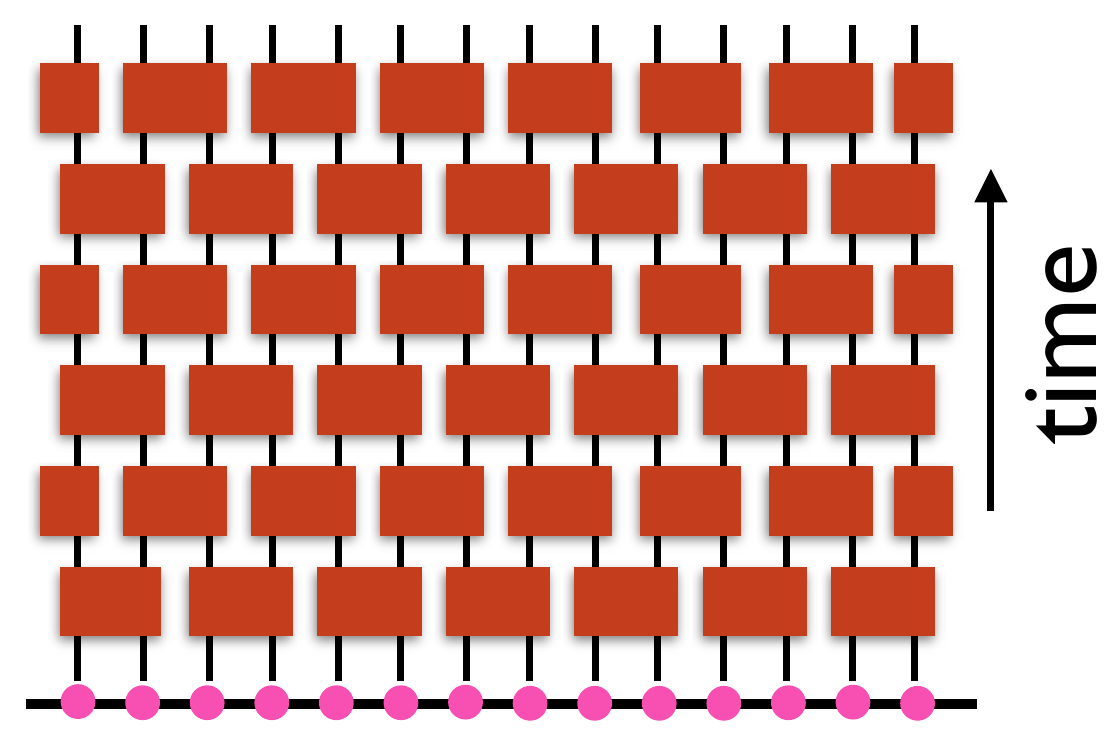}
 \caption{
 {\bf Left: Random unitary circuit in 1+1D.} Each brick represents an independently Haar-random unitary, acting on the Hilbert space of two adjacent `spins' of local Hilbert space dimension $q$.}
  \label{fig:one_plus_one_circuit}
\end{figure}

At a given time $t$, the range of $x$ where the commutator $\otc(x,t)$ is significantly larger than zero gives a measure of the size of the operator. This region typically grows ballistically \cite{Roberts_Shocks}, even when local conserved quantities exhibit diffusive transport {\blue \cite{Aleiner, Sachdev_DiffusiveMetals,bohrdt2016scrambling,luitz2017information}}.\footnote{Strongly disordered Hamiltonians in 1+1D provide counterexamples to this ballistic spreading.} The immediate natural questions about $\otc(x,t)$  include:  what is the `butterfly' velocity $v_B$ associated with this ballistic growth? What is the spatial structure of $\otc(x,t)$? Is there a `hydrodynamic' equation for $\otc(x,t)$ at large time and distance scales? Are there important differences between 1+1D and higher dimensions? We will answer all these questions for the case where the time evolution operator $U(t)$ is  a circuit composed of Haar random unitaries, as in Fig.~\ref{fig:one_plus_one_circuit}.

\begin{figure}
        \includegraphics[width=0.8\linewidth]{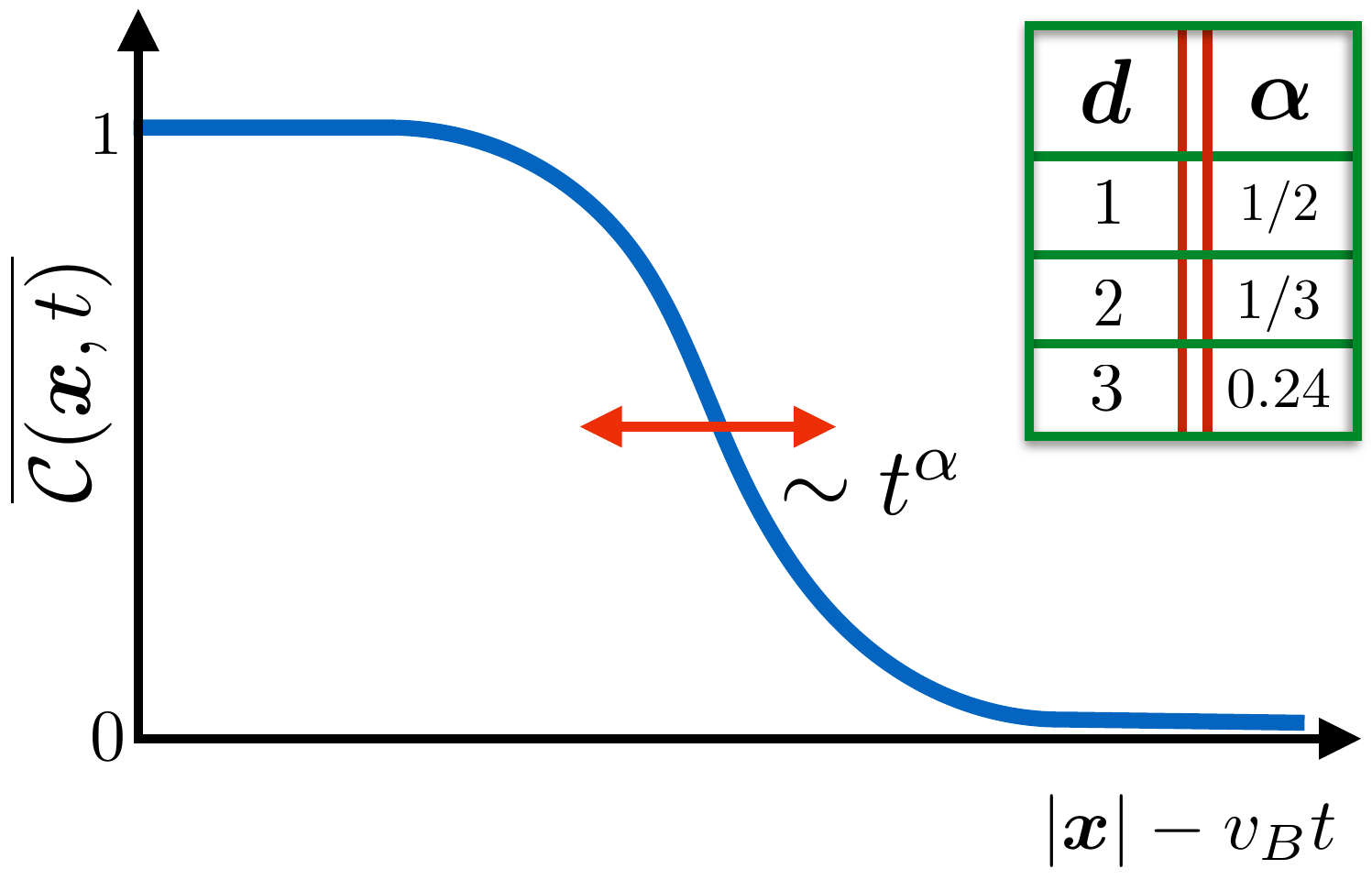}
 \caption{
{\bf Schematic Behavior of the average OTOC:} We find that the average OTOC $\overline{\mathcal{C}}(\boldsymbol{x},t)$  (where the average is over the local unitaries in the quantum circuit) has a front which broadens as $t^{\alpha}$, with the indicated exponents in various spatial dimensions $d$.}
  \label{fig:OTOC_Schematic}
\end{figure}

We demonstrate that, both in 1D and in higher dimensions, operator spreading and the growth of the OTOC can be mapped to classical stochastic growth models.  We show via an exact calculation that  operator spreading in 1+1D can be understood in terms of an equation involving diffusion and drift. The `front' of the operator propagates at a finite velocity $v_B$. However the front also \textit{broadens} diffusively, so that its width is proportional to $\sqrt{t}$ (Fig.~\ref{fig:OTOC_Schematic}). We conjecture that this physics also occurs in  generic (nonintegrable) 1D systems undergoing deterministic Hamiltonian dynamics.  For  random circuit dynamics, we must also consider how the averaged correlator $\overline{\otc}$ differs from the correlator $\otc$ within a given realization of the random circuit. We argue that fluctuations between realizations are small:  typical variations in the front position between different realizations are $O(t^{1/4})$, so negligible in comparison with the $\sqrt{t}$ broadening of the front.

\begin{figure}[t]
   \includegraphics[width=0.9\linewidth]{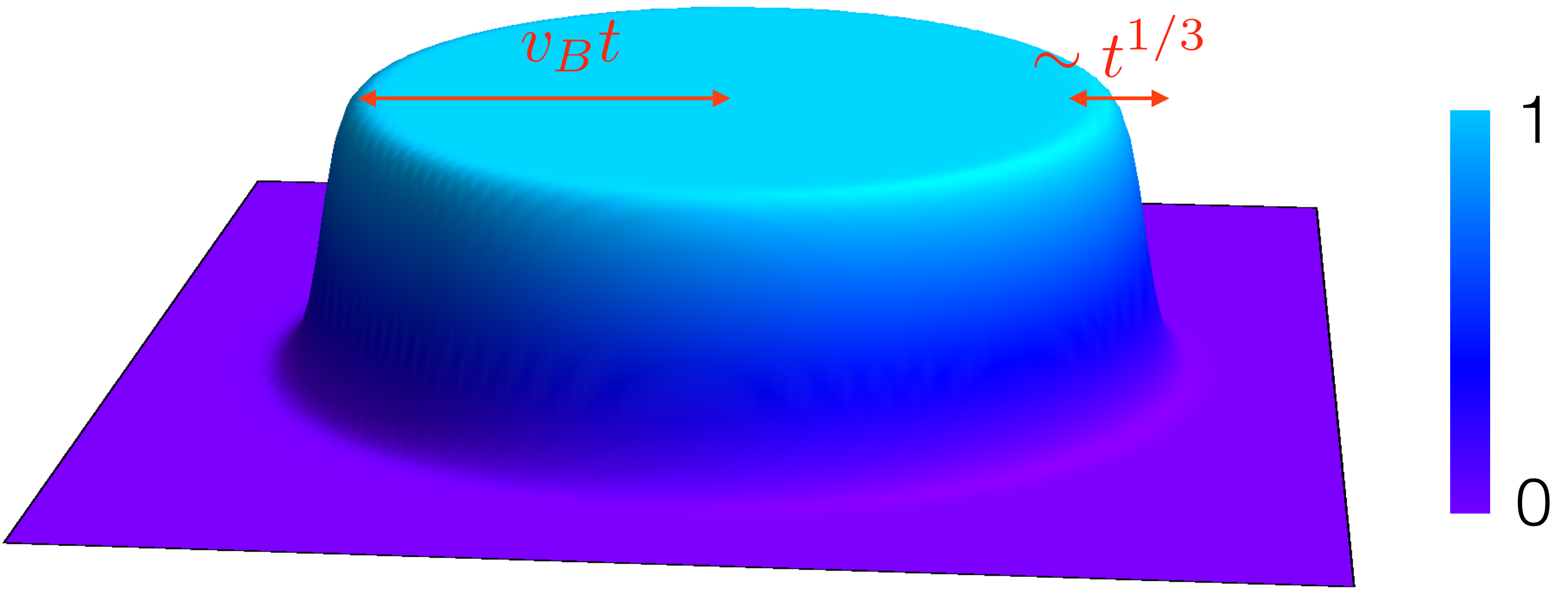}
 \caption{{\bf Cartoon for the form proposed here for the OTOC in two spatial dimensions,} when lattice anisotropy can be neglected. The functional form is given by the Tracy Widom distribution $F_2$.}
  \label{fig:OTOC_Schematic_2D}
\end{figure}

Turning to higher dimensions, we show by an exact mapping that there is a remarkable relationship with a classical droplet growth problem in the Kardar--Parisi--Zhang universality class \cite{kpz_paper}. {\blue (To avoid confusion, we note that this is \textit{not} related to the connection between entanglement growth and KPZ introduced in \cite{NRVH2016}.)  We use this relationship to quantify the broadening of the `front' of a growing operator in a higher dimensional random circuit. In 2+1D the front broadens like $t^{1/3}$ \cite{kpz_paper}, and in 3+1D like $t^{0.240}$ \cite{pagnani_parisi}.\footnote{The phase diagram of the KPZ equation in higher dimensions \cite{kpz_paper} indicates that in 4+1D and above, two distinct universality classes may be  possible for operator growth in a random circuit, one with a growing front width and one without.}   In the two-dimensional case, and in the absence of lattice anisotropies, recent breakthroughs in the theory of interface growth \cite{johansson2000shape,prahofer2000universal,CalabreseLeDoussalRosso2010,Dotsenko2010,SasamotoSpohn2010A,SasamotoSpohn2010B,SasamotoSpohn2010C,GideonCorwinQuastel2011,CalabreseLeDoussal2011,ProlhacSpohn2011,LeDoussalCalabrese2012,ImamuraSasamoto2012,ImamuraSasamoto2013} 
also translate to an exact form for the OTOC, in terms of the celebrated Tracy-Widom distribution (Fig.~\ref{fig:OTOC_Schematic_2D}). The broadening of the front of the OTOC is summarized in Fig. \ref{fig:OTOC_Schematic}. } Again, we conjecture that these universal scaling forms extend to nonintegrable models with time-independent Hamiltonians, although we note that a previous calculation in a different setting has instead found a front that does not broaden with time, and is governed by a traveling wave equation \cite{Aleiner} (see also \cite{chowdhury2017onset, Sachdev_DiffusiveMetals}). (A traveling wave equation arises from our mappings  if we make  a certain mean field treatment, Appendix.~\ref{app:meanfield}. But this mean field is not valid in physical dimensionalities.)

In higher dimensions the \textit{shape} of the spreading operator%
\footnote{That is, the shape of the spatial region in which $\otc$ has saturated. 
We can neglect here the broadening of the front, 
since at late times the length scale associated with this broadening is parametrically smaller than the size of the operator.
}
is also of interest. 
At first sight one might expect the shape of the operator to be asymptotically spherical at late times. 
Instead, we argue that in  systems with an underlying lattice,  which have only discrete spatial symmetries, 
the spreading operator will not become spherical. 
Its asymptotic shape is determined by a model-specific velocity function $v(\hat n)$, 
the speed of the front depending on the local normal vector $\hat n$. 
We verify this for random circuits by simulation in 2+1D.

{
The results above are for random circuits composed of {generic} (Haar-random) unitary matrices. It is interesting to compare with random circuits composed of unitaries from the Clifford group, a discrete subgroup which leads to efficiently simulable dynamics \cite{gottesman1998heisenberg, zimboras2010}. In this case the dynamics of the operator is much simpler \cite{NRVH2016}, and randomness-induced fluctuations are much more severe. But remarkably the results for the \textit{averaged} OTOC $\overline{\otc}$ coincide with the results for generic unitaries. This is a consequence of the fact that the Clifford group is a unitary 2-design~\cite{Divincenzo2001}.

{\blue
In one dimension we give a complementary exact calculation of the averaged OTOC, using a mapping to the partition function of a classical Ising model.  These Ising degrees of freedom have a similar origin to those found in calculations in random tensor networks \cite{HaydenEtAl2016}. We show that special structure arising from the unitarity of the quantum circuit means that this partition function is exactly calculable for any value of the local Hilbert space dimension.

Another important question is how the speed $v_B$ associated with operator spreading 
relates to the speed $v_E$ which can be associated with entanglement growth following a quench in 1D \cite{CardyCalabrese2005EntanglementEvolution,CardyCalabrese2009EntanglementFieldTheory,CalabreseCardyQuenchReview,KimHuse2013,LiuSuh2014,Asplund,CasiniLiuMezei2015,Ho,mezei2016entanglement, NRVH2016}. Refs.~\cite{NRVH2016,mezei2016entanglement}  pointed out that in general $v_E$ is smaller than $v_B$, unlike the situation in a 1D conformal field theory \cite{CalabreseCardyQuenchReview}. 
We extend this here, showing that arbitrarily small values of $v_E/v_B$ can be achieved without any fine-tuning. (In a system with quenched, i.e. time-independent, disorder it is even possible to have $v_B>0$ but $v_E=0$ \cite{ForthcomingGriffiths}.) 

We also use the Ising mapping described above to give an exact calculation of the average entanglement purity  (the exponential of the second Renyi entropy) for a random circuit, complementing the scaling picture, in terms of a coarse-grained minimal cut, of Ref.~\cite{NRVH2016}.
}

\tableofcontents

\section{Operator dynamics in 1+1D}
\label{sec:operator-dyn-1d}

We begin by defining the random circuit dynamics which we consider in 1+1D, 
and describing the `hydrodynamic' continuum picture we propose for the OTOC in 1+1D.
In Sec.~\ref{sec:exact calculation} 
we give an alternative exact calculation of the OTOC, 
confirming and extending the results below.

\subsection{Hydrodynamic equation for averaged OTOC}
\label{hydrodynamic section}

We consider time evolution by a quantum circuit  on an infinite 1D spin chain 
where each spin (qudit) has local Hilbert space dimension $q$. The structure of the quantum circuit is shown in Fig.~\ref{fig:OTOC_Schematic}a.
Two-site unitaries are applied to `even' bonds on even time steps and `odd' bonds on odd time steps 
(a `running bond' layout in the language of bricklaying). 
Each two-site unitary is drawn independently from the uniform distribution on the two-site unitary group $\mathbb{U}(q^2)$. 
Formally, our time evolution operator is ${U(t)=U(t,t-1) U(t-1,t-2) \cdots U(1,0)}$, 
where a single layer of the circuit is given by
\begin{align}
U(t',t'-1) =
\begin{cases}
\bigotimes_{x \in 2\mathbb Z} U_{x,x+1}(t',t'-1) & \text{if $t'$ is even,}\\
\bigotimes_{x \in 2\mathbb Z +1} U_{x,x+1}(t',t'-1) & \text{if $t'$ is odd.}\\
\end{cases}
\end{align}
Each two-site unitary $U_{x,x+1}(t',t'-1)$ is Haar random and independent of all of the others.

Given an operator $\mathcal{O}$, 
we write $\mathcal{O}(t) = U(t)^\dag \mathcal{O} U(t)$. 
We will evaluate the following out-of-time order correlator with respect to this time evolution:
\begin{align}
\otc(x,t) & \equiv \half \Tr \rho_\infty [X_0(t), Y_x]^\dagger [X_0(t), Y_x] \notag  \\
& = - \half \Tr \rho_\infty [X_0(t), Y_x]^2 .
\end{align}
Here, $\rho_\infty$ is the infinite temperature Gibbs state, 
i.e., the mixture of all possible spin configurations with equal weights.
$X_0$ is a Hermitian operator located at the origin of the spin chain,
and $Y_x$ is a Hermitian operator located at site $x$. 
We take both $X$ and $Y$ to be traceless,
and normalized such that $\Tr X^2 = \Tr Y^2 = q$. 
For example if $q=2$ (the spin--1/2 chain)
we can take $X$ and $Y$ to be Pauli matrices at sites $0$ and $x$, respectively.

Since the unitaries in the circuit are random, 
we must distinguish between averaged quantities 
(denoted by $\avg_U$, or whenever unambiguous by overline $\overline{[\cdots]}$)
and quantities within a given realization of randomness.
However, we will argue that fluctuations induced by the random circuit are small in a certain sense, 
meaning that the spatial profile of $\otc(x,t)$ in a given realization of the circuit is, at large times, 
parametrically close to the average value $\overline{\otc(x,t)}$.

Fig.~\ref{fig:OTOC_Schematic} 
is a schematic of the spatial profile we will show for $\overline{\otc(x,t)}$ at fixed large time. 
The `size' of the operator is determined by a butterfly speed which is 
\begin{align}
v_B(q) = \f{q^2 -1}{q^2 +1}. \label{butterfly speed}
\end{align}
Within a region of size $\sim 2 v_B(q) t$ the commutator $\otc(x)$ 
has saturated to a value very close to unity.  
Note that for finite $q$ the butterfly velocity is smaller than the `naive' speed limit of unity, 
which is set by the geometry of the quantum circuit, 
while in the limit $q\rightarrow \infty$ they coincide. 
The  `front' of the operator, i.e. the region in which $\otc$ varies between 0 and 1, 
broadens diffusively. 
The width of the front is proportional to
\begin{align}
\sigma(q,t) = \frac{2q }{q^2+1} \sqrt{t}. \label{sigma fmla}
\end{align}
More precisely, 
letting $\Phi$ denote the cumulative density function of the Gaussian distribution,  
${\Phi(y) = \frac{1}{\sqrt{2\pi}}\int_{-\infty}^y e^{-x^2 /2} \rd x}$ 
(which tends to zero for $y\ll 0$ and to $1$ for $y\gg 0$), 
we have
\begin{align}
\overline{\otc(x,t)}
\simeq \Phi\left(\frac{v_B t + x}{\sigma(t)}\right) \Phi\left(\frac{v_B t - x}{\sigma(t)}\right).
\label{eq:exact_result_for_1D_summary}
\end{align}
In  Sec.~\ref{sec:exact calculation}  we will see that Eq.~\eqref{eq:exact_result_for_1D_summary} 
is the partition function of an Ising-like statistical mechanics problem, and will derive an exact formula on the lattice (without any continuum approximation):
\begin{align}
\overline{\otc(x,t)}  = & \label{eq:exactsolution}  \\
 (1-\xi) ~&g\left(t-1,\frac{t-x-3}{2},p\right)~ g\left(t-1,\frac{t+x-3}{2},p\right)\nonumber \\
 +\,\xi ~&g\left(t-1,\frac{t-x-1}{2},p\right)~ g\left(t-1,\frac{t+x-1}{2},p\right), \nonumber
\end{align}
where
\begin{align}
 p &= \frac{1}{q^2+1}, &
  \xi &= \frac{q^4}{q^4 -1},\nonumber 
\end{align}
and
\begin{align}
 g(n,a,p) &= \sum_{k=0}^a \binom{n}{k} (1-p)^{n-k} p^k .\nonumber
\end{align}
Here we  show how Eq.~\eqref{eq:exact_result_for_1D_summary} 
can be related to a continuum hydrodynamic equation which is asymptotically accurate at large times. For $x$ near the operator's right hand front, $\otc$ is related to a diffusing conserved density $\rho$:
\begin{align}
\overline{\otc}(x,t) &= \int_x \dd x' \overline{\rho}(x',t), &
\end{align}
\begin{align}
\partial_t \overline{\rho}(x,t) &= v_B(q) \partial_x \overline{\rho}(x,t) + D(q) \partial_x^2 \overline{\rho}(x,t).
\end{align}
We will explain the quantity $\rho$ below.

\newcommand{\cS}{\mathcal{S}}

To begin with, focus on the spin--1/2 chain ($q=2$).  
At time $t$ we may write the operator in the basis of products of Pauli matrices \cite{Roberts_Shocks,Ho,mezei2016entanglement,Dahlsten,Znidaric},
\begin{align}
X_0(t)= \sum_{\cS} a_\cS (t) \cS \label{eq:operator_string_expansion}
\end{align}
where $X_0(t=0)$ is a single site operator.
Here the `string' $\cS$ can be any product of Pauli matrices on distinct sites. 
The number of strings in the sum generically grows exponentially with time  
(at the naive lightcone velocity, set by the geometry of the circuit). 
The $\cS$ are normalized as
\begin{align}
\Tr \rho_\infty \cS \cS' = \delta_{\cS \cS'},
\label{eq:orthonormalBasis}
\end{align}
and $X_0$ is also normalized so $\Tr \rho_\infty X_0^2=1$, implying
\begin{align}
\sum_\cS a_\cS(t)^2 = 1.
\end{align}
It is useful also to introduce $\rho(x,t)$, the `fraction' of strings ending at $x$:
\begin{align}\label{rho redistribution}
\rho(x,t) &= \sum_{\substack{\text{strings $\cS$}\\\text{ending at $x$}}} a_\cS(t)^2,
&
\sum_x \rho(x,t) &=1.
\end{align}

We observe that the OTOC is determined by $a_\cS(t)^2$ as follows.
Let $Y_x$ be the Pauli matrix $\sigma^y$ at site $x$. 
(This choice does not sacrifice generality due to Haar randomness of  the circuit.)
Since distinct Pauli matrices anti-commute, we see
\begin{align}
[X_0(t), Y_x]^2 
&= \left( \sum_\cS a_\cS(t) [\mathcal S, Y_x] \right)^2  \notag \\
&= \Bigg( \sum_{\substack{\cS : \cS_x = \sigma^y, \sigma^z}} 2 a_\cS(t) \cS Y_x \Bigg)^2.
\end{align}
Due to the orthonormality in Eq.~\eqref{eq:orthonormalBasis},
we have
\begin{align}
\otc = -\frac12 \Tr \rho_\infty [X_0(t), Y_x]^2 = \sum_{\substack{\cS : \cS_x = \sigma^y, \sigma^z}} 2 a_\cS(t)^2.
\end{align}

This tells us that if we determine the evolution of $\overline{a_\cS(t)^2}$,
then the averaged OTOC is also determined.
The dynamics of $\overline{a_\cS(t)^2}$ turns out to be remarkably simple, as shown in Refs.~\cite{Dahlsten, Znidaric}.
It is best understood if we first consider a system of just two sites (rather than an infinite chain)
over which a Haar random unitary is applied at time $t$.
It is straightforward to calculate (see Appendix~\ref{sec:String_Distribution}) that for arbitrary $q$
\begin{align}
\overline{a_{\cS'} (t+1)^2} &= \sum_\cS W_{\cS' \cS}\ \overline{a_\cS(t)^2}
\label{eq:Markov}
\end{align}
where
\begin{align}
 W_{\cS' \cS} &= \delta_{\cS', I} \delta_{\cS, I} + \frac{1}{q^4-1} (1 - \delta_{\cS',I})(1- \delta_{\cS,I}).
\end{align}
Note two features. First, the result is linear in $\overline{a_\cS(t)^2}$.
Second, $\cS'$ must be the identity if and only if $\cS$ is, 
but otherwise $\overline{a_{\cS'}(t+1)}$ is a constant for all $\cS' \neq I$.
In other words, the random unitary introduces a (fictitious) Markov process on the
probabilistic ensemble $\{ (\cS, \overline{a_\cS(t)^2} ) \}$ of strings \cite{Dahlsten,Znidaric}. This Markov process 
describes a single string $\cS$ which is stochastically updated over time. If $\cS$ is nontrivial, each update maps it to any nontrivial string, with uniform probability.
The generalization to multiple spins is immediate: for each pair of spins that
interact in a given timestep, the  stochastic update is applied to 
the corresponding two-site substring of  $\cS$.
This Markov process will also be used in higher dimensional setting below,
as it is not specific to the 1+1D setting. Note that the fictitious stochastic dynamics, which 
involves a single evolving string, is entirely different from the stochastic dynamics of the operator 
$X_0(t)$ itself (which is a superposition of exponentially many strings).

Returning to the average of the OTOC,
we realize that it only matters whether or not the string component of $X_0(t)$ at the site $x$ is the identity. 
In the ensemble $\{ (\cS,  a_\cS(t)^2 ) \}$,
the fraction 
\begin{align}
\mu(x,t) = \sum_{\cS : \cS_{x} \neq I} a_\cS(t)^2
\end{align}
of strings that occupy the site $x$, 
may fail to commute with $Y_x$. 
There are $q^2-1$ possible nontrivial operators at the the site,
which are all equally probable in the ensemble of string components of $X_0(t)$.
In the present case of $q=2$,
this yields%
\footnote{
For general $q$, one has to start with an operator basis 
that obeys our normalization condition in Eq.~\eqref{eq:orthonormalBasis}.
It is easy to construct such a basis.
Define $X = \sum_{k \in \mathbb{Z}_q} \ket{k+1}\bra{k}$ and 
$Z = \sum_{k \in \mathbb Z_q} e^{2\pi i k/q} \ket k \bra k$.
Then, the discrete group generated by these two matrices
contains exactly $q^2$ elements up to unimportant phase factors.
These are not hermitian, but no problem arises if one considers $|a_\cS|^2$.
Over Haar random unitaries, one easily obtains $\overline{\otc(x,t)} = \frac{q^2}{q^2-1} \overline{\mu(x,t)}$.
}
\begin{align}
\overline{\otc(x,t)} = \frac{q^2}{q^2-1} \ \overline{\mu(x,t)}
\end{align}

In turn, the average occupation number ${\mu(x,t)}$ can be related to the endpoint density $\rho$, assuming that $x$ is far to the right of the left-hand front of the operator:
\begin{align}
\overline{\mu(x,t)} &= \mu_0 \sum_{x' \ge x} \overline{\rho(x',t)}, 
&
\mu_0 & = \frac{q^2 - 1}{q^2}
\end{align}
The constant of proportionality $\mu_0$ has been determined by assuming local equilibration of the structure of the strings.\footnote{To find $\mu_0$, make the ansatz that each $\mu(x)$
is independent from $\mu(x')$ for $x \neq x'$.
Under this ansatz,
the probability that a pair of sites is partially or fully occupied is $1-(1-\mu_0)^2$,
and such an occupied pair evolves to fill one of the pair with probability $1-p$.
Therefore, setting  $\mu_0 = (1-p)(2 \mu_0 - \mu_0^2) = \frac{q^2 - 1}{q^2}$ yields the stationary state.
}
Therefore 
\begin{align}\label{otc_and_rho}
\overline{\otc(x,t)} =\sum_{x' \ge x} \overline{\rho(x',t)}.
\end{align}
It is natural to conjecture that local equilibration of the strings, together with the exponentially large number of strings contributing to $\rho$, will make this identity  valid asymptotically even without the average.

It remains to analyze the dynamics of $\overline{\rho}(x,t)$. The above Markov process implies a simple autonomous dynamics for $\overline{\rho}$:
\begin{align}
\overline{\rho(t+1,x)} &= p \left[ \overline{\rho(x,t)}+ \overline{\rho(t,x+1)} \right],  \notag \\
\overline{\rho(t+1,x+1)} &=(1-p) \left[ \overline{\rho(x,t)}+ \overline{\rho(t,x+1)} \right], 
\label{eq:rho_average_dynamics}
\end{align}
where
\begin{align}
p= \frac{q^2-1}{q^4 -1 } = \frac{1}{q^2+1} \label{step probability}
\end{align}
is calculated by counting the non-identity two-site operators $S$ that have the identity at $x+1$,
and the overline denotes averaging over unitaries applied up to a given time.

Recalling that unitaries are applied on even and odd bonds alternately, 
Eq.~\ref{eq:rho_average_dynamics} gives a complete description of the dynamics of $\overline{\rho(x,t)}$.
This is a lattice diffusion equation for the conserved density $\overline{\rho}$. 
Formally, $\overline{\rho}$ behaves like the probability density for a random walker who starts at the origin, 
and who prefers to travel to the right since $p<\frac 1 2$.
In the continuum (i.e. at long timescales) $\overline{\rho}$ satisfies a simple diffusion equation, 
\begin{align}\label{eq:simple_diffusion_eq}
\partial_t \overline{\rho(x,t)} &= v_B(q) \partial_x \overline{\rho(x,t)} + D(q) \partial_x^2 \overline{\rho(x,t)},
\end{align}
whose drift and diffusion constants are determined in Appendix.~\ref{sec:constantslatticediffusion}:
\begin{align}
v_B(q) & = \f{q^2-1}{q^2+1}, &
D(q) & = \f{2 q^2}{(q^2+1)^2}.
\end{align}
The peak in $\rho$ corresponds to the front of the spreading operator $X_0(t)$. 
It travels at speed $v_B(q)$ and broadens as $\sigma(q,t)$ (Eq.~\ref{sigma fmla}).
We emphasize that this fictitious random walker should {not} 
be thought of as `the endpoint' of the operator $X_0(t)$, which is a superposition of many strings with different endpoints.

From (\ref{otc_and_rho}), or in the continuum 
\begin{align}
\overline{\otc(x,t)} = \int_x \dd x' \overline{\rho(x',t)}
\end{align}
we see that $\overline{C}$ obeys the same equation as $\overline{\rho(x,t)}$ but with different boundary conditions,
\begin{align}\label{eq:otc_diffusion_eq}
\partial_t \overline{\otc(x,t)} = v_B(q) \partial_x \overline{\otc(x,t)}+ D(q) \partial_x^2 \overline{\otc(x,t)}.
\end{align}
Taking into account the similar behaviour at the left hand front gives (\ref{eq:exact_result_for_1D_summary}).

Above we had to make two (very natural) assumptions.
One was that we can ignore the interaction between the left end and the right end,
and the other was that the occupation density 
$\overline{\mu(x,t)}$ reaches its equilibrium value. 
In  Section~\ref{sec:exact calculation}  
we give an exact calculation of the averaged OTOC
 (including exact results for finite $t$ and $x$, not necessarily large) without making any approximation.

\subsection{Hydrodynamic description including fluctuations}

Having determined the averaged OTOC, the key question is about the fluctuations between different realizations of the random circuits. From the point of view of exact results this is a much harder question (it is possible to obtain bounds in regions far from the front: we return to this in
Sec.~\ref{sec:fluctuationBounds}). However, we conjecture that the universal physics of fluctuations in $\rho(x,t)$ can be obtained by upgrading Eq.~\ref{eq:simple_diffusion_eq} to a \textit{stochastic} diffusion equation for the random quantity $\rho(x,t)$. This description indicates that fluctuations are strongly suppressed at late times. Since the diffusive broadening is present in a single realization (i.e. is not an artefact of disorder averaging) it is natural to conjecture that it will also be present  in generic non-random 1D many-body systems. 

Microscopically we expect noise in both the diffusion constant and the drift, but we restrict to noise in the latter since it is more relevant in the RG sense:
\ba \label{hydrodynamic_with_fluctuations}
\partial_t \rho(x,t) &=\partial_x \left( v_B+ \eta(x,t) \right) \rho(x,t) + D \partial_x^2 \rho(x,t).
\end{align}
Here $\eta(x,t)$ is white noise, uncorrelated in space and time. 

The statistical properties of this equation are easy to analyze. In the absence of the noisy drift term, $\rho(x,t)$ forms a `wavepacket' whose width grows like $\sqrt t$ and whose center of mass is at  $x_\text{cm}=v_B t$. When the noisy drift is turned on, it induces statistical fluctuations in $x_\text{cm}$ whose magnitude scales with time as 
\begin{align}
\Delta x_\text{cm} \sim t^{1/4}.
\end{align}
A quick way to see this is to ask what the  drift velocity has been in a given realization, 
\textit{averaged} over the spacetime region visited by the wavepacket. 
The wavepacket visits a spacetime volume of order $\int^t \dd t' \sqrt {t'} \sim t^{3/2}$. 
Averaging the drift velocity $\eta(x,t)$ over this spacetime volume yields $v_\text{av} \sim t^{-3/4}$.
The typical random displacement of the wavepacket is thus of order 
$\Delta x_\text{cm}\sim v_\text{av} t \sim t^{1/4}$. 
A standard perturbative calculation in Appendix~\ref{noisy_diffusion_equation_appendix} 
reproduces this exponent $1/4$,
which also characterizes the spreading of directed waves in random media~\cite{saul1992directed}. 

The quantity $\Delta x_\text{cm}$ is parametrically smaller than $\sqrt t$, 
the width of the front of the averaged commutator. 
Therefore this heuristic argument indicates that the front profile of the averaged OTOC 
also applies to the OTOC within a given instance of the random circuit.
This is somewhat surprising.
To see why, let us contrast the above Haar random dynamics with Clifford dynamics for $q=2$.

\subsection{Comparison with Clifford circuit dynamics}
\label{1d clifford}

The Clifford group is a discrete subgroup of the unitary group,
defined by the property that any Pauli matrix is mapped to a product of Pauli matrices.
When the quantum circuit consists of Clifford operators,
an initial Pauli matrix remains a single string (rather than evolving into a superposition of  exponentially many strings as for dynamics with generic unitaries) and the endpoint density $\rho(x,t)$ is localized on a single site for all times.

However, uniformly random Clifford circuits have a crucial relationship with Haar random circuits.
Under a uniformly random Clifford update on a pair of sites, a nontrivial operator is mapped with equal probability to any of the nontrivial operators,
and thus the dynamics satisfies the master equation of the Markov process in Eq.~\eqref{eq:Markov}~\cite{NRVH2016}.
As a result, the \textit{averaged} quantities such as the average end point density $\overline{\rho(x,t)}$, 
the average occupation number $\overline{\mu(x,t)}$,
and, most importantly, the average OTOC $\overline{\otc(x,t)}$,
obey exactly the same dynamics as the Haar random case.
Formally, this is a consequence of the fact that random Clifford operators form a unitary 2-design \cite{Divincenzo2001};
see Appendix~\ref{app:Clifford} for the definition of design and a proof for  random Clifford.
One may say that Clifford dynamics realizes the {\it a priori}-fictitious Markov process in a physical system.%
\footnote{
All the statements here hold for any prime power $q$ such as $q=2,3,4,5,7,8,9,11,13,16,\ldots$.
}

Despite the equivalence of averaged quantities, 
the quantities \textit{within} a realization are entirely different. 
In the Clifford case the endpoint density $\rho$ and the OTOC $\otc$ are strongly fluctuating, 
while we have argued that for generic unitaries they are self-averaging (fluctuations are parametrically small).

\section{Higher dimensions}
\label{sec:higherD}

We now address the structure of the out of time order correlator $\otc(x,t)$ in spatial dimensions greater than one, by exploiting the relationship between the averaged OTOC and a fictitious classical Markov process (Sec.~\ref{hydrodynamic section}).  We show that this process is a classical droplet growth problem whose universal physics can be understood in terms of the Kardar--Parisi--Zhang equation \cite{kpz_paper}. By taking appropriate averages, we then obtain exact universal exponents and scaling forms for the OTOC in a circuit composed of Haar random unitary matrices.   We conjecture that these scaling forms also apply to more realistic Hamiltonian dynamics in non-integrable lattice models and field theories.

Somewhat surprisingly, we show that the `shape' of the spreading operator at late times does not become spherical, unless the microscopic dynamics has symmetry under continuous spatial rotations. In a lattice model, the spreading operator remembers forever that the lattice has only discrete point group symmetries. Our argument for this is not specific to  random circuit dynamics. The point is simply that `the' butterfly velocity $v_B$, which sets the speed at which the operator's front moves, generically depends on the front's orientation, resulting in an anisotropic profile for the spreading operator at long times.  Another surprising outcome, given previous work in the context of many-body perturbation theory including Ref.~\cite{Aleiner}, is that for the dynamics considered here the averaged OTOC $\overline{\otc}$ does not satisfy a local differential equation.

In 2+1D, when lattice anisotropy is absent (e.g. in an appropriate continuum model) or negligible,  recent results in KPZ theory \cite{johansson2000shape,prahofer2000universal,CalabreseLeDoussalRosso2010,Dotsenko2010,SasamotoSpohn2010A,SasamotoSpohn2010B,SasamotoSpohn2010C,GideonCorwinQuastel2011,CalabreseLeDoussal2011,ProlhacSpohn2011,LeDoussalCalabrese2012,ImamuraSasamoto2012,ImamuraSasamoto2013} 
yield the the full functional form of $\overline{\otc(x,t)}$ as a function of position and time. For an initially localized operator, this is expressed in terms of the GUE Tracy Widom distribution  \cite{SasamotoSpohn2010A,GideonCorwinQuastel2011} (which  describes the extremal eigenvalue statistics for the  Gaussian Unitary Ensemble of Hermitian matrices \cite{Tracy_Widom,Forrester}).

\subsection{Higher dimensions: setup and mapping to classical growth}
\label{highDsetup}

\begin{figure}
 \includegraphics[width=0.95\linewidth]{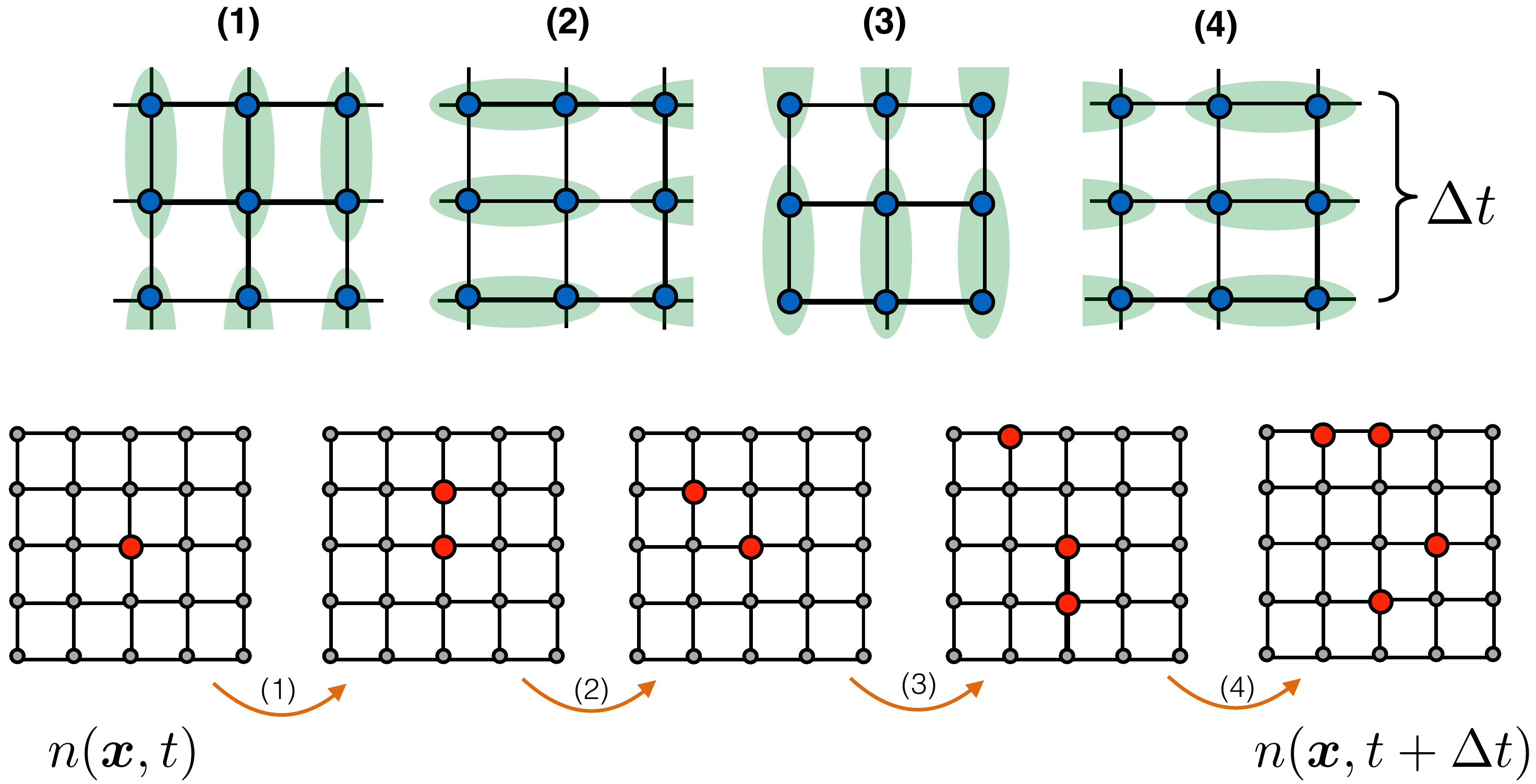}
 \caption{{\bf Top: 2+1D Haar-Random Quantum Circuit:} We consider unitary dynamics in which two-site Haar-random unitaries are applied on the bonds of a two-dimensional square lattice, in the columnar dimer configurations shown in (1-4).  {\bf Bottom:} allowed updates in the corresponding stochastic process.}
  \label{fig:Dimer}
\end{figure}
\begin{figure}[b]
 \includegraphics[trim = 0 0 0 0, clip = true, width=0.37\textwidth, angle = 0.]{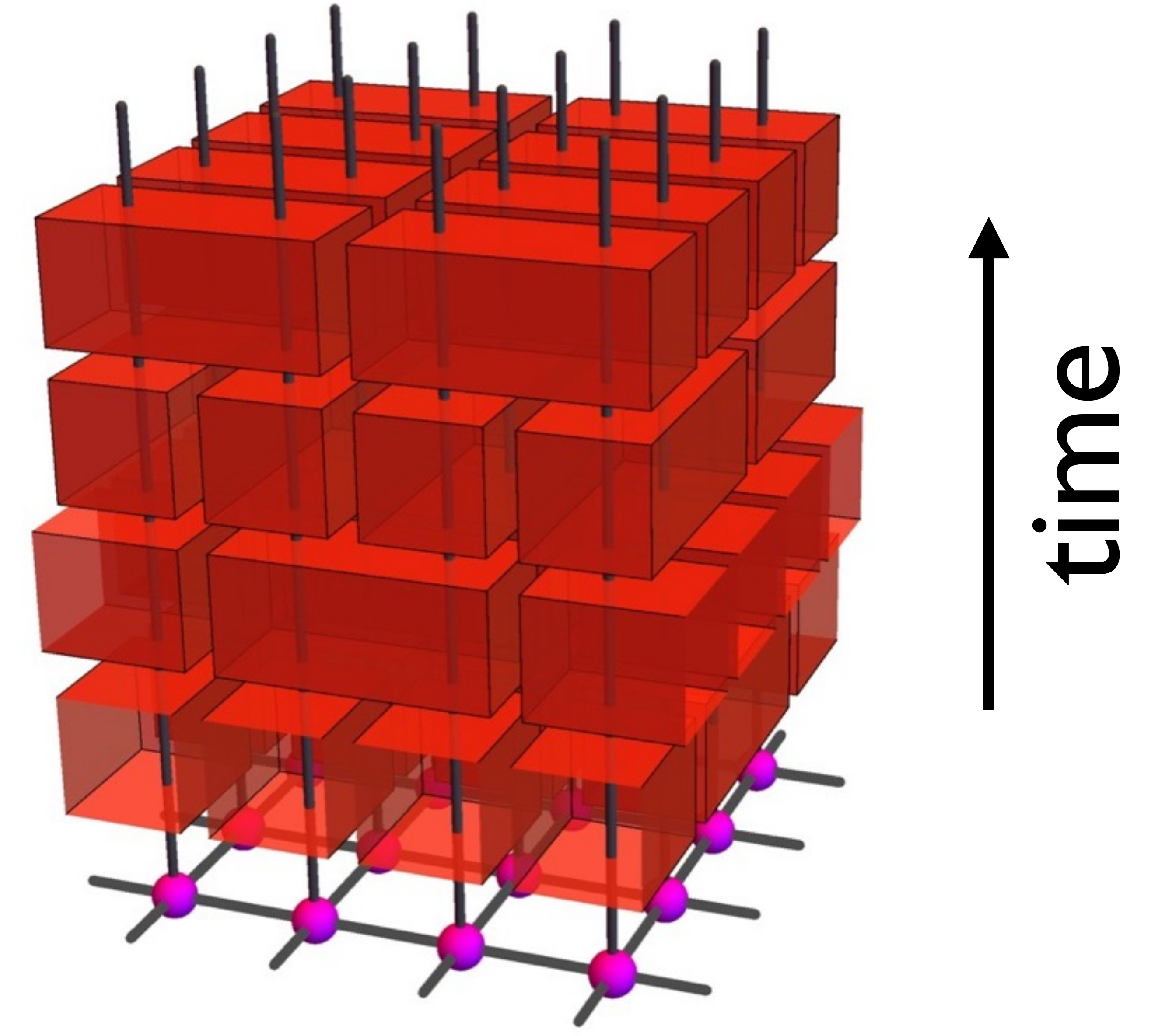}
 \caption{{\bf Geometry of $2+1$D Haar-random circuit.}}
  \label{fig:3DDimer}
\end{figure}

We now describe the unitary dynamics for which we wish to study operator spreading and the OTOC. We  choose a circuit where in each timestep Haar-random two-site unitaries are applied to bonds of a $d$-dimensional cubic lattice in a manner that generalizes the 1+1D protocol. We describe the 2+1D case for concreteness; the generalization to higher dimensions is immediate. The periodicity of the circuit is 4 layers. Four successive layers cycle through the four columnar `dimer coverings' of the square lattice as shown schematically in Fig. \ref{fig:Dimer} and Fig.~\ref{fig:3DDimer}, so that the site at the origin interacts sequentially with its neighbours at $\boldsymbol{x}=(0,1)$, $(-1,0)$, $(0,-1)$, $(1,0)$.

The reduction to a classical stochastic process in terms of the fictitious occupation numbers
\ba
n(\boldsymbol{x}) & =\, 0 \text{ or } 1 
\end{align}
proceeds just as in 1D (Sec.~\ref{sec:operator-dyn-1d} and Appendix.~\ref{sec:String_Distribution}). Consider two adjacent sites $\boldsymbol{x}$ and $\boldsymbol{y}$ which undergo a joint update in a given timestep. If both sites are initially empty ($n(\boldsymbol{x})=n(\boldsymbol{y})=0$) they remain so after the update. If at least one of the sites is initially occupied ($n(\boldsymbol{x})=1$ or $n(\boldsymbol{y})=1$ or both) then the configuration after the update can be $n(\boldsymbol{x})=1$, $n(\boldsymbol{y})=0$ with probability $p$, or  $n(\boldsymbol{x})=0$, $n(\boldsymbol{y})=1$ with the same probability, or $n(\boldsymbol{x})=n(\boldsymbol{y})=1$ with probability $1-2p$, where as before
\begin{align}
p& = \frac{1}{q^{2}+1}.
\end{align}
If we consider the OTOC for a spreading operator which is initially localized at a single site, then the corresponding classical model is initialized with $n=1$ at the origin and $n=0$ everywhere else.  A possible evolution of $n(\boldsymbol{x})$ in a single timestep is shown in Fig. \ref{fig:Dimer}.

Recall that the Haar-avaraged OTOC is related to the mean occupation number for this Markov process at time $t$ by the relation
\be \label{eq:C_n_relation}
\overline{\otc(\boldsymbol{x},t)} = \frac{q^{2}}{q^{2}-1}\< {n(x,t)} \>_\text{classical}
\ee
as illustrated schematically in Fig. \ref{fig:Droplet}. The averages on the two sides of the above equation have different meanings. On the left, the bar denotes an average over realizations of a unitary circuit, and $\otc$ is a  correlator for this  quantum dynamics. On the right, the angle brackets denote an average in a classical stochastic process. The real number $\otc$ and the integer $n$ are only related after averaging. As we noted above, the fictitious Markov process can be physically realized by random Clifford dynamics,
whenever $q$ is a prime power.

\begin{figure}
 \includegraphics[width=0.95\linewidth]{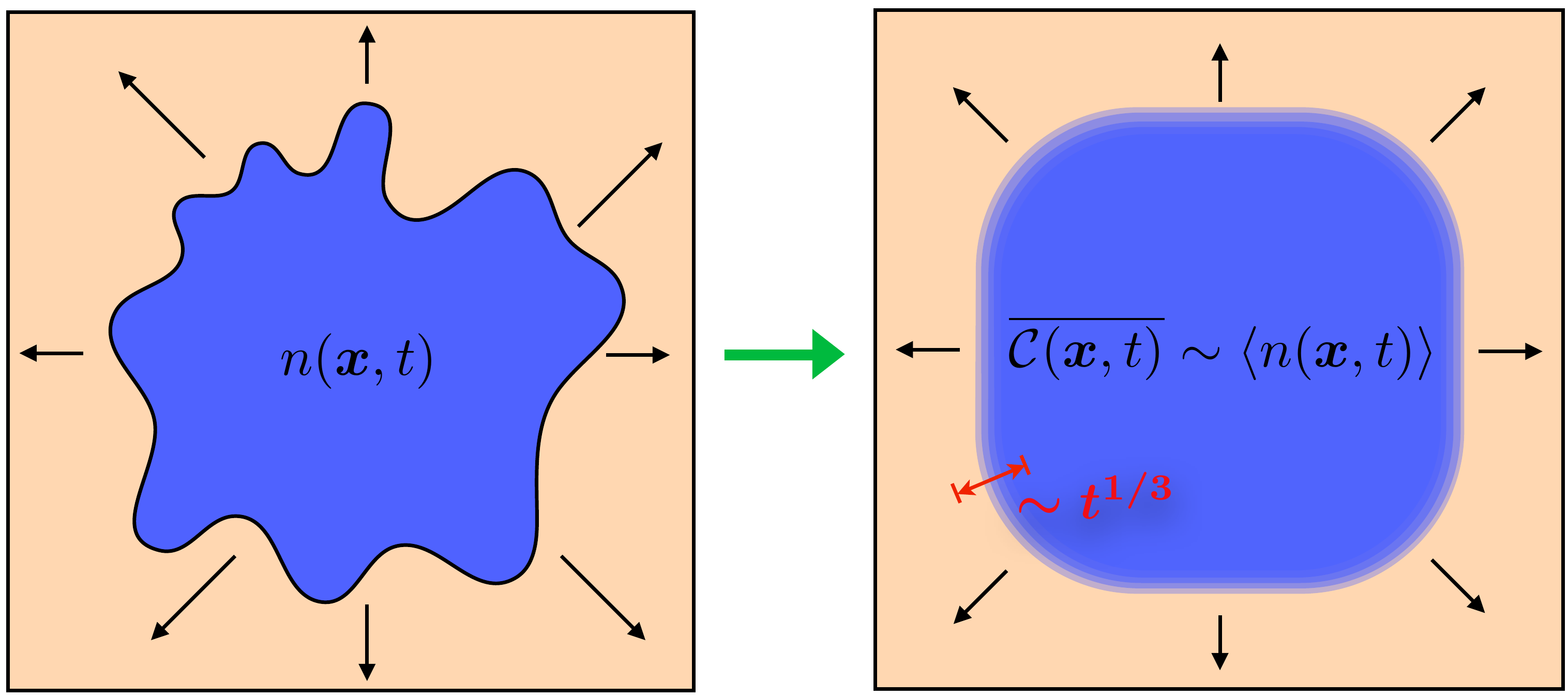}
 \caption{{\bf Growth of a Classical Droplet and the OTOC:} We relate the behavior of the OTOC (averaged over the unitaries in the circuit) to a classical stochastic process for the growth of a droplet in two spatial dimensions.  A given configuration of the classical droplet is specified by a binary occupation number $n(\boldsymbol{x},t)$ as shown the left.  Remarkably, the average droplet profile $\langle n(\boldsymbol{x}, t)\rangle$ precisely reproduces the averaged OTOC. }
  \label{fig:Droplet}
\end{figure}

\begin{figure*}
$\begin{array}{ccc}
 \includegraphics[trim = 65 200 115 190, clip = true, width=0.288\textwidth, angle = 0.]{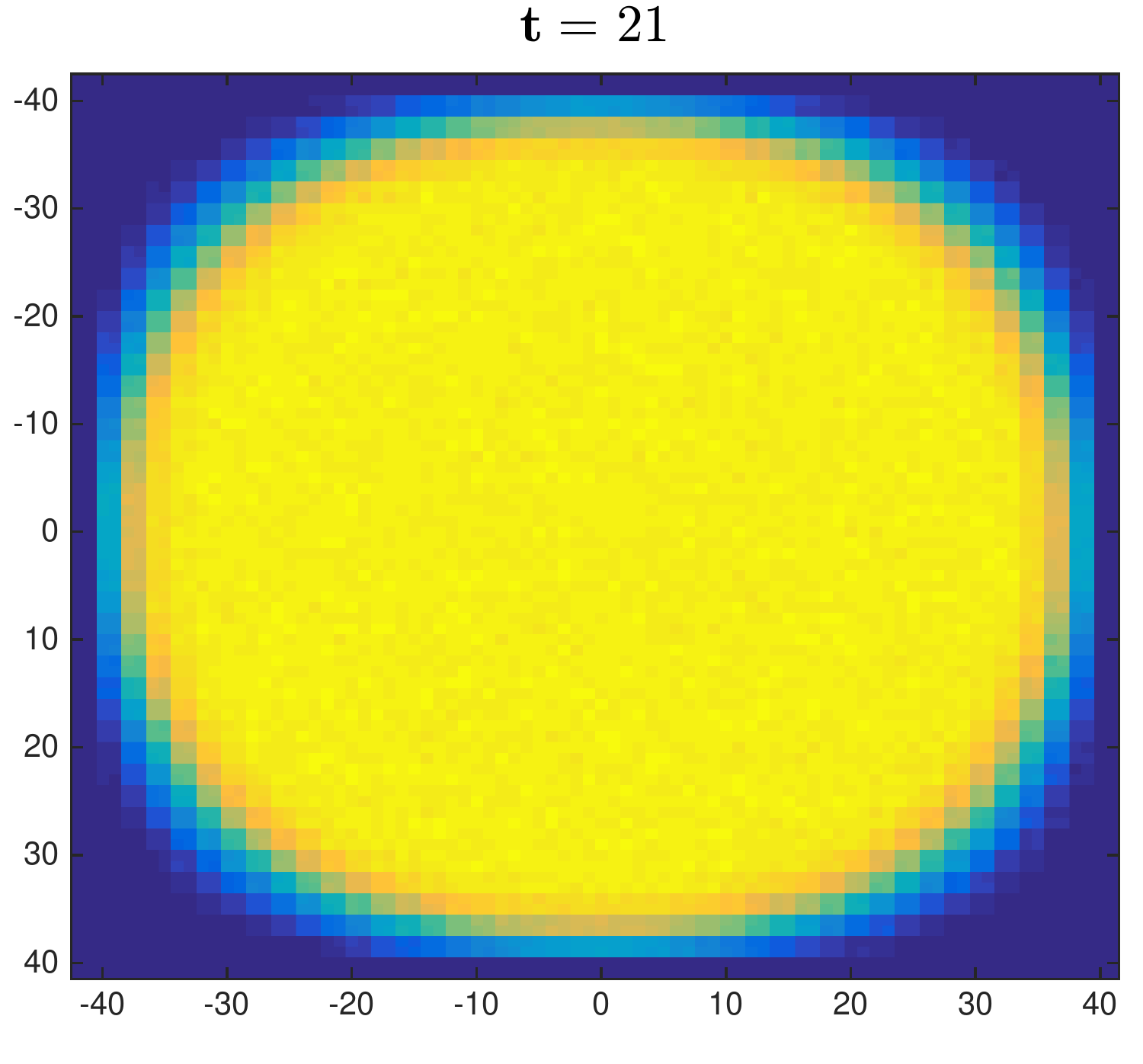} &
\includegraphics[trim = 61 200 115 190, clip = true, width=0.288\textwidth, angle = 0.]{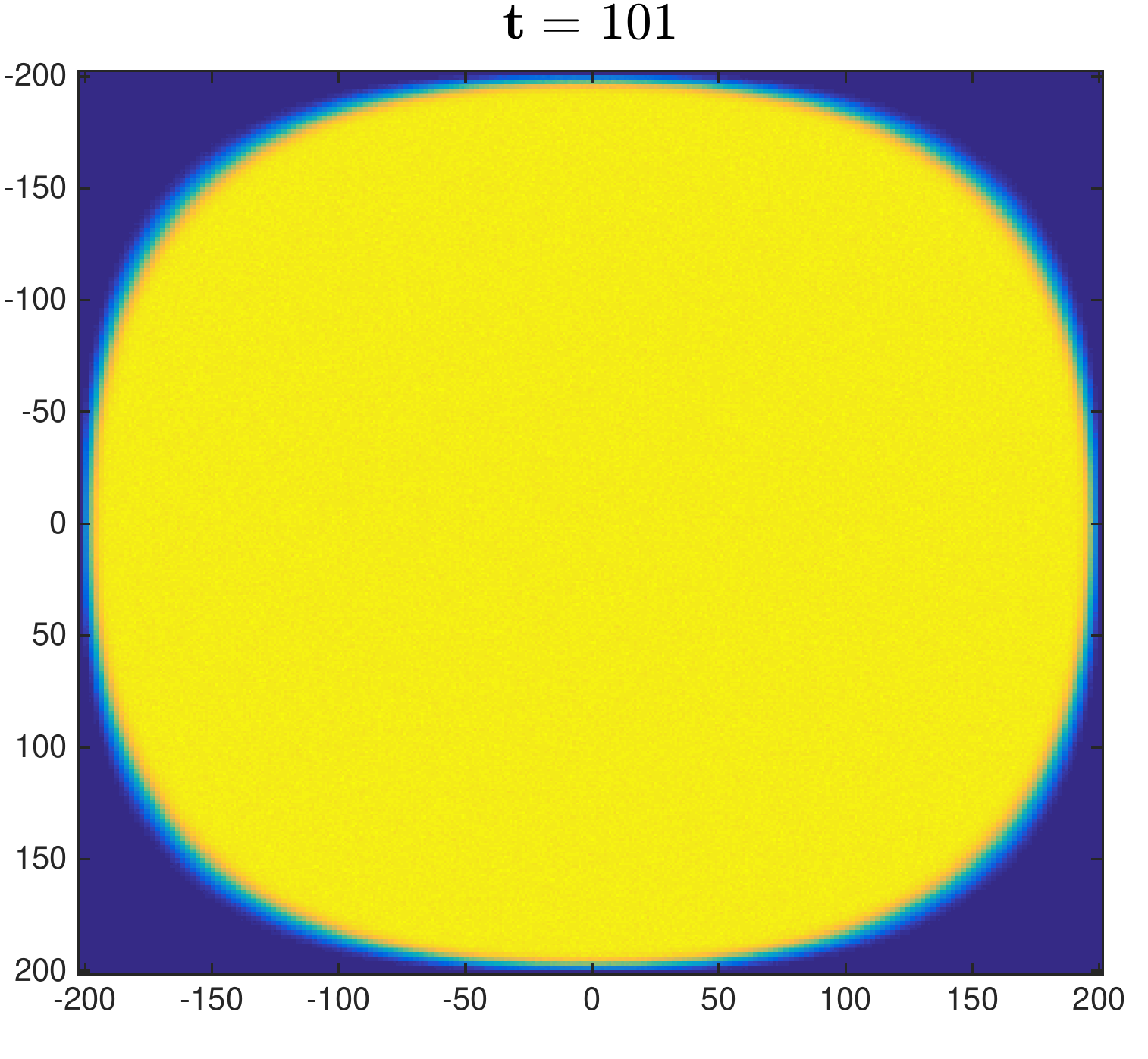} &
\includegraphics[trim = 61 200 40 190, clip = true, width=0.342\textwidth, angle = 0.]{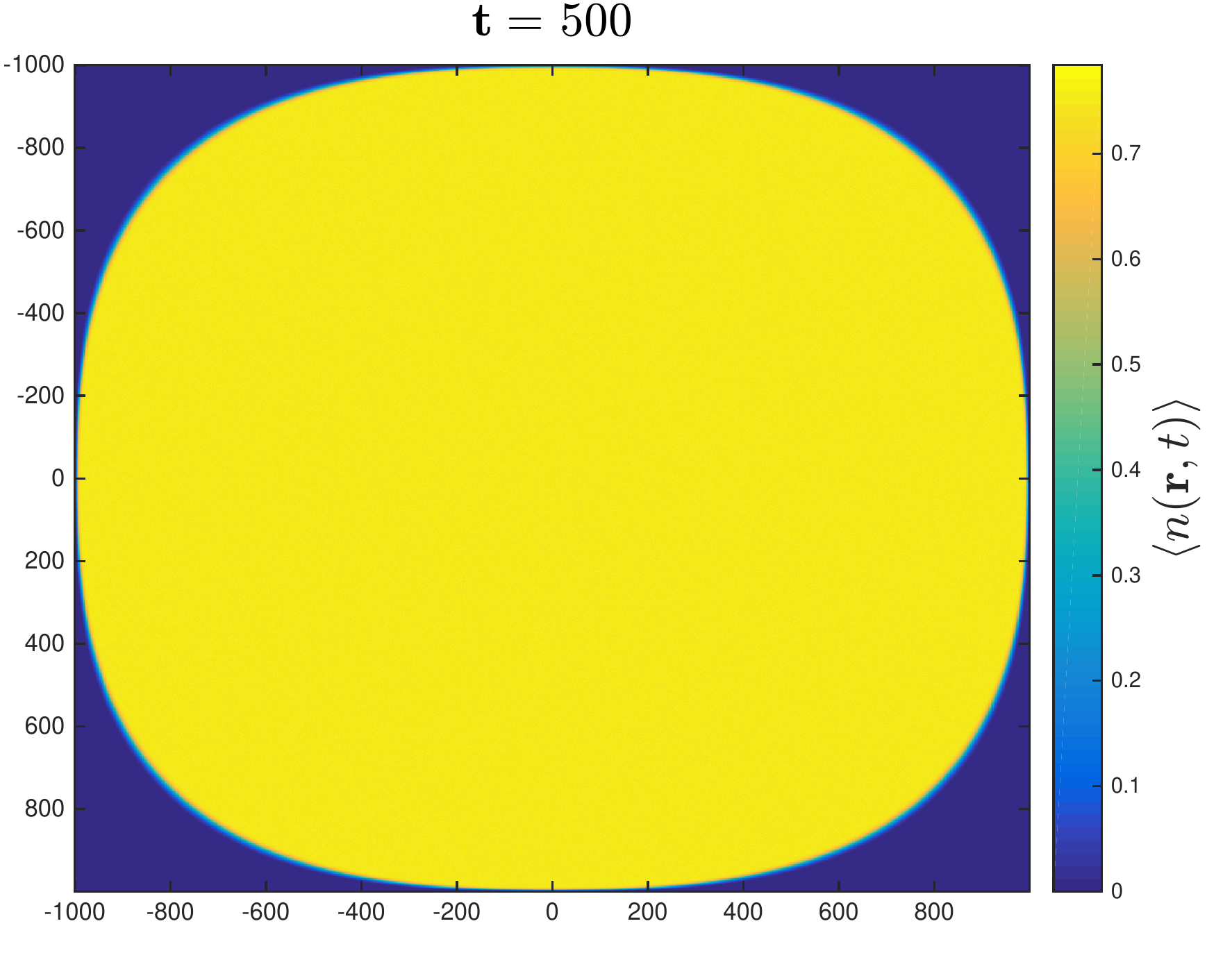}
 \end{array}$
 \caption{{\bf Growth of a 2D Cluster ($q=2$):} We determine the behavior of the averaged OTOC by simulating the stochastic growth of a two-dimensional cluster over $M = 2\times 10^{3}$ realizations, with local updates applied at each timestep, as described in the text.  The average occupation number for the cluster $\langle n(\boldsymbol{x}, t)\rangle$ is shown for the indicated times in the evolution as it approaches its asymptotic shape. }
  \label{fig:Cluster}
\end{figure*}

\subsection{Classical model in 2+1D: \\ analytical and numerical results}

The `seed' at the origin grows to produce a cluster of linear size $\sim t$. 
In the interior of this cluster the state equilibrates rapidly to a state in which nearby sites are essentially uncorrelated, with average occupation  $\< {n(x,t)} \>_\text{classical} = (q^{2}-1)/q^{2}$. In a given realization there is an interface between the occupied and unoccupied regions which is sharp on length-scales of the order of the lattice spacing. The evolution of the droplet is very similar to well-studied growth models such as the Eden model \cite{eden1961}, and reduces to the stochastic growth of this one-dimensional interface. The size of the occupied region grows linearly in time,  with statistical fluctuations in the shape of the interface. (The average shape in our 2D model is not circular, but has only four-fold rotational symmetry; we discuss this in Sec.~\ref{operator shape}.) 

Typically such growth processes are in the universality class of the Kardar-Parisi-Zhang (KPZ) equation \cite{kpz_paper}. Consider a section of the interface and let $\xi$ be a coordinate parallel to the interface and $h$ its height in the perpendicular direction. The KPZ equation is
\begin{align}
\partial_t h = c + \nu \partial_\xi^2 h + \f{\lambda}{2} (\partial_\xi h)^2 + \zeta(x,t),
\end{align}
where $\zeta$ is uncorrelated spatiotemporal noise.  The constant $c$ contributes to the average normal  growth rate for the interface, while the $\nu$ term describes diffusive smoothing of sharp features. Finally, the non-linear $\lambda$ term  encodes the dependence of average growth rate on the slope. This equation renormalizes to a nontrivial fixed point. One of its most basic properties is the fact that the fluctuations in the height at a given position $\xi$ grow with time as $t^{\beta}$, with $\beta=1/3$.

Let us write the shape of the droplet as a parameterized curve in polar coordinates, with $R(\theta)$ the radius at angle $\theta$ from the origin. (As mentioned above, the interface is sharp on an $O(1)$ lengthscale, and therefore $R(\theta)$ is well defined up to an $O(1)$ uncertainty; this is sufficient since the properties we discuss below are on parametrically larger lengthscales when $t$ is large.) From KPZ scaling we would expect
\begin{align}\label{meanandfluctuations}
\<R(\theta,t)\> & = r(\theta)\,t  - A(\theta) t^\beta + \cdots \\
\sqrt{ \< R(\theta,t)^{2} \> - \< R(\theta,t)\>^2 } & = C(\theta)\,t^{\beta} + \cdots
\end{align}
with the exactly-known exponent $\beta = 1/3$. We will discuss the nonuniversal function $r(\theta)$ below and in Sec.~\ref{operator shape}, and we will discuss more detailed universal properties in the next section.

We have examined the growth of the droplet for spin-1/2 degrees of freedom ($q=2$) on the square lattice, by tracking the average, evolving support of a cluster over $M = 2\times 10^{3}$ realizations of the classical dynamics up to time $t = 1000$.  We store only the density $\langle{n}(\boldsymbol{x},t)\rangle$, averaged over all $M$ realizations, as a function of position and time, as this is the quantity with a direct interpretation in the quantum setting. We have also investigated smaller values of $q$: these do not have an  interpretation in the quantum circuit, but in the classical model decreasing $q$ simply corresponds to increasing the probability $p$ in the update. At each time slice, the form of $\langle{n}(\boldsymbol{x},t)\rangle$   is fitted, along cuts through lattice symmetry axes, to extract the cluster size and the the width of the front (where $\<n(\boldsymbol{x},t)\>$ is appreciably different both from zero and from its $t\rightarrow\infty$ value).    We observe linear growth of the size as expected. Note that the fluctuations in the second equation of (\ref{meanandfluctuations}) imply that the width of the front region  is expected to scale like $t^{1/3}$.

Fig. \ref{fig:Numerics} (top) shows the growing width of the front for cuts along the diagonal, $\theta = \pm \pi/4$. There, at the largest times we can access, the fitted exponent is $\beta = 0.3305 \pm 0.0269$, extracted from a fit to the blue data points in Fig. \ref{fig:Numerics}. As expected, this value is consistent with the KPZ value $\beta = 1/3$. 

A slight surprise is that the behaviour along the axis, e.g. at $\theta=0$ is rather different: see Fig.~\ref{fig:Numerics} (bottom), which does not show KPZ growth. Generically the only stable fixed point for the growth of a 1D interface is believed to be the KPZ fixed point. However anomalous growth is possible in this  model, for $q$ greater than a critical value $q_c\lesssim 2$, when the direction of the front's local normal vector is fine-tuned to coincide with one of the axes, as occurs at $\theta=0$. In this regime, a front with normal parallel to a lattice axis moves at a speed exactly equal to the naive light-cone speed, $v_B=2$, and does not roughen. This is a known phenomenon in  various lattice growth models in discrete time which have synchronous parallel updates, and can be understood by a relationship with directed percolation \cite{richardson1973random,durrett1981shape,SavitZiff,kertesz1989anomalous,krug1990growth}: see Appendix.~\ref{appendix_depinning} for an explanation. While interesting, this phenomenon is an artefact of the specific  discrete spacetime geometry we have chosen, which could be eliminated by modifying  this geometry,\footnote{The effect  disappears for smaller $q$. For example for $q=1.4$  we see clear KPZ growth both at $\theta=0$ (fitted exponent value $\beta=0.3223 \pm 0.0199$) and at $\theta=\pi/4$ ($\beta = 0.3304 \pm 0.0149$).} and we certainly do not expect it to be relevant to continuous time dynamics. (It would be interesting to look for this effect in appropriate deterministic Floquet dynamics, however.) It has an effect on the shape of the droplet, which we discuss in Sec.~\ref{operator shape}.

We now discuss the OTOC scaling that results from the KPZ mapping, neglecting effects of lattice anisotropy (which we will return to in Sec.~\ref{operator shape}).

\begin{figure}
 \includegraphics[trim = 40 190 70 210, clip = true,width=0.8\linewidth]{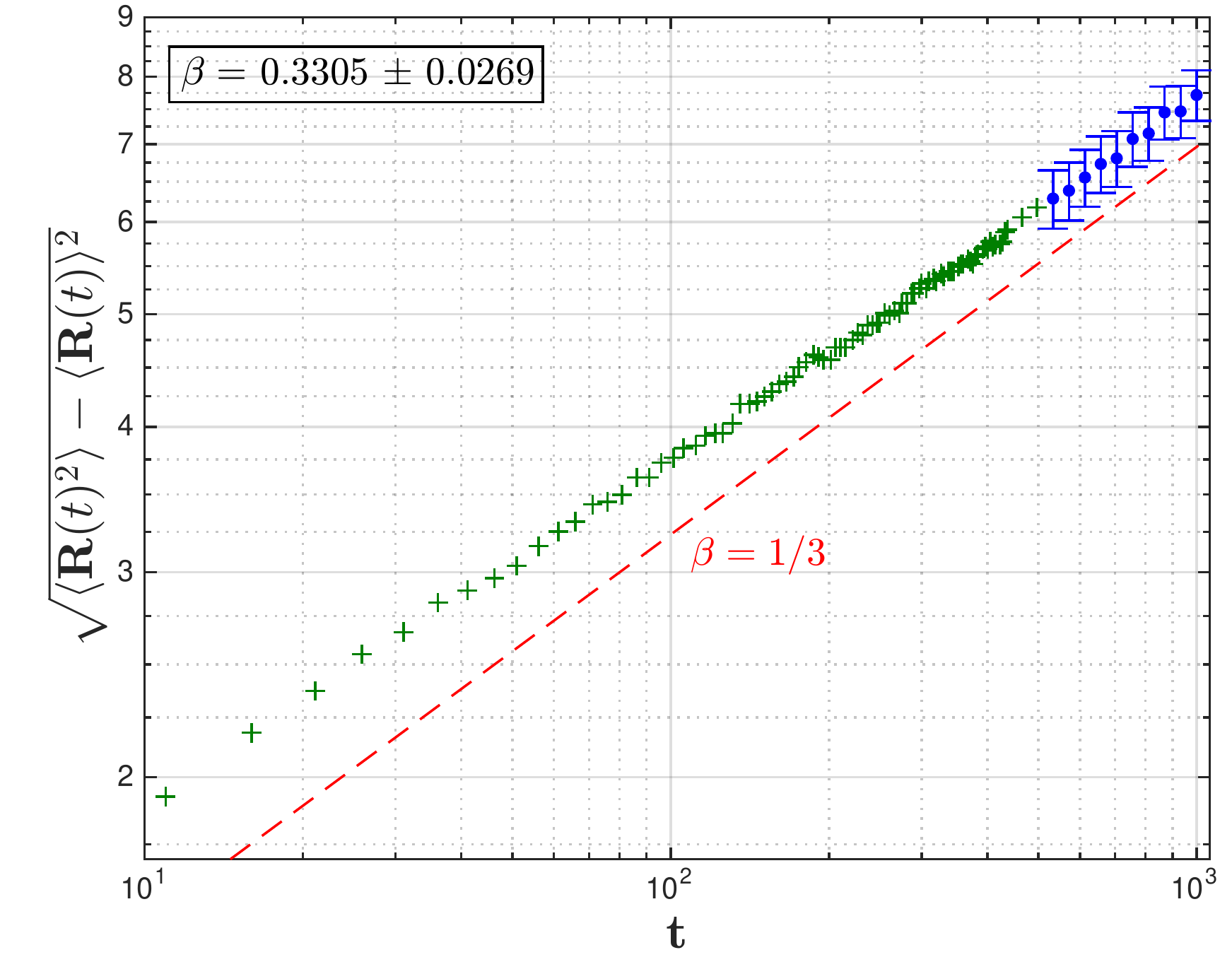}
  \includegraphics[trim = 40 190 70 210, clip = true,width=0.8\linewidth]{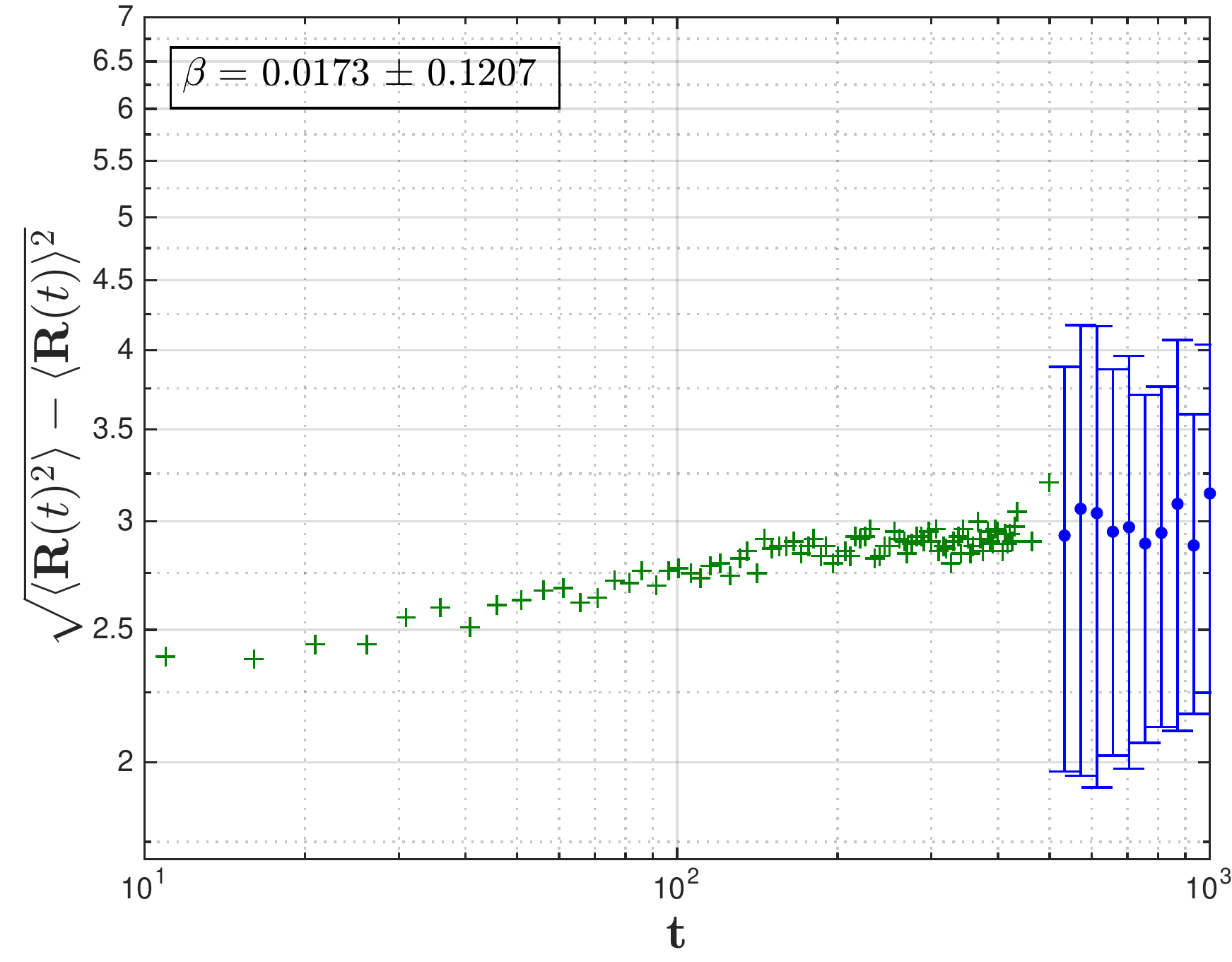}
 \caption{{\bf Fluctuation Exponent $\beta$:} We fit the profile of the evolving droplet for $q=2$ along the $\theta = \pm \pi/4$ directions (top) to extract the mean operator size and magnitude of the fluctuations about the mean. The fluctuations exhibit power-law growth with exponent $\beta = 0.3305 \pm 0.0269 $, consistent with the KPZ value $\beta = 1/3$. When fitting the profile along $\theta = 0$ (bottom), we observe no appreciable growth of the fluctuations; we argue in Appendix \ref{appendix_depinning} that this occurs for sufficiently large $q$ when the front's local normal vector is precisely aligned with a lattice axis (as a result of the specific circuit geometry).}
  \label{fig:Numerics}
\end{figure}

\subsection{Scaling of the OTOC in 2+1D}

We have already mentioned the basic consequence of KPZ growth, which is  the $t^{1/3}$ broadening of the front. But, unusually for a nontrivial fixed point, not only the exact critical exponents but also certain exact scaling functions are known for the growth of an interface in 1D \cite{johansson2000shape,prahofer2000universal,CalabreseLeDoussalRosso2010,Dotsenko2010,SasamotoSpohn2010A,SasamotoSpohn2010B,SasamotoSpohn2010C,GideonCorwinQuastel2011,CalabreseLeDoussal2011,ProlhacSpohn2011,LeDoussalCalabrese2012,ImamuraSasamoto2012,ImamuraSasamoto2013} (see  \cite{sasamoto20101,KriecherbauerKrug,CorwinReview,Halpin-Healy2015} for reviews). We can now apply this information to the OTOC to obtain scaling functions which we propose are generic.

To simplify things let us consider a case where lattice anisotropy is absent or very weak, so that the spreading operator is circular and the OTOC depends only on a radial coordinate and time. Weak anisotropy could certainly be engineered in an appropriate random circuit. More importantly, we conjecture that the scaling form below captures  universal scaling in realistic rotationally invariant many-body systems and field theories. 

For the growth of a droplet, the probability distribution of the interface radius is given by the GUE Tracy Widom distribution \cite{SasamotoSpohn2010A,GideonCorwinQuastel2011} (which has been  been observed experimentally in striking experiments on the growth of a turbulent domain in liquid crystals \cite{Takeuchi2011,Takeuchi2012}). Following convention we write 
\be
R(\theta,t) = v_B t + c t^{1/3} \chi(\theta,t),
\ee
where the non-universal constants $v_B$ and $c$ are of order one, and $\chi(\theta,t)$ is a random variable whose mean and variance are of order one at large times. Focussing on a fixed value of $\theta$, the cumulative probability distribution of $\chi$ at a fixed time is $t$--independent at large times and given by the Tracy Widom distribution $F_2$:
\be
P(\chi < s) = F_2(s).
\ee
Remarkably, this allows us to fix the full functional form of $\overline{\otc(\boldsymbol{x},t)}$ in two dimensions, in the case where lattice anisotropy is absent.  In polar coordinates $(r,\theta)$, and in the continuum, Eq.~\ref{eq:C_n_relation} is  
\be
\overline{\otc(r,\theta,t)} = \< \Theta\left[ R(\theta,t) - r \right] \>_\text{classical},
\ee
where $\Theta$ is the Heavyside step function. The right hand side is precisely the probability that $\chi$ is greater than $(r- v_{B}t) / c t^{1/3}$. We suppress the $\theta$ dependence since we are assuming rotational symmetry:
\be
\overline{\otc(r,t)} = 
1 - F_2 \lf   \f{r - v_B t}{c \, t^{1/3}} \ri.
\ee
The form of $\overline{\otc(r,t)}$ is shown in Fig. \ref{fig:TW}. The asymptotic behaviour near the trailing  edge, close to saturation, i.e. for $[v_{B}t-r]/ct^{1/3}\gg 1$, is \cite{Deift,Baik} 
\be
\overline{\otc(r,t)} =
1 - b_{1} \frac{c^{1/8}\,t^{1/24}}{|r-v_{B}t|^{1/8}}\exp\left[\frac{(r-v_{B}t)^{3}}{12 c^{3} t}\right] + \cdots
 \ee
 where $b_{1} = 2^{1/24}e^{\zeta'(-1)}$ with $\zeta'(-1) \approx -0.165$, the derivative of the Riemann zeta function. Near the leading edge, $[r-v_B t]/ct^{1/3}\gg 1$,
\be
\overline{\otc(r,t)} =  \frac{c^{3/2}\,t^{1/2}}{16\pi(r - v_{B}t)^{3/2}}\,\,\exp\left[-\frac{4(r - v_{B}t)^{3/2}}{3c^{3/2} t^{1/2}}\right] + \cdots.
\ee
The former asymptotic expansion of $F_2$ was achieved only recently \cite{Deift,Baik}.

\begin{figure}[b]
 \includegraphics[trim = 50 190 70 210, clip = true, width=0.7\linewidth, angle = 0.]{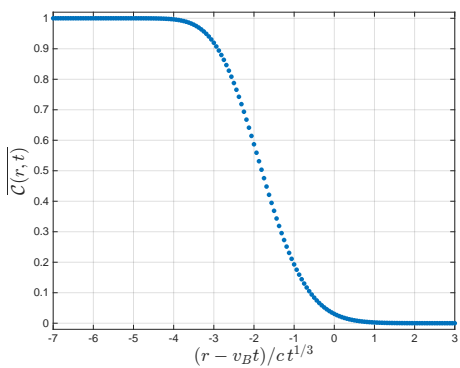}
 \caption{{\bf The OTOC in (2+1)D:} Plot of the front of the averaged OTOC $\overline{\otc(r,t)}$ in two spatial dimensions and in the absence of lattice anisotropy, as determined  from the exact expression in terms of the Tracy-Widom distribution in the main text.  }
  \label{fig:TW}
\end{figure}

One can also consider operator spreading with other initial conditions. For example we can initialize an operator in a half-plane so that $\otc(x,t)$ has a straight, rather than a circular, front. The scaling form for $\otc(x,t)$ will then be given by the Tracy Widom distribution of the Gaussian orthogonal ensemble, denoted $F_1$. The objects $F_1$ and $F_2$ are of fundamental importance in a broad range of mathematical and physical problems and it would be very interesting to see whether any of these connections shed light on operator growth.

\subsection{Scaling of the OTOC in $3+1$D and above}

The basic features of the 3+1D case are very similar to those in 2+1D. The KPZ equation extends to an interface of arbitrary dimensionality \cite{kpz_paper}. For the  the 3+1D quantum problem, the dimensionality of the interface is two and the critical exponent $\beta$ relevant to the width is  $\beta \simeq 0.240$ (Ref.~\cite{pagnani_parisi} and references therein). The analogue of $F_2$ which yields the universal form of $\overline{\otc}$ is not known analytically, but has been determined numerically \cite{Halpin-Healy2015}. Numerical simulations for the 3+1D random circuit, along the lines of those above, would be feasible.

Dimensions equal to or higher than 4+1 are of course inaccessible experimentally, but they are nonetheless interesting because in these high dimensionalities the KPZ equation yields a phase transition as a function of the strength of nonlinearity \cite{kpz_paper}. Both a rough phase, in which fluctuations grow as $t^\beta$ with $\beta>0$, and a smooth phase, where fluctuations remain of order one as $t\rightarrow \infty$, exist. It would be interesting to know whether both phases are accessible in appropriate many-body systems.

\subsection{Shape of the operator at late times}
\label{operator shape}

It is interesting to consider the shape of the spreading operator at late times --- 
i.e. the shape of the growing spatial region in which the OTOC  $\overline{\otc(x,t)}$ 
has already saturated to its late time value (to within an exponentially small correction).  
Rescaling distances by a factor of $t^{-1}$ gives a `droplet' of $O(1)$ size, 
which we expect to reach a fixed asymptotic shape. 
In this scaling limit the width of the front is negligible, 
so the front can be treated simply as a curve. 
What is its asymptotic shape?

At first glance, one might expect that the asymptotic shape is a circle in two spatial dimensions 
and a sphere in higher dimensions. 
For example, this would be expected 
if the OTOC satisfied a local nonlinear differential equation 
in which derivatives higher than 2 could be neglected, as 
 discrete lattice symmetries would ensure that 
such an equation had symmetry under continuous spatial rotations.
Instead, we argue that for many body systems on the lattice the shape of the operator 
is model-dependent and retains information about the discrete symmetries of the lattice, 
even at arbitrarily late times. For the random circuit model this follows immediately from the mapping to domain growth processes, for which anisotropy is a well-known feature \cite{Freche_1985, Hirsch_1986,PhysRevA.38.418,WolfWulff,krug1991solids}. Figure.~\ref{fig:Cluster} shows the shape of the droplet in the present model for various values of $q$.

\begin{figure*}
$\begin{array}{ccc}
 \includegraphics[trim = 61 215 40 190, clip = true, width=0.31\textwidth, angle = 0.]{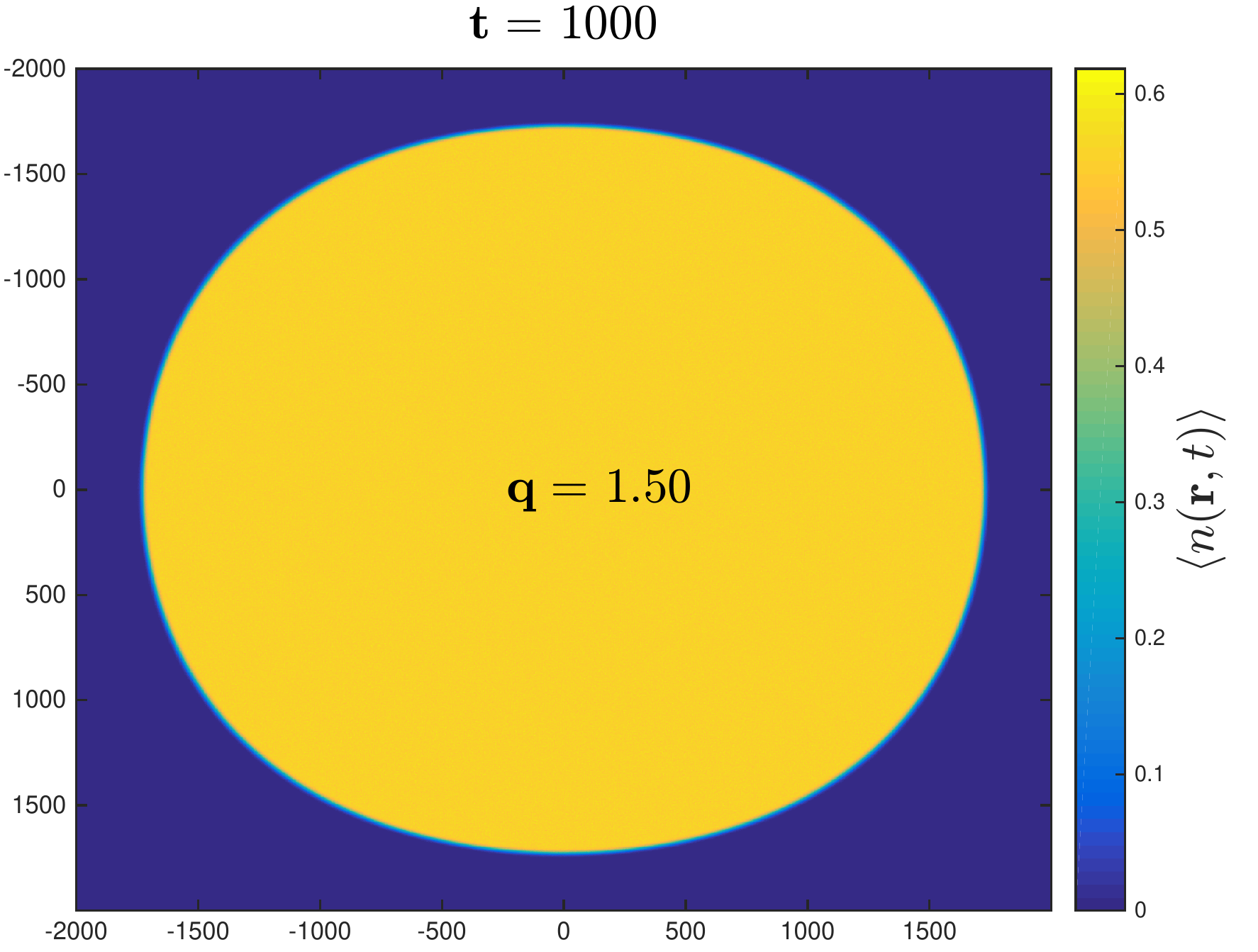} &
\includegraphics[trim = 61 215 40 190, clip = true, width=0.31\textwidth, angle = 0.]{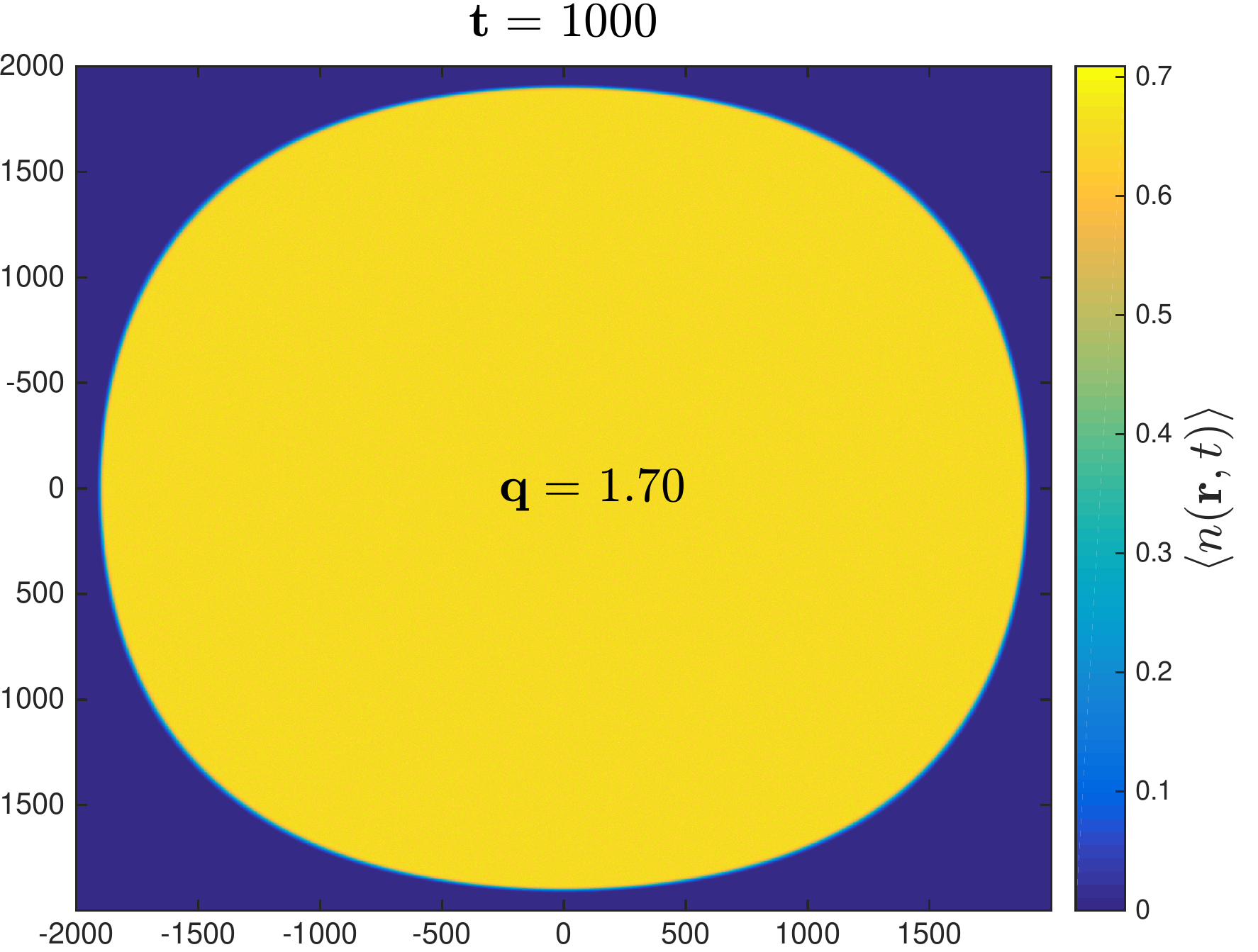} &
\includegraphics[trim = 61 215 40 190, clip = true, width=0.3\textwidth, angle = 0.]{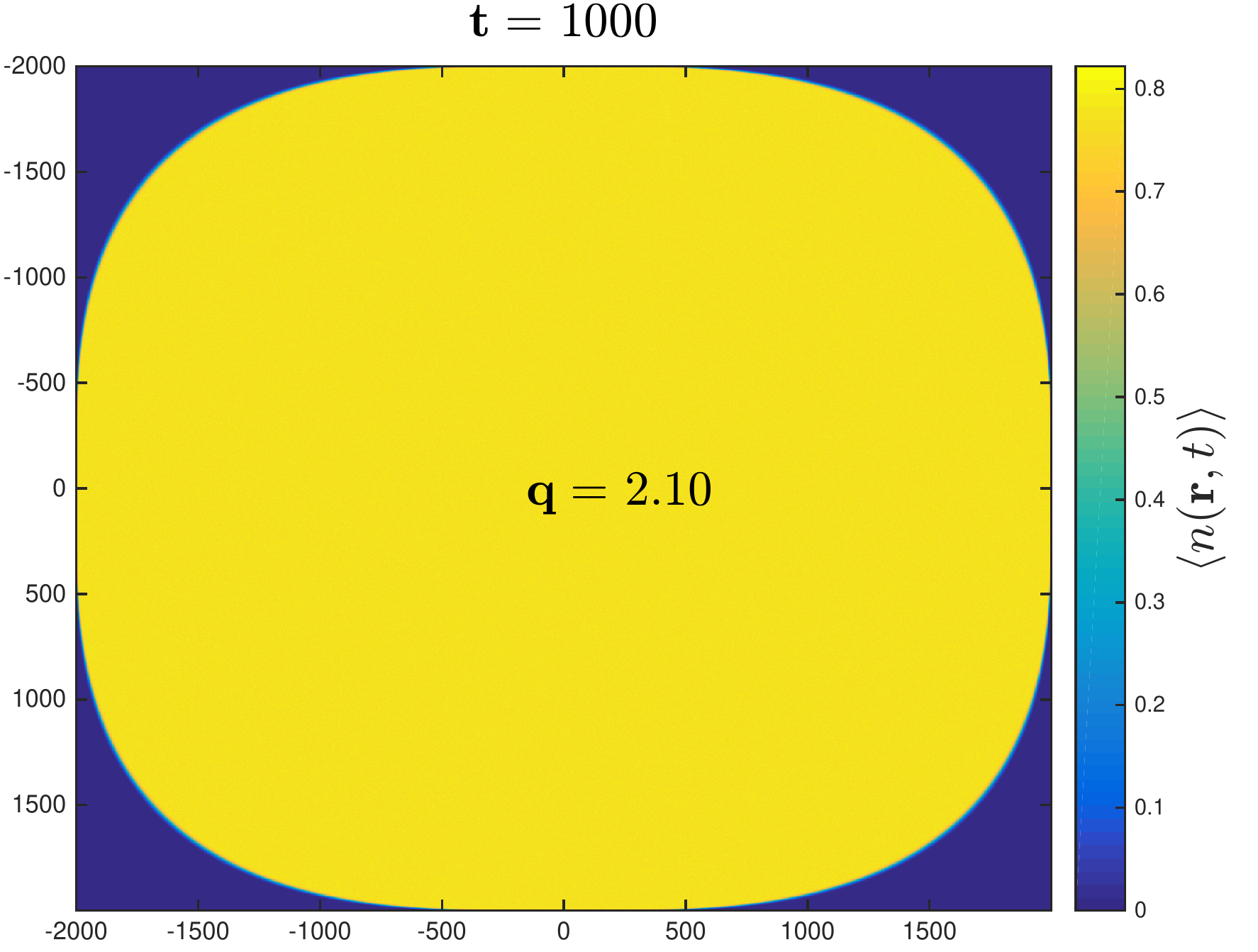}
 \end{array}$
 \caption{{\bf ``Faceting" of the Cluster:} Shown are the cluster shapes at fixed time $t=10^{3}$, for the indicated values of $q$.  When $q$ is sufficiently large (third panel), the cluster develops ``facets" along the $\theta = 0$, $\pi$ directions, where the normal growth speed is the maximum possible given the circuit geometry. The region shown is the naive light cone.}
  \label{fig:Cluster}
\end{figure*}

For concreteness consider the 2D case (similar statements hold in higher dimensions). 
The asymptotic droplet shape is described by a radius $R(\theta)$ depending on the polar coordinate $\theta$.  
Since the size of the operator is large at large times, 
the curvature of the front is parametrically small, 
except possibly at isolated $\theta$ values where $R(\theta)$ is not smooth. 
Away from such isolated points, 
the local \textit{velocity} of the front, in the direction of its normal vector, 
can depend only on the orientation of this local normal vector.
This dependence is captured by a velocity function $v_B(\phi)$, 
where $\phi=\phi(\theta)$ is the angle of the local normal vector to the $x$-axis.
A priori $v_B(\phi)$ is constrained only by lattice symmetry; for example on the square lattice 
\begin{align}
v_B(\phi) = v_0 + v_1 \cos 4\phi + v_2 \cos 8 \phi + \ldots.
\end{align}
It is evident that the asymptotic shape cannot be a circle except when $v(\phi)$ is a constant function. 
Since the front of the operator advances by $v(\phi) \mathrm{d}t$ in the normal direction $\hat n$,
the distance between the front and the origin grows by $v(\hat n) \mathrm{d}t / \hat n \cdot \hat r$,
which must be equal to $h(\hat r) \mathrm{d} t$. 
Expressing the normal vector in terms of $h$, one obtains
\begin{align}
v(\phi(\theta)) = \frac{h(\theta)^2}{\sqrt{h(\theta)^2 + (\partial_\theta h(\theta))^2}}
\end{align}
where $\phi(\theta)$ is the angle of the normal at polar position $\theta$ on the interface. This equation is solved by a geometrical construction described in Ref.~\cite{WolfWulff,krug1991solids}:
$h(\theta) = \min_\phi \f{v_B (\phi)}{\cos(\phi-\theta)}$.
When the effect of lattice anisotropy is weak 
(as is likely to be the case in many realistic situations when the relevant degrees of freedom are long-wavelength modes), we expect $v_B(\phi)$ to be a smooth, weakly varying function, and we may also solve for the shape perturbatively  in ${w(\phi) = v_B'(\phi)/v_B(\phi)}$, as described in  Appendix.~\ref{rthetaappendix}. Restoring the time dependence, 
\begin{align}
R(\theta) = v_B(\theta) \, t \, \exp \lf 
 - \f{1}{2} w(\theta)^2 + \f{1}{6} \partial_\theta w(\theta)^3 + \ldots
\ri.
\end{align}
However when $v_B(\phi)$ varies sufficiently strongly, the asymptotic shape $R(\theta)$ can  include sharp corners or straight segments on the boundary: in this regime the  perturbative solution above is no longer appropriate.

For many-body systems in continuous time we expect $v(\phi)$ to be analytic. In the present lattice model, $v(\phi)$ is analytic for $q<q_c$ ($q_c\lesssim 2$) while for $q>q_c$ this function is nonanalytic near $\phi=0$ as a result of the anomalous behaviour of a lattice-aligned front: $v(\phi) \simeq 2 + \text{const.}\, |\phi|$ \cite{krug1990growth}. This leads to flat facets near $\theta=0$ in the asymptotic shape \cite{krug1990growth}.  This change in the surface morphology as $q$ is varied is shown in Fig.~\ref{fig:Cluster}.

\begin{figure}[b]
 \includegraphics[trim = 0 0 0 0, clip = true, width=0.7\linewidth, angle = 0.]{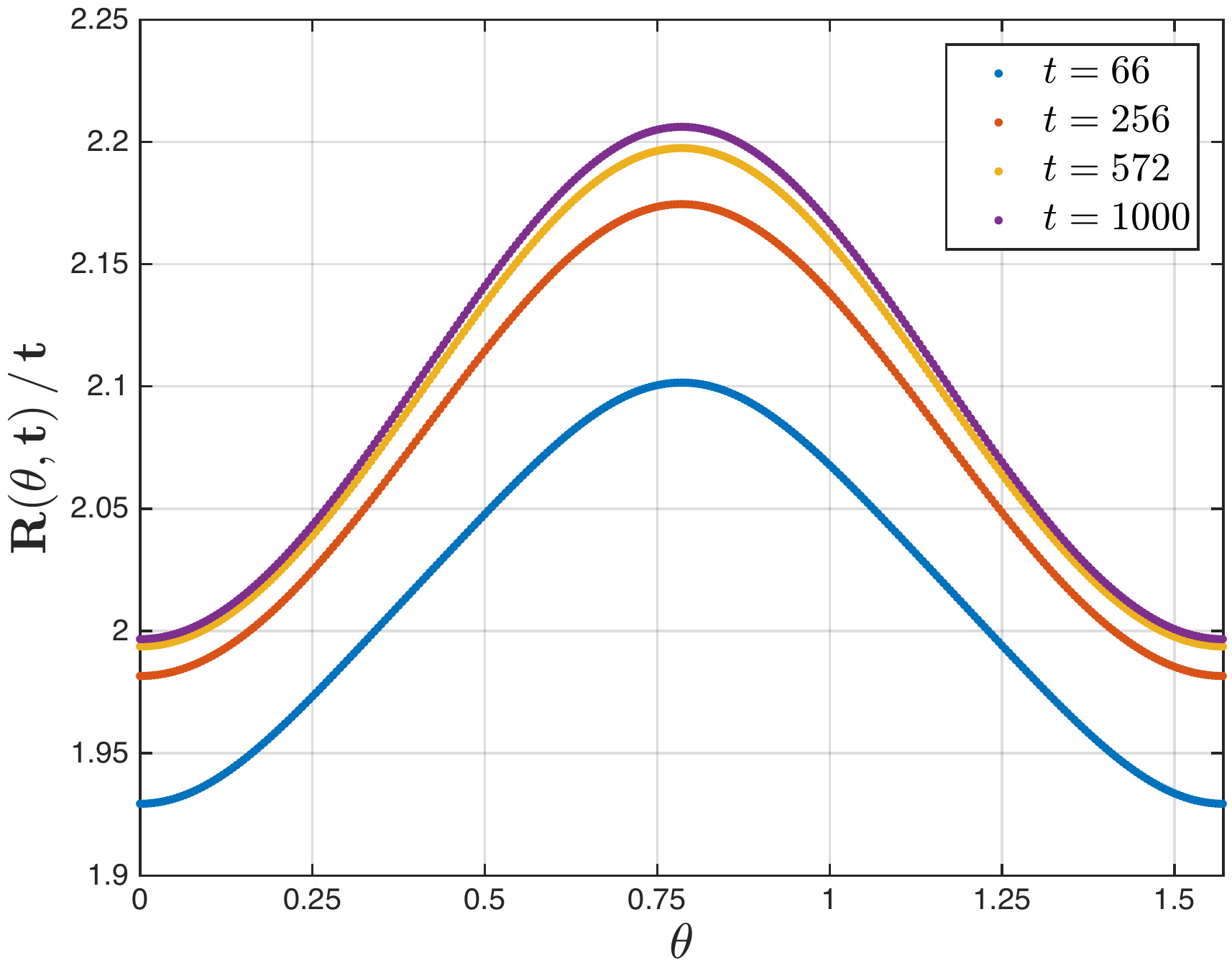}
 \caption{{\bf Anisotropy in the Cluster Profile:} Numerically determined anisotropy in the average shape of the 2D cluster $R(\theta, t)/t$, at the indicated times. The anisotropy in the cluster shape grows in time, and appears to asymptote to a non-trivial steady-state shape.}
  \label{fig:Anisotropy}
\end{figure}

For the random circuit model, it is straightforward to determine $v_B(\theta)$ in the extreme limit $q=\infty$ where growth becomes deterministic. The propagation of the front in this limit is similar to that of the `next nearest neighbour' deterministic Eden model introduced in Ref.~\cite{KrugSpohnDeterministic} and has the same nonanalytic angular dependence of the velocity\footnote{This is seen by following  the evolution of a front through the four layers comprising the time step.} \cite{KrugSpohnDeterministic}:
\be
v_{q=\infty}(\phi) = 2 \lf |\cos \phi| + |\sin \phi | \ri.
\ee
In this limit, the growing operator is simply a square.

Fig.~\ref{fig:Anisotropy} shows the angular dependence of the radius for the 2+1D random circuit dynamics at $q=2$, for several values of the time, showing a clear anisotropy. Note also that $R(\theta=0) \rightarrow 2t$ at late times. In the light of our 1D results where, for finite $q$, $v_B$ is always less than the speed associated with the naive lightcone, it is remarkable that in a  higher dimensional  circuit it is possible for the OTOC front to propagate at the maximal speed in some directions. However we emphasize that this effect relies on the specific discrete spacetime geometry.

 \subsection{Formal viewpoint}
  
Before returning to 1D, we restate the higher-dimensional results of Sec.~\ref{highDsetup} in a more formal language which parallels our discussion in 1D. We introduce a density on clusters, $C$, where $C$ is a collection of sites: 
\ba
\rho(C) &= \sum_{\substack{\mathcal{S}; \,\, \mathrm{supp}(\cS) = C}} a_\mathcal{S}^2,
&
\sum_C \rho(C) & = 1.
\end{align}
Here $\mathrm{supp}(\cS)$ is the support of $\mathcal{S}$. After coarse-graining, we can represent $C$ by a closed surface of spherical topology, namely the boundary of the coarse-grained cluster. Therefore $\rho(C)$ is the natural analogue of the `endpoint density' $\rho(x)$ in 1D. The surface growth picture implies that the effective dynamics of ${\rho}(C)$ are the dynamics of the probability distribution of a growing interface. Therefore, when this is KPZ, $\rho(C)$ satisfies the Fokker-Planck equation corresponding to the KPZ equation. We will discuss this further elsewhere.

\section{Exact calculation of OTOC in `spacetime' picture}
\label{sec:exact calculation}

We now given an analytical treatment of the OTOC from a `spacetime' point of view. 
This leads to connections with domain walls in an effective Ising model.
Similar Ising degrees of freedom have appeared 
in work on random tensor networks~\cite{HaydenEtAl2016}.
Here the effective Ising model looks complicated at first sight, 
but turns out to be much simpler than those encountered in random (non-unitary) tensor networks,
due to special structure arising from unitarity. 

This spacetime picture may be much more generalizable than the dynamical point of view above.
In Sec.~\ref{sec:purity} we will use it to calculate an entanglement--related quantity.
In the future, we hope that the tools introduced in this section 
will be generalizable to higher moments of the OTOC which capture fluctuations ($\overline{\otc^2}$ etc.), 
or higher powers of the commutator, 
or to a direct calculation of the von Neumann entropy.

Our exact result for the OTOC for arbitrary $x$ and $t$ (not necessarily large) is
\begin{align}
\overline{\otc}(t,x) =&  \label{eq:exactsolution}\\
 (1-\xi) ~&g\left(t-1,\frac{t-x-3}{2},p\right)~ g\left(t-1,\frac{t+x-3}{2},p\right) \nonumber \\
 +\,\xi ~&g\left(t-1,\frac{t-x-1}{2},p\right)~ g\left(t-1,\frac{t+x-1}{2},p\right), \nonumber
\end{align}
where
\begin{align}
 p &= \frac{1}{q^2+1}, & \xi &= \frac{q^4}{q^4 -1},\nonumber 
 \end{align}
 and
 \begin{align}
 g(n,a,p) &= \sum_{k=0}^a \binom{n}{k} (1-p)^{n-k} p^k .\nonumber
\end{align}
There are no approximations in Eq.~\eqref{eq:exactsolution}.
Approximating $g$ by the cumulative density function 
${\Phi(y) = \frac{1}{\sqrt{2\pi}}\int_{-\infty}^y e^{-x^2 /2} \rd x}$ 
of the Gaussian distribution,
we reproduce Eqs.~\ref{butterfly speed},~\ref{sigma fmla},~\ref{eq:exact_result_for_1D_summary} above.
This approximation is valid when $t$ is large.

Although the spin chain is spatially infinite in both directions and so is our quantum circuit,
the time evolved operator $U^\dag(t) X_0 U(t)$ is supported only on the interval $[-t,t-1]$ of length $2t$.
Therefore, it suffices to consider  an observable $Y$ inserted in this interval, 
and our correlator becomes the trace of a $q^{2t} \times q^{2t}$ matrix.
The infinite temperature Gibbs state reduces to the identity matrix divided by the dimension $q^{2t}$.
Expanding the commutator, we see
\begin{align}
 -&\half \Tr  \rho_\infty ( [U^\dagger(t) X_0 U(t), Y_x] )^2  \notag \\
&= q^{-2t} \lf \Tr[ U^\dagger(t) X_0^2 U(t) Y_x^2 ] - \Tr [ X_0(t) Y_x X_0(t) Y_x] \ri \notag \\
&=: q^{-2t} \Tr[ U^\dagger(t) X_0^2 U(t) Y_x^2 ] - F 
\label{eq:introductionOfF}
\end{align}
The Haar average of the first term is easy to evaluate.
The observable $X_0^2$ is conjugated by a unitary $U_{-1,0}(1,0)$
 and after taking the Haar average becomes proportional to the identity.
The constant of proportionality is fixed by the trace-preserving condition.
By the normalization convention, $\Tr X_0^2 = q = \Tr Y_x^2$,
and therefore the Haar average of the first term is equal to $q^{-2t} \Tr I = 1$.
The second term $F$ contains all the complexity.

Observe that the local unitaries form a square lattice that is rotated by $45^\circ$.
It is thus natural to introduce null coordinates as
\begin{align}
u &:= (t + x +1)/2, & v &:= (t - x+1)/2 ,\\
t &= u+v-1, & x &= u - v .
\end{align}
Due to the cylic property of the trace,
the only unitaries in the circuit that could affect the correlator
are those in the intersection (a rectangle) 
of the future light cone of $X_0$ and the past light cone of $Y_x$. 
From now on, let us use $\ell_u$ and $\ell_v$ 
to denote the linear sizes of this intersection along $u$- and $v$-direction, respectively.
There are $\ell_u \ell_v$ local unitaries contained in the intersection of the lightcones.

\subsection{Reduction to Ising spins}
\label{sec:reduction_Ising_spins}

For each local unitary $U$ the expression $F$ contains two $U$s and two $U^\dagger$s.
We will see that averaging over the local unitaries allows us to 
express $F$ as a partition function for a set of classical Ising spins. 
To see why such Ising spins arise, 
consider the standard expression for the Haar average of a single unitary matrix in $\mathbb U(n)$:
\begin{align}
\avg_{U \in \mathbb U(n)}  ~ U_{a'a} & U^*_{b'b} U_{c'c} U^*_{d'd} =  \\
 \frac{1}{n^2-1} & \Big[
\delta_{a'b'}\delta_{c'd'} \times \delta_{ab} \delta_{cd} 
+
\delta_{a'd'}\delta_{b'c'}\times\delta_{ad}\delta_{bc} 
\notag \\
-&\frac{1}{n} (
\delta_{a'b'}\delta_{c'd'}\times\delta_{ad}\delta_{bc}
+
 \delta_{a'd'}\delta_{b'c'}\times \delta_{ab}\delta_{cd}
)
\Big].\notag
\end{align}
(See Appendix~\ref{sec:HaarAverage} for a self-contained derivation of this formula.)  
It is convenient to regard the above expression 
as a matrix whose rows are labelled by the multi-index $(a', b',c',d')$ 
and whose columns are labelled by $(a,b,c,d)$. 
Note that two types of contraction appear for the unprimed indices, namely  $\delta_{ab} \delta_{cd}$ and $\delta_{ad} \delta_{bc}$, and similarly for the primed ones. 
Correspondingly,  in bra--ket notation the above matrix can be written 
in terms of two vectors which we denote $\ket{I_\uparrow}$ and $\ket{I_\downarrow}$ 
(the reason for the notation will become clear below):
\begin{align} 
&\overline{U \otimes U^* \otimes U \otimes U^*}=   \label{eq:generalHaarAverage}  \\
& \frac{n^2}{n^2 -1} 
\left( 
\left( \ket{I_\uparrow}\bra{I_\uparrow} + \ket{I_\downarrow}\bra{I_\downarrow} \right) 
- \frac{1}{n} \left( \ket{I_\uparrow}\bra{I_\downarrow} + \ket{I_\downarrow}\bra{I_\uparrow} \right) 
\right).\notag
\end{align}
In the natural basis, these vectors are
\begin{align}
\langle a b c d \ket{I_\uparrow}& = \frac 1 n \delta_{ab}\delta_{cd}, 
&
\langle a b c d \ket{I_\downarrow} &= \frac 1 n \delta_{ad} \delta_{cb}.
\end{align}
With this definition, we may write (\ref{eq:generalHaarAverage}) as
\begin{align} 
\overline{U \otimes U^* \otimes U \otimes U^*} &= \f{n^2}{n^2-1} \sum_{s, s' = \uparrow, \downarrow} 
w(s,s') \ket{I_s} \bra{I_{s'}},
\label{eq:generalHaarAverageIsing}
\end{align}
with
\begin{align}
w(s,s') &= \begin{cases}
1 & \text{if $s=s'$}\\
-\frac 1 n & \text{if $s\neq s'$.}\\
\end{cases}
\end{align}
We see that the unitary may be associated with a \emph{pair} of classical Ising `spins' $s$ and $s'$. 

For the application of interest to us 
the unitaries are two-site unitaries acting on the $q^2$-dimensional space 
associated with spins $i$ and $i+1$. 
In this case it is easy to see that 
\begin{align}
\ket{I_\uparrow} &= \ket{\uparrow}_i \ket{\uparrow}_{i+1},
&
\ket{I_\downarrow} &= \ket{\downarrow}_i \ket{\downarrow}_{i+1},
\label{eq:separateKets}
\end{align}
with 
\begin{align}
\langle \alpha \beta\gamma\delta \ket{\uparrow}& = \frac 1 q \delta_{\alpha \beta}\delta_{\gamma\delta}, 
&
\langle  \alpha \beta\gamma\delta \ket{\downarrow} &= \frac 1 q \delta_{\alpha \delta} \delta_{\gamma\beta},
\end{align}
and now $\alpha,\ldots, \delta$ run over the $q$ basis vectors 
associated with a given spin. 
The vectors $\kup$ and $\kdown$ have norm 1 and satisfy
\begin{align}
	\braket{\uparrow|\downarrow} =
	 \frac{1}{q} .
\end{align}

When we consider $F$, the spins arising from each unitary in the circuit will form an interacting network. 
The interactions between spins from the same unitary will be given by $w(s,s')$, 
while the interactions between spins from different unitaries 
will arise from the inner products of kets $\ket{\uparrow, \downarrow}_i$ associated with a given spin.

As a final piece of notation, we generalize the definition of $\ket{I_\uparrow}$ and $\ket{I_\downarrow}$. 
Given an operator $\mathcal{O}_{ab}$ on the $n$-dimensional space, 
we define $n^4$-dimensional vectors $\ket{\mathcal{O}_\uparrow}$ and $\ket{\mathcal{O}_\downarrow}$ via
\begin{align}
\langle a b c d \ket{\mathcal{O}_\uparrow}& = \frac{\mathcal{O}_{ab}\mathcal{O}_{cd}}{\Tr \mathcal{O}\mathcal{O}^\dag}, 
&
\langle a b c d \ket{\mathcal{O}_\downarrow} &=\frac{\mathcal{O}_{ad} \mathcal{O}_{cb}}{\Tr \mathcal{O}\mathcal{O}^\dag}.
\end{align}
Choosing $\mathcal{O}$ to be the identity gives the vectors $\ket{I_{\uparrow,\downarrow}}$.

Before we evaluate $F$ for arbitrary $\ell_u$ and $\ell_v$,
let us consider the simplest case where $\ell_u=\ell_v=1$.
\begin{align}
\overline{ F} 
&= q^{-2} \avg_U \Tr U X U^\dagger Y U X U^\dagger Y \notag\\
& = q^{-2} \avg_U  \hspace{-5pt}  \sum_{i_1,\ldots,i_8 = 1}^{q^2}  \hspace{-5pt} 
 U_{i_1 i_2} X'_{i_2 i_3} U^*_{i_4 i_3} Y'_{i_4 i_5} U_{i_5 i_6} X'_{i_6 i_7} U^*_{i_8 i_7} Y'_{i_8 i_1}\notag \\
& = q^{-2} \avg_U\hspace{-5pt}  \sum_{i_1,\ldots,i_8 = 1}^{q^2}  \hspace{-5pt}
 Y'_{i_8 i_1} Y'_{i_4 i_5} U_{i_1 i_2} U^*_{i_4 i_3} U_{i_5 i_6} U^*_{i_8 i_7} X'_{i_2 i_3} X'_{i_6 i_7} \notag \\
& = q^{2} \bra{(I\otimes Y)_\downarrow} \overline{U \otimes U^* \otimes U \otimes U^*} \ket{(I\otimes X)_\uparrow}.
 \label{eq:expressionF}
\end{align}
In the second line, $X' = I \otimes X$ and $Y' = I \otimes Y$.
The third line is a trivial rearrangement of the second,
and the fourth employs the formal correspondence between matrices and normalized vectors introduced above.

The Haar average of the tensor product of four unitaries is given by \eqref{eq:generalHaarAverage} with $n=q^2$.
To complete the evaluation of $\overline{F}$ we observe that 
{$\braket{(I\otimes X)_\uparrow|(I_{q^2})_\uparrow} = q^{-4} (\Tr X)^2 (\Tr I_q)^2 = 0$},
{$\braket{(I\otimes X)_\uparrow|(I_{q^2})_\downarrow} = q^{-4} (\Tr X^2) (\Tr I_q) = q^{-2}$},
{$\braket{(I\otimes Y)_\downarrow|(I_{q^2})_\uparrow} = q^{-4} (\Tr Y^2) (\Tr I_q) = q^{-2}$}, and
{$\braket{(I\otimes Y)_\downarrow|(I_{q^2})_\downarrow} = q^{-4} (\Tr Y)^2 (\Tr I_q)^2 = 0$}.
This gives $\overline{F} = -1/(q^4-1)$.

When $\ell_u,\ell_v > 1$, 
we map the layout of local unitaries to a partition function for the spins $s$, $s'$ in Eq.~\ref{eq:generalHaarAverageIsing}. 
To facilitate the mapping, we decompose the input bra $\bra{I_{s'}}$ and  output ket $\ket{I_s}$ into  separate `legs'
corresponding to the two physical spins, as in Eq.~\eqref{eq:separateKets},
\begin{align} 
\overline{U \otimes U^* \otimes U \otimes U^*} &= \frac{q^4}{q^4-1} \sum_{s, s' = \uparrow, \downarrow} 
w(s,s') \ket{s}\ket{s} \bra{s'}\bra{s'}.
\label{eq:DecomposedHaarAverage}
\end{align}
Similarly, the vectors encountered above for the case $\ell_u=\ell_v=1$ can be decomposed as 
$\ket{(IX)_\uparrow} = \kup \ket{X_\uparrow}$
and $\ket{(IY)_\downarrow} = \kdown \ket{Y_\downarrow}$,
which satisfy
\begin{align}
 \braket{\uparrow|Y_\downarrow} &= q^{-2} \Tr Y^2 = \frac{1}{q}, & \braket{\downarrow|Y_\downarrow} &= q^{-2} (\Tr Y)^2 = 0, \nonumber \\
 \braket{\uparrow|X_\uparrow} &= q^{-2} (\Tr X)^2 = 0, & \braket{\downarrow|X_\uparrow} &= q^{-2} \Tr X^2 = \frac{1}{q}.
\end{align}
The expression Eq.~\eqref{eq:expressionF} is now depicted as in Fig.~\ref{fig:uv1expressionF}.
\begin{figure}
 \includegraphics[trim = 0 0 0 0, clip = true, width=0.3\linewidth, angle = 0.]{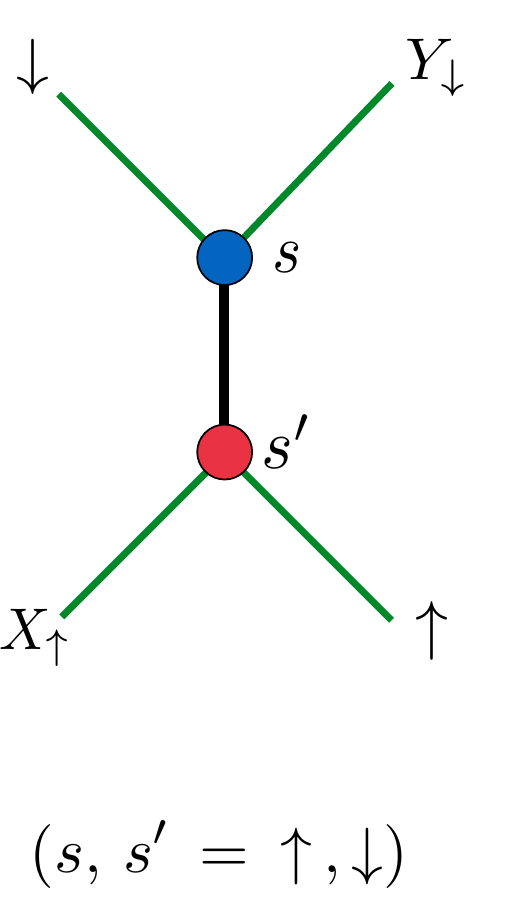}
 \caption{Elementary tensor for computation of $\overline F$.
The  boundary conditions in the top-left figure are for $\ell_u = \ell_v = 1$. }
\label{fig:uv1expressionF}
\end{figure}

It is now clear that for general $\ell_u$, $\ell_v$ we may regard the array of unitaries 
as a tensor network composed of tensors of the form \eqref{eq:DecomposedHaarAverage}. 
The boundaries of this tensor network 
--- i.e. the external legs of the array of $\ell_u \times \ell_v$ unitaries ---  
involve inner products with fixed vectors. 
Two of the boundary legs are dressed with $q \ket{X_\uparrow}$ and $q \bra{Y_\downarrow}$; 
see Fig.~\ref{fig:uv1expressionF}. 
Apart from these, the external legs on the top boundary are dressed by states $q \bdown$, 
while those on the bottom boundary are dressed with $q \kup$.
In addition $\overline{F}$ includes an overall dimension factor $q^{-2t}$ 
coming from the infinite temperature Gibbs state.
For convenience, we absorb the overall dimension factor $q^{-2t}$ into the vectors on the lower boundaries;
these vectors are taken to be normalized, 
whereas the boundary bras in the top boundaries have norm $q$.

We may now interpret $\overline{F}$ as a partition function for the Ising spins $s_{u,v}$ and $s'_{u,v}$ 
which according to Eq.~\eqref{eq:DecomposedHaarAverage} are associated with 
the unitary at position $(u,v)$. 
These spins take the values $\uparrow$, $\downarrow$. 
The weight associated with the `bond' between $s_{u,v}$ and $s'_{u,v}$ 
comes from the single--unitary Haar average and is $\frac{q^4}{q^4-1}$ if $s_{u,v} = s'_{u,v}$,
and $\frac{-q^2}{q^4-1}$ if $s_{u,v} \neq s'_{u,v}$.
The leg of the tensor network connecting the unitary at $(u,v)$ to that on its lower right at $(u,v-1)$
yields an interaction between $s'_{u,v}$ and $s_{u,v-1}$ 
which comes simply from the inner product  $\braket{s'_{u,v-1} | s_{u,v}}$.
This gives weight $1$ if $s'_{u,v} = s_{u,v-1}$,
and weight $\frac{1}{q}$ if $s'_{u,v} \neq s_{u,v-1}$.

\begin{figure}
 \includegraphics[trim = 0 0 0 0, clip = true, width=0.7\linewidth, angle = 0.]{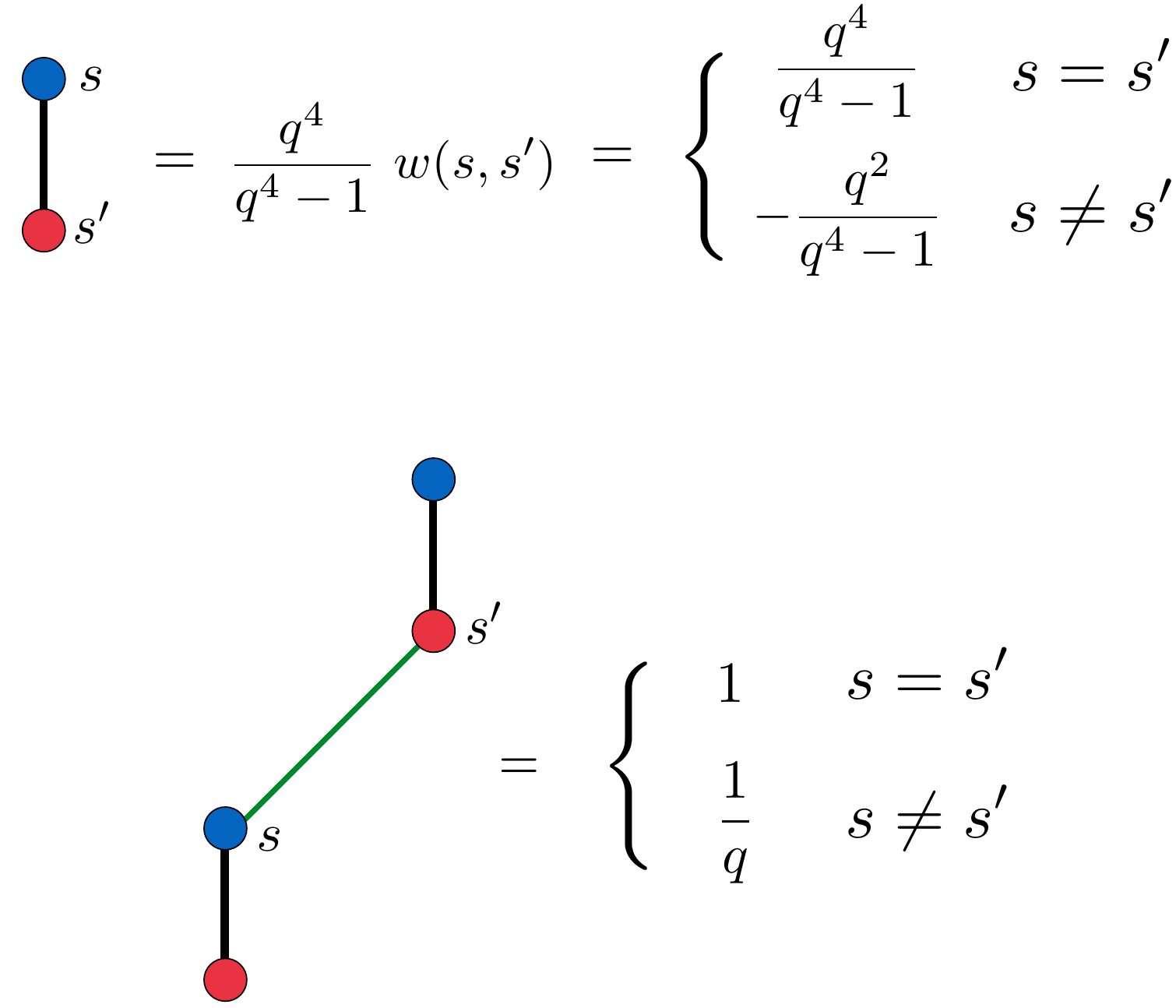}
 \caption{Weights due to the interaction between adjacent Ising variables arising from the same unitary (top) and from unitaries at adjacent time steps (bottom). 
After integrating out the `bra' Ising variable $s'$,
we obtain the weights shown in Fig.~\ref{fig:wtTable}. }
\label{fig:isingweights}
\end{figure}


We have thus mapped the Haar average of the out-of-time correlator 
to a partition function for Ising degrees of freedom 
(with the $q$--dependence residing in the interactions on the bonds). 
At first sight, this may appear to be a formidable problem. 
Note in particular that some configurations have negative weight. 
However, a simplification is possible, as a result of the unitarity of the underlying dynamics. 
A hint that a simplification is possible comes from the fact that the expression for $F$, 
Eq.~\eqref{eq:introductionOfF}, 
becomes trivial if one of the operators $X_0$ and $Y_x$  is the identity operator. 
In the Ising language this corresponds only to a slight change of boundary conditions.

The simplification is effected by integrating out the `bra' variable $s'_{u,v}$ from each unitary. 
This generates a three-spin interaction among the `ket' variables $s_{u,v}$, $s_{u-1,v}$, and $s_{u,v-1}$.
The calculation is straightforward and yields the table of weights in Fig.~\ref{fig:wtTable}.
For example, if $s_{u,v} = s_{u,v-1} = s_{u-1,v}$, then the weight is
$\frac{q^4}{q^4-1} \cdot 1 \cdot 1 + \frac{-q^2}{q^4-1} \cdot \frac 1 q \cdot \frac 1 q = 1$.


\begin{figure}[t]
 \includegraphics[trim = 0 0 0 0, clip = true, width=0.8\linewidth, angle = 0.]{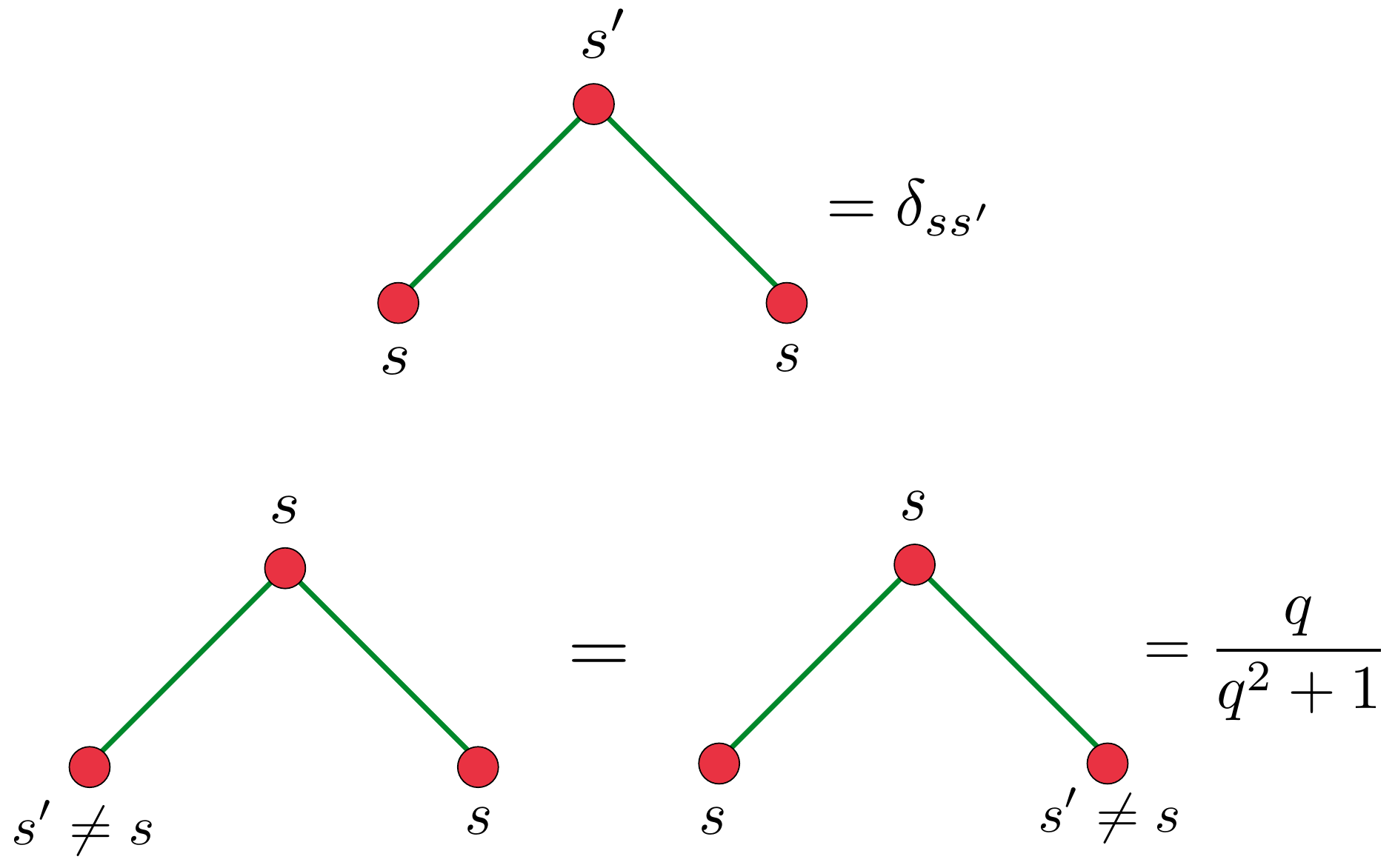}
\caption{Weights for the 3-body interaction which arises after integrating out half of the Ising variables (the bra variables).}
\label{fig:wtTable}
\end{figure}

The fact that the weight is zero for two of the configurations means 
that only a very restricted subset of Ising configurations are allowed. 
We will show that these can be summed exactly by viewing the configurations in terms of domain walls.

Let us specify the new boundary conditions. 
The above rules apply along the bottom boundaries due to our normalization convention for the boundary kets,
except for the site where the observable ket $\ket{X_\uparrow}$ is dangling.
\begin{align}
 \raisebox{4ex}{\xymatrix{
  & \uparrow & \\ 
  \uparrow_\text{fixed} \ar@{--}[ru]& & X_\uparrow \ar@{--}[ul] 
 }}
 = \frac{-1}{q^4 -1} , \\
 \raisebox{4ex}{\xymatrix{
  & \downarrow & \\ 
 \uparrow_\text{fixed} \ar@{--}[ru] &  & X_\uparrow \ar@{--}[ul] 
 }}
 = \frac{q^2}{q^4 -1}
 \label{eq:wtFromX}
\end{align}
The top boundary bras, which have norm $q$,
follow the rule that
\begin{align}
 \raisebox{4ex}{  \xymatrix{
 \downarrow_\text{fixed} \ar@{--}[d] \\
 \uparrow
 }}
 = 1,\qquad
  \raisebox{4ex}{\xymatrix{
 \downarrow_\text{fixed} \ar@{--}[d] \\
 \downarrow
 }}
 = q, \qquad
 \raisebox{4ex}{\xymatrix{
Y_\downarrow \ar@{--}[d] \\
\uparrow
 }} 
 = 1, \qquad
 \raisebox{4ex}{\xymatrix{
Y_\downarrow \ar@{--}[d] \\
\downarrow
 }} 
 = 0 
 \label{eq:upperBoundaryRule}
\end{align}

\subsection{Partition function for two directed paths}

Now the problem is reduced to a partition function of Ising variables 
with the three-body interaction and the boundary interaction. 
We first simplify the partition function by relating it to 
one with modified boundary conditions as follows. 
We denote the weight of a given configuration by $W_{X,Y}(s)$, 
where the subscripts indicate the dependence on the boundary conditions induced by the operators $X$ and $Y$. 
Because of the last rule in Eq.~\eqref{eq:upperBoundaryRule}, 
the spin at the site where $Y_x$ is attached --- null coordinate $(\ell_u,\ell_v)$ --- 
has to be $s_{\ell_u,\ell_v} = \,\uparrow$.
As a result we can replace $Y_\downarrow$ with $\downarrow_\text{fixed}$, 
which according to Eq.~\eqref{eq:upperBoundaryRule} gives the same weight when $s_{\ell_u,\ell_v} = \,\uparrow$.
Let us denote the weight of a configuration $s$ under this modified top boundary condition by $W_X(s)$,
dropping the subscript $Y$. 
We may then write the desired quantity $\overline{F} = \sum_s W_{X,Y}(s)$ as
\begin{align}
\overline{F}
 &=\sum_{\substack{s~: \\ s_{\ell_u,\ell_v} = \uparrow}}W_X(s)\nonumber \\
 &= \sum_s W_X(s) - \sum_{\substack{s~: \\ s_{\ell_u,\ell_v} = \downarrow}} W_X(s) .
\end{align}
We claim that the first term is equal to 1, and thus
\begin{align}
-\half  \Tr \rho_\infty  \overline{[U(t) X_0 U(t)^\dagger, Y_x]^2}
& = 1- \overline{ F} \\
& = \sum_{\substack{s~: \\ s_{\ell_u,\ell_v} = \downarrow}} W_X(s).
\end{align}
The claim can be shown in two ways.
First, the out-of-time correlator ${1-\overline{F}}$ must vanish if the operator $Y$ is replaced by the identity. The boundary  vector $\ket{Y_\downarrow}$ then becomes $\ket{\downarrow}$, and the partition function for $\overline{F}$ becomes precisely $\sum_s W_X(s)$. Therefore ${1-\sum_s W_X(s)=0}$. The other way to show the claim is 
by directly integrating out the Ising variables inductively,
starting from the top line
with respect to the all--$\downarrow$ boundary condition along top boundary.
This is a nontrivial consistency check on our reduction.

Now we focus on $W_X(s)$ with the variable at null-coordinate $(\ell_u,\ell_v)$ fixed to be $\downarrow$.
If the bottom variable where $X$ is attached is $\uparrow$, then the second rule in Fig.~\ref{fig:wtTable}
together with the boundary condition along the bottom boundary
dictates that all the bulk variables be $\uparrow$.
This cannot be fulfilled for the top-right variable, 
implying that the weight is zero.

Hence, we have fixed two Ising variables in the bulk to be $\downarrow$ where the observables $X_0$ and $Y_x$ are attached.
Let us think of domain walls instead of spins. 
The key point is the first rule  in Fig.~\ref{fig:wtTable}, which leads to the domain walls
being directed, drastically simplifying the partition function.
(Since the top boundaries have a different interaction,
let us speak of domain walls to mean 
disagreeing edges in the south and west of a square encompassing an Ising variable,
with the two bottom edges of the square at null coordinate $(1,1)$ excluded.)
If we follow a domain wall from the  top to the bottom,
it should always go down-left or down-right.
This implies that there are two non-intersecting domain walls 
extending from the bottom to the top boundary. The starting vertices of the right and left domain walls have null coordinates $(1,0)$ and $(0,1)$, respectively.
Each domain wall has length $t-1$,
giving the weight factor 
\begin{align}
 \left(\frac{q}{q^2+1}\right)^{t-1}
 \label{eq:wtFromDomainWall}
\end{align}
from the weight table in Fig.~\ref{fig:wtTable}.

A domain wall can be deformed without changing the weight to the partition function.
There is essentially one local deformation of the domain wall.
One can easily see from Fig.~\ref{fig:wtTable}
that the weights of the two configurations in Fig.~\ref{fig:shiftingDomain} are the same.
\begin{figure}
\begin{align*}
\fbox{\raisebox{6ex}{\xymatrix@!0{
\downarrow & & \downarrow && \uparrow && \uparrow \\
& \downarrow && \downarrow &\ar@{=}[ul]& \uparrow & \\
\downarrow && \downarrow &\ar@{=}[ur]& \uparrow && \uparrow
}}}
& &\\
\fbox{\raisebox{6ex}{\xymatrix@!0{
\downarrow & & \downarrow && \uparrow && \uparrow \\
& \downarrow &\ar@{=}[ur]& \uparrow && \uparrow & \\
\downarrow && \downarrow &\ar@{=}[ul]& \uparrow && \uparrow
}}}
\end{align*}
\caption{
Two configurations with the same weight. 
This implies that the domain walls can fluctuate freely in the bulk.
}
\label{fig:shiftingDomain}
\end{figure}
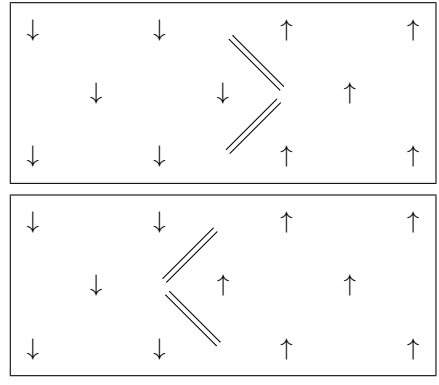
However, the end position of the domain wall at the top boundary 
{\em does} affect the weight,
and we will need to count the number of domain walls for each ending position.
For the right domain wall, the end position can be at a vertex $(t-v,v)$ in null coordinates
for some $v = 0,1,\ldots, \ell_v -1$.
Likewise, the end position of the left domain wall can be $(u,t-u)$ for some $u = 0,1, \ldots, \ell_u-1$.
The weight from the top boundary interaction is then 
\begin{align}
q^{2t-2u - 2v}
\label{eq:wtTopBoundary}
\end{align}
by Eq.~\eqref{eq:upperBoundaryRule}.

It remains to count the number of domain walls given their end positions.
The right domain wall connects $(1,0)$ to $(t-v,v)$
while the left domain wall connects $(0,1)$ to $(u,t-u)$,
with the constraint that they must not intersect.
To handle the constraint, we use a reflection trick.
Regard the domain walls as random walkers.
The right random walker randomly chooses between up-left and up-right edges,
to proceed from $A = (1,0)$ to $B = (t-v,v)$.
Similarly, the left random walker proceeds from $C=(0,1)$ to $D=(u,t-u)$.
Any pair of paths $A \to B$ and $C \to D$ that have a common point
can be viewed as a pair of paths $A \to D$ and $C \to B$.
Conversely, any pair of paths $A \to D$ and $C \to B$, which must meet at at a point,
can be viewed as a pair of paths $A \to B$ and $C \to D$ with a common point.
Therefore, the number of pairs of paths from $A \to B$ and $C \to D$ without intersection
is the number of all unrestricted pairs from $A \to B$ and $C \to D$,
minus the number of all unrestricted pairs from $A \to D$ and $C \to B$.
The number of our domain wall configurations is therefore
\begin{align}
 \binom{t-1}{v} \binom{t-1}{u} - \binom{t-1}{v-1}\binom{t-1}{u-1}
 \label{eq:numDomainWalls}
\end{align}
where the second factor vanishes when $u=0$ or $v=0$.

Finally we combine the results above:
\begin{align}
 & \sum_{\substack{s~: \\ s_{\ell_u,\ell_v} = \downarrow}} 
  W_X(s) =\\
&\quad\quad
\underbrace{\frac{q^2}{q^4-1}}_\text{Eq.~\eqref{eq:wtFromX}}
\underbrace{\left( \frac{q}{q^2+1}\right)^{2t-2}}_\text{Eq.~\eqref{eq:wtFromDomainWall}}
\sum_{u=0}^{\ell_u -1} \sum_{v=0}^{\ell_v-1} 
 \underbrace{q^{2t-2u-2v}}_\text{Eq.~\eqref{eq:wtTopBoundary}}  \notag\\
&\quad\quad\quad\quad \times \underbrace{\left[\binom{t-1}{v} \binom{t-1}{u} - \binom{t-1}{v-1}\binom{t-1}{u-1}\right]}_%
\text{Eq.~\eqref{eq:numDomainWalls}}. \notag
\end{align}
This correctly reproduces the answer $q^4/(q^4-1)$ when $\ell_u = \ell_v = 1$.
This can be conveniently rewritten as
\begin{align}
\overline{\otc(t,x)} & = \frac{(1-p)^2}{1-2p} ~g(t-1,\ell_v-1)~ g(t-1,\ell_u-1)  \notag \\
  &- \frac{p^2}{1-2p} ~g(t-1,\ell_v-2)~ g(t-1,\ell_u-2)
\end{align}
where
\begin{align}
t &= \ell_u + \ell_v -1,
&
x &= \ell_u - \ell_v,
&
p &:= \frac{1}{q^2+1},
\end{align}
\begin{align}
g(n,a) &:= \sum_{k=0}^a \binom{n}{k} (1-p)^{n-k} p^k . 
\end{align}
Further simplification is possible since ${g(t-1,a)} \simeq {g(t-1,a-1)}$ for large $t$.
\begin{align}
\overline{\otc(t,x)}
& \simeq g(t-1,\ell_v-1,p) g(t-1,\ell_u-1,p)\\
& \simeq \Phi\left( \frac{v_B t + x}{\sigma} \right) \Phi\left( \frac{v_B t - x}{\sigma} \right)
\end{align}
where
\begin{align}
 v_B = \frac{q^2-1}{q^2+1}, \quad \sigma = \frac{2q\sqrt t}{q^2+1}, 
 \quad \Phi(y) = \frac{1}{\sqrt{2\pi}}\int_{-\infty}^y  e^{-x^2/2} \mathrm{d} x .
\end{align}

\subsection{Bounds on Fluctuations}
\label{sec:fluctuationBounds}
Here we estimate the fluctuation of $\otc(t,x)$ due to the {\it randomness of the unitaries}.
One might wish to calculate this fluctuation directly,
using a similar technique that we employ for the average of $\otc(t,x)$,
but we found the exact computation unwieldy as it involves high powers of unitaries.
Nevertheless, we can argue that the fluctuations are negligible in two regimes.

Since the random variable $\otc(t,x)$ takes values between 0 and 2,
the variance is upper bounded by $2 \bar \otc$.
Therefore, the standard deviation is upper bounded by
\begin{align}
 \sqrt{\overline{ \otc(t,x)^2 }- \overline{\otc(t,x)}^2} &  \le \sqrt{2\overline{\otc(t,x)}} \\
 & \simeq O(1) \exp\left(-\frac{1}{4}\left(\frac{|x|- v_B t}{\sigma}\right)^2\right) \notag.
\end{align}
This bound is valid for any $t,x$, but only meaningful when $|x| \gg v_B t$.
This basically says that there is almost no ``leakage'' of operators beyond the lightcone defined by $v_B$.
(In passing, we note that one can also use 
Markov inequality $\Pr[X \ge a] \le a^{-1} \avg X $ 
which holds for any positive random variable $X$ and a positive number $a$
to have a probability tail bound.)

In the opposite regime where $|x| \ll v_B t$,
we have shown that the average $\overline{\otc(t,x)}$ is almost 1;
the discrepancy is upper bounded by $O(1) \exp( - (v_B t - |x|)^2/2\sigma^2)$.
Thus, in this regime the fluctuation is basically given by
\begin{align}
 \avg_U F^2(U) = N^{-2} \avg_U (\Tr UXU^\dagger Y UXU^\dagger Y )^2 \ge 0
\end{align}
where $F$ is defined in Eq.~\eqref{eq:introductionOfF}
and $N$ is the dimension of the Hilbert space of spins where $UXU^\dagger Y$ is supported on.
Here $U$ includes all the local unitaries in the evolution quantum circuit.

To estimate the fluctuation, we consider a slightly different system where $2ct$ spins form a ring,
where $c$ is some absolute constant that depends on $q$ only.
If $c > 1$, this does not modify the dynamics at all,
since the evolved operator $U X_0 U^\dagger$ is supported on $2t$ spins.
For $c < 1$,
while we do not insist that this allows us to compute the fluctuation rigorously, 
we anticipate that qualitative conclusions from this modified setting 
carry over to the original open chain system.

In Appendix~\ref{app:fluctuationBound}, we show that
if $c = M_q^{-1} = \tilde O(q^{-2})$, then
for all $|x| < ct$
\begin{align}
\avg_{U} F^2(U) \le 11 q^{-4 ct}.
\label{eq:fluctationDeepInLightcone}
\end{align}
That is, deep in the lightcone,
there is a region in spacetime bounded by a nonzero speed
where the fluctuation of $\otc(t,x)$ is suppressed exponentially in $t$.
It is likely that this is only a bound, rather than a tight estimate of fluctuation.
Eq.~\eqref{eq:fluctationDeepInLightcone} is proved
using previous results on approximate unitary designs~\cite{BHH2016},
and estimates for $\avg_U F^2(U)$ when $U$ is truly Haar random~\cite{Hastings2007}.

\section{Entanglement growth}
\label{sec:purity}

Entanglement can by quantified in various ways,
but perhaps the simplest measure is the entanglement purity 
$\mathcal P = \Tr\,(\Tr_{A^c} \ket \psi \bra \psi )^2 \le 1$,
where $A$ is some region.
A pure state $\ket \psi$ on $A \cup A^c$ is entangled 
if and only if the purity is not equal to 1.
The logarithm of the entanglement purity is the Renyi-2 entropy
\begin{align}
 S_2(A) = -\log \Tr \, \lf \Tr_{A^c} \ket \psi \bra \psi \ri^2.
\end{align}
In this section, we  calculate exactly  the average purity
of `half' of the infinite chain,  for arbitrary $t$, 
under the evolution protocol in Section~\ref{sec:operator-dyn-1d},
with an initial product state. (Previous work has obtained a bound on the saturation time for $q=2$ \cite{Znidaric}.)
The calculation technique will be very similar to the OTOC calculation;
the difference is only in the boundary conditions.

Let $A$ be the left half of our chain, and $B$ be the right half.
The initial pure density matrix is 
$\rho(t=0) = \cdots \otimes P_{-1} \otimes P_0 \otimes P_1 \otimes \cdots$,
where $P_i$ is a projector $\ket i \bra i$, the state at site $i$.
If $U$ is the full time-evolution unitary consisting of local unitaries,
then the entanglement purity $\mathcal P $ across the cut between $A$ and $B$ is
\begin{align}
 \overline{\mathcal P(t) }
&=
 \sum_{a,a'=1}^{q^{|A|}} \sum_{b,b'=1}^{q^{|B|}}
 \overline{\bra{ab} U \rho(0)U^\dagger \ket{a'b} \bra{a'b'} U \rho(0) U^\dagger \ket{a b'}} \notag \\
&=
 q^{|A|+|B|}  \times  \notag \\
 &   \bra{\uparrow^{\otimes |A|} \downarrow^{\otimes |B|}} ~
 \overline{U \otimes U^* \otimes U \otimes U^*}~
 \ket{\cdots (P_0)_\uparrow (P_1)_\uparrow \cdots }
 \label{eq:purityAsInnerProduct}
\end{align}
The notation $\ket \uparrow, \ket \downarrow$ is the same as in Sec.~\ref{sec:reduction_Ising_spins}.
For a one-dimensional projector $P$ on $q$-dimensional space,
the $q^4$-dimensinal vector $\ket{P_\uparrow}$ satisfies
\begin{align}
 \braket{\uparrow| P_\uparrow } &= \frac 1 q =\braket{\downarrow| P_\uparrow } .
 \label{eq:projector-innerproduct}
\end{align}

The expression for the purity can be thought of as a partition function for
classical Ising spins as in Sec.~\ref{sec:reduction_Ising_spins}.
There are two Ising spins associated with each local unitary; 
see Eq.~\eqref{eq:generalHaarAverageIsing}.
Due to Eq.~\eqref{eq:projector-innerproduct},
for any configuration of the Ising spins,
the weight factor from the bottom boundary is $q^{-|A|-|B|}$,
which cancels the factor $q^{|A|+|B|}$ in  front of Eq.~\eqref{eq:purityAsInnerProduct}.
Hence, the average purity is simply the sum of weights from the domain wall in the bulk (e.g. see Fig. \label{fig:shiftingDomain}).

In Sec.~\ref{sec:reduction_Ising_spins},
we first integrated out the `bra' Ising variables $s'$,
but here we find it simpler to integrate out the `ket' Ising variables $s$.
The transition rules of Fig.~\ref{fig:wtTable} are now upside down,
but otherwise the same.
Then, we have a single domain wall starting from the top boundary
to reach the bottom.
Any domain wall has length exactly $t$, giving rise to weight $\left(\frac{q}{q^2+1}\right)^t$.
The domain wall can choose between left-down or right-down moves
as it proceeds from the top, and therefore there are $2^t$ domain walls.
We conclude that
\begin{align}
 \overline{\mathcal P(t)} = \left( \frac{2q}{q^2+1}\right)^t.
\end{align}
We may  define the `purity speed'
\begin{align}
 \overline{\mathcal P(t)} 
&\equiv q^{-v_P t}, &
 v_P &= \log_q \f{q^2+1}{2q}.
\end{align}
This quantity gives a bound on the growth rate of the second Renyi entropy:
\begin{align}
 \overline{ S_2(\rho(t)_A)} = \overline{-\log_q \mathcal P(t)} 
 \ge -\log_q \overline{\mathcal P(t)} = v_P t
\end{align}
The inequality is because the function $f(x)=-\log x$ is convex. 
Note that this expression bounds the growth rate of $S_2$ but does not fix it. 
The distribution of $S_2$ fluctuates in a window of small size compared to its mean~\cite{NRVH2016},%
\footnote{
Ref.~\cite{NRVH2016} argued that the width of the distribution scales as $t^{1/3}$. 
The mean value is of order $t$.
}
but since $S_2$ appears in the exponential in ${\overline{q^{-S_2}}}$, 
this does not rule out the possibility that this quantity is affected 
by rare anomalously small values of $S_2$, 
making it very different from $q^{-\overline{S_2}}$.

The von Neumann entropy $S_{vN}$ is always greater than or equal to $S_2$, 
so the growth rate $v_E$ of $S_{vN}$ is also bounded by $v_P$:
\begin{align}
 v_E \ge v_P = \log_q \frac{q^{2}+1}{2q} = 1 - \frac{\log 2}{\log q} + O\left( \frac{1}{q^2 \log q} \right)
\end{align}
where the expansion is for large $q$.

In Ref.~\cite{NRVH2016} 
we argued that the universal fluctuations of the entanglement in random circuit dynamics 
may be understood in terms of a coarse-grained minimal cut, of random shape, through the random circuit. 
This picture may be contrasted with the domain wall calculation of the averaged purity, 
which reduces to a statistical mechanics problem without quenched randomness. 
This is reminiscent of the difference between a quenched and an annealed average 
in the statistical mechanics of disordered systems~\cite{cardy1996scaling}. 
A direct exact calculation of $\overline{S_2}$ 
(not to mention  $\overline{S_\text{vN}}$, or of the fluctuations in the entropy) 
for finite%
\footnote{
In the limit $q\rightarrow\infty$ it is easy to show that $v_E=v_P$ in the present model.
}
$q$ would be much more difficult than the calculation above, 
as a replica-like limit~\cite{cardy1996scaling} would be required to handle the logarithm. 
However structure arising from unitarity might make this calculation tractable. 
This is an interesting task for the future.

The scaling limit of the representation obtained in this section, where we take length and time scales to be large and of the same order, yields a `deterministic' domain wall configuration. This is simply  a vertical line for the infinite geometry considered here.\footnote{This is because the $\sqrt{t}$ fluctuations in the transverse position of the domain wall are negligible compared to $t$; compare \cite{NRVH2016} where the minimal cut configuration is also deterministic in the scaling limit.} We expect that extending the calculation to higher dimensions will give, in the scaling limit,  a formula for $ -{\log \overline{ \mathcal{P}}}$ as the `energy' of a minimal surface  (representing the Ising domain wall) which has a deterministic coarse-grained geometry, obtained from an effective elastic energy minimization problem. This is precisely the scaling picture proposed in Ref.~\cite{NRVH2016} for the growth of entanglement in higher-dimensional systems.

\begin{figure}
 \includegraphics[trim = 0 0 0 0, clip = true, width=0.9\linewidth, angle = 0.]{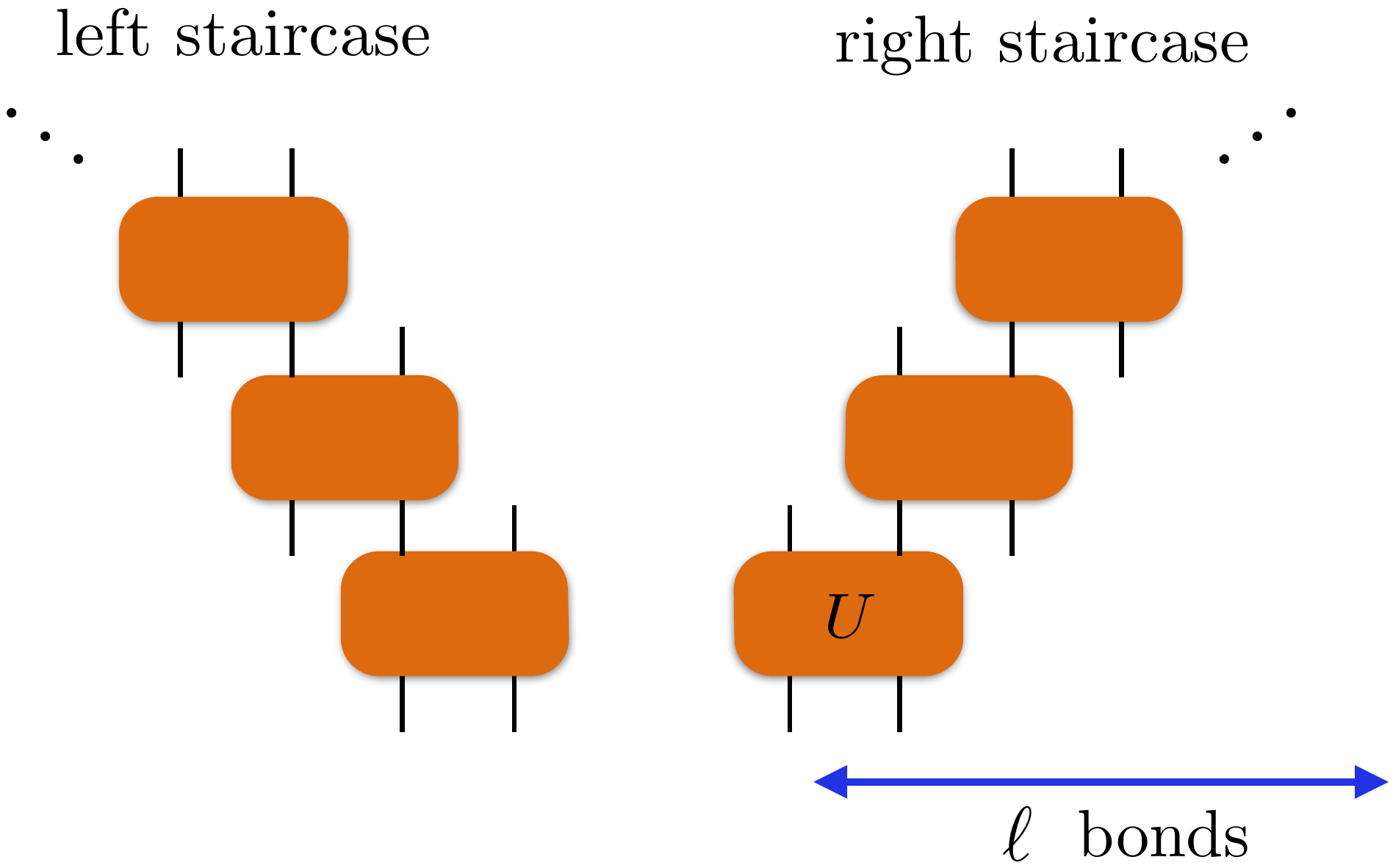}
 \caption{{\bf Random Circuit built from ``Staircase" Unitaries:} We use ``left" and ``right" staircases --- built from random two-site unitary operators as shown, and extending over $\ell$ bonds --- as the building blocks for a random quantum circuit in which the ratio of the entanglement and butterfly velocities $v_{E}/v_{B}$ may be made arbitrarily small.  }
  \label{fig:Staircase}
\end{figure}

\subsection{Nonuniversality of the ratio $v_E/v_B$}

In Ref.~\citep{NRVH2016}, see also Ref.~\cite{mezei2016entanglement},
we showed that the speed $v_E$ associated with entanglement growth 
is in general smaller than the operator growth speed $v_B$,  
and gave explicit models displaying a ratio $v_E/v_B < 1$. 
In these models\footnote{
This was determined analytically for a certain large $q$ model, distinct from that here, 
and numerically for various circuits composed of Clifford gates.
}  this ratio happened to be $1/2$. Values close to 1/2 were also found numerically in Ref.~\cite{mezei2016entanglement} and Ref.~\cite{leviatan2017quantum}. These results might lead one to wonder whether this value is in some sense generic. 
Here we show that it is not, by constructing random circuit dynamics, involving interactions of large but finite range, 
which give \textit{arbitrarily small} values of $v_E/v_B$ without any fine tuning. 
The construction uses random unitaries made up of `staircases' of length $O(q^2)$ 
which are made up of smaller random unitaries (Fig.~\ref{fig:Staircase}).  
When $q$ is large, we obtain a ratio $v_E/v_B$ which is at most of order $1/q^2$. 
(In a deterministic spin chain with quenched spatial disorder it is even possible to have $v_E/v_B = 0$~\cite{ForthcomingGriffiths},
but here we insist on \textit{statistical} translational invariance: 
i.e. we insist that the probability distribution for the circuit is invariant under translations.)

Consider quantum circuit dynamics 
in which `staircase unitaries' are applied at random locations 
and at random times in a Poissonian fashion. 
A staircase is a collection of 2--site unitaries arranged as in Fig. \ref{fig:Staircase}. 
Left and right-oriented staircases are applied with equal probability. 
The staircase acts on $\ell$ bonds and we take $\ell$ large but finite, 
satisfying $\ell/q^2 \gg 1$. 
Let $r$ be the rate at which staircases are dropped at a given location.

A single staircase can increase the entanglement  across a given bond by at most 2 units,
implying $v_E \leq 2 r \ell$. 
On the other hand a single staircase can move the endpoint of an operator a long way 
when $\ell \gtrsim q^2 \gg 1$. 
The random walk picture of Sec.~\ref{hydrodynamic section} 
shows that in the limit of large $\ell/q^2$, 
a staircase advances the front of the OTOC by an average distance $\sim q^2/2$. 
This involves an average over the two staircase orientations, 
only one of which is effective in advancing the front a long distance. 
The large value is because, when $q$ is large, 
the small value of $p = 1/({q^2+1})$ (Eq.~\ref{step probability}) means 
the random walker can `run' a long way up a rightward-oriented staircase before falling off. 
The previous implies $v_B \simeq q^2 r \ell /2$ at leading order in $\ell$. 
This yields a ratio $v_E/v_B \lesssim  4/q^2$  in this regime, 
which can be made arbitrarily small by taking $q$ (and hence $\ell$) to be large.

\section{Outlook}
\label{sec:outlook}

We have argued 
that universal scaling forms for the out-of-time-order correlator 
can be obtained using mappings to paradigmatic problems in classical statistical mechanics. 
In one dimension we gave an extremely simple hydrodynamic picture in terms of diffusion. 
In higher dimensions we gave a mapping to classical surface growth and the KPZ equation.%
\footnote{
Ref.~\cite{NRVH2016}
obtained a connection between entanglement growth in 1+1D random unitary circuits and the KPZ equation.
To avoid confusion we emphasize that 
the connection with KPZ discussed here is physically entirely distinct from that one,
and is not even in the same number of spatial dimensions.
}

These mappings were derived exactly for random unitary circuits, 
which are natural `least structured' models for chaotic quantum dynamics in situations 
where conserved quantities are not playing an important role.
We have conjectured that the universal scaling forms found here also apply to OTOCs at asymptotically late times 
in generic, nonintegrable many body systems and quantum field theories.
It will be interesting to test this conjecture in other situations where calculations are possible.

This picture differs from that obtained in a number of previous calculations 
using many-body perturbation theory \cite{Aleiner, Sachdev_CriticalFermiSurface, Sachdev_DiffusiveMetals, chowdhury2017onset}, 
and it will be interesting to understand the reasons for these differences. 
Ref.~\cite{Aleiner} found an operator front that did not broaden in time, 
whereas here we find a broadening front in all dimensions below $4+1$. 
Additionally, in Ref.~\cite{Aleiner} the OTOC was found to obey a local, nonlinear traveling wave equation, 
which is unlike what we found for random circuits. 
In 1D we obtained a \textit{linear} hydrodynamic equation, 
while in higher dimensions $\overline{\otc}(x,t)$ in a random circuit 
is not governed by a local differential equation at all,  contrary to standard lore about OTOCs.

Interestingly, a \textit{mean field} approximation to the classical growth process would yield a local differential (or rather difference) equation for the OTOC, of traveling wave form. This is discussed in Appendix.~\ref{app:meanfield}. However, the mean field approximation is not valid in physical dimensionalities.

Assuming that the results here do indeed have applications to realistic systems 
with Hamiltonians that are fixed in time, 
it will be interesting to consider extensions of the present coarse-grained pictures which take conserved quantities into account.

We have also given exact results for entanglement growth in 1+1D which  support the scaling ideas put forward in \cite{NRVH2016}, as discussed in Sec.~\ref{sec:purity}. In this picture (in any D) entanglement growth is determined 
by a minimal surface in spacetime,
whose geometry becomes  well-defined\footnote{But model dependent above 1+1D} in an appropriate scaling limit 
and is determined essentially by an elastic minimization problem.
Furthermore, it was argued in that paper and in Ref.~\cite{mezei2016entanglement} that generically $v_E < v_B$, 
where $v_E$ is the speed characterizing the growth of entanglement.  
Here we have shown that it is possible to have arbitrarily small $v_E/v_B$ in a random quantum circuit.

The effective Ising partition functions for calculating the OTOC 
and the purity turned out to have interesting structure, 
making them drastically simpler than they appeared at first sight, and much simpler than the analogous partition function for a non-unitary tensor network \cite{HaydenEtAl2016}.
It would be very interesting to explore whether similar simplifications occur 
when the averaging involves higher powers of the unitary circuit.
If so this would permit calculations of, say, modified versions of the OTOC involving higher powers of the commutator, 
or a direct calculation of the fluctuations. 
Even more interesting would be a direct calculation of the von Neumman entropy, which would have to use a replica limit to handle the logarithm.

OTOCs involving higher powers of the commutator are important for comparison with Lieb-Robinson bound.
The OTOC considered here can be thought of as the squared Frobenius norm of the commutator divided by the Hilbert space dimension,
whereas the Lieb-Robinson bound is on the operator norm of the commutator.
The two norms are related as the operator norm is always upper bounded by the Frobenius norm,
but our results do not put any nontrivial bound on the operator norm, due to the large dimension factor.
The exact relation of the two quantities is yet to be determined.

In addition to exploring implications for realistic many-body systems,
interesting questions remain that are specific to the random circuit context. 
(Note that, at the most basic level, 
our results show that operator growth saturates the naive causal lightcone of the quantum circuit as $q\rightarrow \infty$, but not for finite $q$.)
The randomness in the circuit necessarily implies 
statistical fluctuations in all observables including $\otc(x,t)$. 
We have argued that these statistical fluctuations are (perhaps counterintuitively) 
a subleading effect at late times.  
We have shown this in regimes far from the front of the OTOC by giving inequalities,
and we have given a heuristic argument for it in the region near the front. 
This argument was based on a phenomenological extension of the hydrodynamic equation for $\overline{\otc(x,t)}$ 
in the 1D case to allow for statistical fluctuations in $\otc(x,t)$ (Eq.~\ref{hydrodynamic_with_fluctuations}). 
It would be desirable to give a microscopic derivation of Eq.~\ref{hydrodynamic_with_fluctuations}.
(For the entanglement entropy, statistical fluctuations were investigated in Ref.~\cite{NRVH2016}.)
It also remains to characterize the classical growth problem in Section~\ref{sec:higherD} more fully,
for example by obtaining the nonuniversal constants via an approximate analytic treatment.

The KPZ equation is connected to a remarkable array of topics 
in classical statistical mechanics \cite{HALPINHEALY1995}, including the directed polymer in a random medium \cite{HuseHenley1985} and one-dimensional hydrodynamics \cite{ForsterStephenNelson},
and has beautiful experimental applications \cite{Takeuchi2011,Takeuchi2012}.
Through the Tracy-Widom distribution \cite{Tracy_Widom},
it is also connected to random matrix theory and an array of combinatorial problems 
(for example the longest increasing subsequence problem and the statistics of random permutations \cite{baik1999distribution, tracy2001distributions}).
It will be interesting to explore which members of this array can shed light on operator growth.

\textit{Related work:} While this manuscript was being finalized, we became aware of related work \cite{curtforthcoming}, to  appear in the same arXiv posting. We also alert the reader to forthcoming numerical work on operator spreading \cite{jonayhuseforthcoming}.

\begin{acknowledgments}
We thank D. Huse, C. Jonay, D. Chowdhury, J. Ruhman and H. Spohn for useful discussions. We are grateful to T. Veness and J. Chalker for helpful discussions regarding droplet shape. JH thanks M.~B. Hastings for useful discussions. AN was supported by the Gordon and Betty Moore Foundation under the EPiQS initiative (grant No.~GBMF4303) and by EPSRC Grant No.~EP/N028678/1.
SV is supported by the DOE Office of Basic Energy Sciences,  Division  of  Materials Sciences  and  
Engineering under Award de-sc0010526 and was also supported for part of this work by the National 
Science Foundation under Grant No. NSF PHY11-25915.
\end{acknowledgments}

\appendix

\section{Evolving Distribution on \\ Operator Strings}
\label{sec:String_Distribution}

In this Appendix we give a more detailed explanation of the relationship between the dynamics of the coefficients $a_\cS$ and a Markov process \cite{Dahlsten, Znidaric} and of the derivation of the diffusion picture. We consider Haar-random, local unitary dynamics.  In an $N$-site system with a $q$-dimensional Hilbert space at each site, a Hermitian operator that has evolved under the unitary circuit $\OO(t) = {U}(t)^{\dagger}\,\OO\,{U}(t)$ may expanded in a basis of SU$(q^{N})$ generators $\{\mS\}$ as 
\begin{align}
\OO(t) = \sum_{\mS}a_{\mS}(t)\mS
\end{align}
Our normalization convention is $\mathrm{Tr}(\mS\mS') = q^{N}\delta_{\mS\mS'}$, so that 
$a_{\mS}(t) = q^{-N}\mathrm{Tr}(\OO(t)\mS)$. 
The squared coefficient $a_{\mS}(t)^{2}$ evolves as
\begin{align}
a_{\mS}(t)^{2} &= q^{-2N}\sum_{\mS',\mS''}a_{\mS'}(t-1)a_{\mS''}(t-1) 
\\ &\qquad\quad\qquad\times \mathrm{Tr}[\mU\,\mS'\mU^{\dagger}
\mS]\,\mathrm{Tr}[\mU\,\mS''\mU^{\dagger}\mS] \notag \\
&= q^{-2N}\sum_{\mS',\mS''}a_{\mS'}(t-1)a_{\mS''}(t-1) \notag
\\ &\qquad\quad\qquad\times
\prod_{\rb}\mathrm{tr}[U_{\rb}\,S_{\rb}'
U^{\dagger}_{\rb}S_{\rb}]\,\mathrm{tr}[U_{\rb}\,S_{\rb}''U^{\dagger}_{\rb}S_{\rb}] \notag
\end{align}
where $\mU$ is a layer of $m$-site unitaries that were applied at time $t-1$.  In the second line, 
we have written $\mU = \prod_{\rb}U_{\rb}$ where $\rb$ is the coordinate of disjoint, $m$-site 
clusters on which the unitary $U_{\rb}\in \mathrm{U}(q^{m})$ acts, and we have also decomposed
$\mS = \prod_{\rb}S_{\rb}$ as a product of basis elements acting on these $m$-site 
clusters.  These operators are normalized according to 
$\mathrm{tr}[S_{\rb}S_{\rb}'] = q^{m}\delta_{S_{\rb},S_{\rb}'}$ and 
$\tr[S_{\rb}] = q^{m}\delta_{S_{\rb},1}$. 
The Haar average of the above expression is given by
\begin{align}
&\overline{\mathrm{tr}[U_{\rb}\,S_{\rb}'U^{\dagger}_{\rb}S_{\rb}]\,
\mathrm{tr}[U_{\rb}\,S_{\rb}''U^{\dagger}_{\rb}S_{\rb}]}
\notag \\
&= \frac{\delta_{S_{\rb}',S_{\rb}''}}{1-q^{-2m}}
\left\{q^{2m}\delta_{S_{\rb},1}\delta_{S_{\rb}',1} + 
1 - \delta_{S_{\rb}',1} - \delta_{S_{\rb},1}\right\}
\end{align}
And so, the Haar-averaged $a_{\mS}(t)^{2}$ evolves \textit{linearly}
\begin{align}
\overline{a_{\mS}(t)^{2}} &= \frac{1}{q^{2N}}\sum_{\mS',\mS''}{a_{\mS'}(t-1)}\,{a_{\mS''}(t-1)}
\notag \\
&\times \prod_{\rb}
\frac{\delta_{S_{\rb}',S_{\rb}''} \left(q^{2m}
\delta_{S_{\rb},1}\delta_{S_{\rb}',1} + 1 - \delta_{S_{\rb}',1}
 - \delta_{S_{\rb},1}\right)  }{1-q^{-2m}}  
\notag \\
&= \sum_{\mS'}W_{\mS\mS'}\,a_{\mS'}(t-1)^{2}
\end{align}
with the real, symmetric matrix 
\begin{align}
{W_{\mS\mS'} = \prod_{\rb}\left[\delta_{S_{\rb},1}\delta_{S_{\rb}',1} + \frac{(1 - \delta_{S_{\rb},1})(1 - \delta_{S_{\rb}',1})}{q^{2m}-1}\right]}
\end{align}
Averaging again over the unitaries applied in the previous timesteps gives an equation for $P_{\mS}(t) \equiv \overline{a_\mS^2(t)}$
\begin{align} \label{eq:strings_process}
P_\mS(t) = \sum_{\mS'}W_{\mS \mS'} P_{\mS'}(t-1)
\end{align}
 which is formally a master equation for a fictitious Markov process \cite{Dahlsten, Znidaric}; at a given time there is a single string $\mS$ which is updated stochastically in each time step, via local updates involving a cluster of $m$ sites.  From the form of $W_{\mS\mS'}$ we see that the local update on $m$ sites is performed by replacing a non-trivial generator on the cluster \emph{randomly} by any one of the $q^{2m}-1$ non-trivial generators.  We emphasize that this fictitious Markov process is {not} the true unitary dynamics of the operator $\mathcal{O}(t)$.  
 
This fictitious classical stochastic process dramatically simplifies through the following observations. We focus here on one spatial dimension  with updates on bonds.  First, observe that the matrix elements $W_{\mS\mS'}$ only depend on the support of the generators $\mS$ and $\mS'$, so that (\ref{eq:strings_process}) gives rise to a simpler Markov process for the binary occupation number $n(x)$, which is 1 if the corresponding generator has support at site $x$ and 0 otherwise (if $\mS$ acts as the identity at $x$).  
Formally, the probability distribution of the occupation numbers is given by
\begin{align}
P[\{ n \} ; t] = \sum_{\mS}\,'\, \,\,\overline{a_\mS(t)^2} 
\end{align}
where the prime indicates that the sum is only over  strings $\mathcal{S}$ that are compatible with the configuration $n(x)$.  Further, the endpoint of the string  observes an \emph{autonomous} Markovian dynamics.  Since $m=2$, updates involving the endpoint  either include the site to the right of it which is empty or that to the  left which may be empty or full. The dynamical rule above implies that the probabilities for the position of the endpoint after the update are independent of whether the leftward site was initially occupied or empty. Formally the probability distribution for the position of the endpoint in this fictitious dynamics is 
\begin{align}
p_\text{end} (x; t) = 
\sum_{\substack{n \,\text{with}\\ 
\text{endpt at $x$} }} P[\{n\}; t]
= \sum_{\mS \,\,\text{ends at}\, x} \overline{a_\mS(t)^2} 
\end{align}
which is precisely $\overline {\rho(x,t)}$, as defined in Sec 2.  Therefore, for an endpoint at $x$ or $x+1$, a single update applied to the sites $x$ and $x+1$ leaves the endpoint at $x$ with probability $p = (q^{2}-1)/(q^{4}-1) = 1/q^{2}+1$ and at $x+1$ with probability $1-p$.  This establishes the claim in Sec. 2 for the evolution of $\overline{\rho(x,t)}$ in a single timestep.

\section{Velocity and diffusion constant for lattice diffusion equation}
\label{sec:constantslatticediffusion}

Our layout of the evolution operator is 
such that the local unitaries alternate between even and odd bonds.
In other words, a bond at time step $t$ is either at the left or right of the bond at $t-1$.
Thus, it suffices to count the left and right moves
to specify the position of the right end-bond of $X_0(t)$.
As described in the main text the probability of a left move is $p$.
Let $u \ge 0$ be the number of right moves, and $v \ge 0$ be the number of left moves.
We have  $u+v=t$, and  $u-v$ (or $u-v \pm 1$) is the spatial coordinate of the right endpoint.
Therefore, the probability distribution of the position of the right end bond is
\begin{align}
 f(u,v) = \binom{u+v}{u}(1-p)^u p^v .
\end{align}
This is correctly normalized since $\sum_{u+v=t} f(u,v) = (1-p+p)^{t} = 1$.
Then,
the probability that a site $x$ is left to the right end of $X_0(t)$ is
\begin{align}
 \sum_{u + v = t, \, u - v \ge x} f(u,v) 
&= 
 \sum_{v=0}^{(t-x)/2} \binom{t}{v} (1-p)^{t-v} p^v
\\
&\simeq
 \Phi\left(\frac{v_B t-x}{\sigma}\right)
\label{eq:PrXisInside}
\end{align}
where $\Phi$ is the cumulative density function of the normal distribution,
and
\begin{align}
 v_B = \frac{q^2-1}{q^2+1}, \quad \sigma = \frac{2q \sqrt{t}}{q^2+1}.
\end{align}

\section{Noisy diffusion equation}
\label{noisy_diffusion_equation_appendix}

Starting with Eq.~\ref{hydrodynamic_with_fluctuations}, WLOG rescale space so $D=1$ and set $v=0$ by going to the moving frame. Let 
$\rho_0$ be the solution without noise, $\rho_0 = (4\pi t)^{-1/2} e^{-x^2/4t}$. In terms of the Green's function
\ba\notag
\rho(x,t) - \rho_0(x,t) &= \hspace{-2pt} \int_{x',t'} \hspace{-5pt} G(x-x',t-t') \partial_{x'} \eta(x',t')\rho(x',t') \\
&=
\int_{x',t'} G'(x-x',t-t')  \eta(x',t')\rho(x',t').
\end{align}
The centre of mass position of the wavepacket within a given realisation is ${x_\text{cm} = \int_x x \rho(x,t)}$, so, if  $\overline{x_{cm}}$ is the centre of mass position averaged over realisations,
\ba\notag
x_\text{cm} -
\overline{x_{cm}} &= \int_{x,x',t'} \hspace{-4pt} x G'(x-x',t-t') \eta(x',t') \rho(x',t')\\
& = - \int_{x,x',t'}  \hspace{-4pt} G(x-x',t-t') \eta(x',t') \rho(x',t'), \notag \\
{\lf x_\text{cm} - \overline{x_{cm}} \ri^2} &= \int_{\substack{x,\tilde x, x', \\ \tilde x', t', \tilde t'}}  \hspace{-2pt}G(x-x',t-t')  G(\tilde x-\tilde x', t-\tilde t') \notag \\
&\qquad\qquad\times   \eta(x',t')\eta(\tilde x',\tilde t') \rho(x',t')\rho(\tilde x',\tilde t').\notag
\end{align}
Averaging over the noise with ${\<\eta(x,t) \eta(x',t')\>} = \lambda {\delta(x-x')\delta(t-t')}$, 
\be
\overline{{[ x_\text{cm} - \overline{x_{cm}} ]^2}} = \hspace{-2pt}
\lambda \hspace{-2pt}
 \int_{\substack{x,\tilde x,\\ x', t'}}\hspace{-2pt}  G(x-x',t-t')G(\tilde x- x',  t- t')  \rho(x',t')^2.
 \ee
For the leading order scaling, we replace $\rho$ with $\rho_0$ on the right hand side. Then dimensional analysis applied to the integral gives 
\be
\sqrt{\overline{{\lf x_\text{cm} - \overline{x_{cm}}\ri^2}}} \propto \lambda^{1/2} t^{1/4} + \ldots
 \ee
This is a statistical variation in $x_\text{cm}$ of order $t^{1/4}$, in agreement with the heuristic argument and with Ref.~\cite{saul1992directed}. This variation is small compared to the width of $\rho$, indicating that $\rho - \rho_0 \ll \rho$ at late times. The typical size of $\partial_x \rho$ near the peak is $O(1/t)$, so ${x_\text{cm} - \overline{x_\text{cm}}}\sim t^{1/4}$ corresponds to  $\rho-\rho_0\sim t^{-3/4}$, as compared with $\rho \sim t^{-1/2}$. The approximation above is therefore self-consistent.

\section{Random Clifford operators}
\label{app:Clifford}

Here we review that the left- and right- invariant probability distribution over 
the Clifford group on $n$ $q$-dimensional qudits is a unitary 2-design when $q$ is a prime number.
In other words, for a qudit of prime power dimension $q^n$, 
the unitary group $U(q^n)$ has a finite subgroup that is a unitary 2-design,
and there is a linear operator basis that remains closed under conjugations by this subgroup.
This is a well-known result~\cite{DiVincenzo2001}, but we include it here for readers' convenience.

To define the Clifford group,
we first need the Pauli group.
Define $X = \sum_{j=0}^{q-1} \ket{j+1 \mod q} \bra{j}$ and $Z = \sum_{j=0}^{q-1} e^{2\pi i j / q} \ket j \bra j$.
Then, the Pauli group is the subgroup of $U(q^n)$ generated by
matrices $X_1,Z_1, \ldots, X_n, Z_n$ where
\begin{align}
 X_j &= I_q ^{\otimes (j-1)} \otimes X \otimes I_q^{\otimes (n-j)},\nonumber \\
 Z_j &= I_q ^{\otimes (j-1)} \otimes Z \otimes I_q^{\otimes (n-j)}.
\end{align}
The Clifford group is defined to be the normalizer of the Pauli group in $U(q^n)$.
The Pauli group quotiented out by its center $\langle \omega = e^{2\pi i/q} \rangle$
is abelian since $X Z X^{\dagger}= \omega^{-1} Z$,
and is isomorphic to the additive group $\mathbb Z_q^{2n}$.
We define $P_v$ for $v \in \mathbb Z_q^{2n}$ to be an element of the Pauli group (Pauli operator)
as
\begin{align}
P_v = X_1^{v_1} X_2^{v_2} \cdots X_n^{v_n} Z_1 ^{v_{n+1}} Z_2 ^{v_{n+2}} \cdots Z_n^{v_{2n}}
\end{align}

The center of the Pauli group is also contained in the center of $U(q^n)$,
and therefore the conjugation action by the Clifford group on the Pauli group
induces an action on $\mathbb Z_q^{2n}$.
It turns out that this group $\mathcal S$ of action
consists precisely of those that preserves the symplectic form
\begin{align}
\lambda_n =
 \begin{pmatrix}
 0 & -I_n \\
 I_n & 0
 \end{pmatrix}
\end{align}
over $\mathbb Z_q$.


A probability distribution $\nu$ of unitary matrices to form a 2-design means that
\begin{align}
 \avg_{U \sim \nu} U \otimes U^* \otimes U \otimes U^* 
 = \avg_{U \sim \mu} U \otimes U^* \otimes U \otimes U^*
\end{align}
where $\mu$ is the Haar probability distribution over $U(q^n)$,
and $U^*$ is the complex conjugate of $U$.
Tautologically, the Haar distribution is a 2-design.
This is equilvalent to having that
\begin{align}
  \avg_{U \sim \nu} U O U^\dagger \otimes U O' U^\dagger
 = \avg_{U \sim \mu} U O U^\dagger \otimes U O' U^\dagger
 \label{eq:conj-avg}
\end{align}
for any $q^n \times q^n$ matrices $O$ and $O'$.
Since Pauli operators (the elements of the Pauli group defined above)
generates over the complex numbers the full operator algebra,
it is enough to have Eq.~\eqref{eq:conj-avg} with $O$ and $O'$ being Pauli operators.

Let $\nu$ be the left-invariant (hence right-invariant)
probability distribution over the Clifford group.
This is the uniform distribution over the finite Clifford group.
Consider a $\mathbb C$-linear map $\Pi_\nu$ on the set of operators
defined by 
\begin{align}
\Pi_\nu : O \otimes O' \mapsto \avg_{U \sim \nu} U O U^\dagger \otimes U O' U^\dagger. 
\end{align}
Since $\nu$ is a left-invariant distribution over a group of unitaries,
$\Pi_\nu$ is a projector (which is hermitian under the Hilbert-Schmidt inner product).
Since the Clifford group includes the Pauli group,
we have for arbitrary $a,b \in \mathbb Z_q^{2n}$
\begin{align*}
\Pi_\nu( P_{a} \otimes P_{b}^\dagger ) 
&= \sum_{x, y \in \mathbb Z_q^{2n}} 
\eta_{x, y}^{a,b} ~
P_{x} \otimes P_{ y}^\dagger \\
 & (\eta_{x, y} \in \mathbb C, \text{Pauli basis expansion})\\
&= P_{c}^{\otimes 2} 
\Pi_\nu \left( P_{a} \otimes P_{b}^\dagger \right) 
(P_{c}^{\otimes 2})^{-1} 
\\ & \text{(for any $c \in \mathbb Z_q^{2n}$ by the left-invariance of $\nu$)}\\
&= \sum_{x, y \in \mathbb Z_q^{2n}} 
\eta_{x, y}^{a,b} ~
\omega^{c^T \lambda_n (x - y)} ~
P_{x} \otimes P_{y}^\dagger
\\ & \text{(commutation relation among Pauli operators)}\\
&= \sum_{x \in \mathbb Z_q^{2n}}
\eta_{x, -x}^{a,b} ~
P_{ x} \otimes P_{x}^\dagger 
\\ &\text{($\mathrm{det}~ \lambda_n = 1$, and $c$ was arbitrary)}.
\end{align*}
The use of inverse $P_b^\dagger$ here instead of $P_b$ is
for notational convenience later.

Now, observe that for any nonzero $x, y \in \mathbb Z_q^{2n}$
there exists a symplectic transformation $S \in \mathcal S$ 
such that $y = S x$.
For this step, it is essential that $q$ is prime.
By the right-invariance of $\nu$ by $S$,
we see $\Pi_\nu( P_x \otimes P_{x}^\dagger ) = \Pi_\nu( P_y \otimes P_{y}^\dagger )$.
This implies that
\begin{align}
\Pi_\nu( P_{ a} \otimes P_{ b} )
&= \eta^{a,b}_{0,0} I + \eta^{a,b} \sum_{x \in \mathbb Z_q^{2n}\setminus \{0\}} P_{x} \otimes P_{x}^\dagger.
\end{align}
We claim that this is a linear combination of the identity operator
and the swap operator $\mathcal F = \sum_{u,v=0}^{q-1} \ket{u}\bra{v} \otimes \ket{v}\bra{u}$.
This is easily verified once we expand $\mathcal F$ in the Pauli operator basis;
using $\Tr(\mathcal F O \otimes O') = \Tr(OO')$ for any $q^n \times q^n$ matrices $O$ and $O'$,
we see that $\mathcal F \propto \sum_{x \in \mathbb Z_q^{2n}} P_x \otimes P_x^\dagger$.

The identity operator and the swap operator commute with $U \otimes U$ where $U \in U(q^n)$.
This implies that $\Pi_\nu ( P_a \otimes P_b^\dagger )$ commutes with $U \otimes U$,
and hence is equal to $\Pi_{\mu_\text{Haar}} \circ \Pi_\nu ( P_a \otimes P_b^\dagger )$.
By the right-invariance of the Haar distribution $\mu_\text{Haar}$,
we conclude that Eq.~\eqref{eq:conj-avg} is proved.

When $q$ is not prime, any probability distribution over the Clifford group
fails to be a unitary 2-design.
Let $n=1$.
Since the image of $\Pi_\mu$ is a linear combination of the identity and the swap,
we must have (see App.~\ref{sec:HaarAverage} below)
\begin{align}
\Pi_\mu(O \otimes O')= \sum_{s = \pm 1} \frac{\Tr(O) \Tr(O') + s \Tr(OO')}{q(q+ s)} \frac{I +s \mathcal F}{2}.
\label{eq:Haar-proj}
\end{align}
When $q=6$, there are non-identity Pauli operators $P$ and $Q$
such that $P^2 = I$ and $Q^3 = I $.
By Eq.~\eqref{eq:Haar-proj}, we have
$\Pi_\mu(P \otimes P^\dagger) = \Pi_\mu( Q \otimes Q^\dagger ) \neq 0$.
However, $\Pi_\nu(P \otimes P^\dagger)$ is a linear combination of Pauli operators,
each of which squares to identity, 
whereas $\Pi_\nu( Q \otimes Q^\dagger)$
is a linear combination of those that cube to identity,
so they cannot be equal.

\section{Anomalous behaviour of the \\ front for $\phi=0$}
\label{appendix_depinning}

 Above we noted that for sufficiently large $p$, $p>p_c$, the lattice growth process which we consider has anomalous behaviour when the front is oriented parallel to a lattice plane. This is a known phenomenon in various lattice growth models in discrete time which have \textit{synchronous parallel updates} and is well understood in terms of directed percolation \cite{richardson1973random,durrett1981shape,SavitZiff,kertesz1989anomalous,krug1990growth}.
   
In the regime $p>p_c$ the lattice-aligned ($\phi=0$) front has a speed $v(\phi=0)=2$ which is precisely the maximum possible speed allowed by causality. In this regime the front is pinned to the `light front' and is not rough (i.e. the width is of order one). (Exactly at $p_c$, the aligned front is logarithmically rough \cite{kertesz1989anomalous}.) For our lattice model it appears that $p_c\lesssim 2$.

This phenomenon is easily understood via a correspondence with directed percolation \cite{SavitZiff}. First consider a straight, lattice-aligned front in  the trivial deterministic limit $p=0$ ($q=\infty$). Apart from possibly on the first time step, this flat front advances by two lattice spacings every period: the front keeps pace with the `light cone' which is the line $x = 2 t$. Let $\tilde n(y,t) = 0,1$ denote the occupation numbers of the column of sites at the lightcone: $\tilde n(y,t)$ is the occupation number of the site at position $(2t, y)$ at time $t$. When $p=0$ we have $\tilde n(y,t) = 1$. We are interested in the  density $\<\tilde n\>$ (averaged over $y$) at late times when $p$ is nonzero. If this density remains finite, that means the front has an $O(1)$ width, and is attached to the light cone. If it instead tends to zero, the front detaches from the light cone, and we expect to recover standard KPZ roughening. Note that, in order to determine $\tilde n$ at time $t+1$, it is sufficient to know only $\tilde n$ at time $t$. The dynamics of the occupation numbers $\tilde n(y,t)$ are as follows. Under a horizontal dimer update (which advances the lightfront) each occupied $y$ has a chance $(1-p)$ of becoming unoccupied. Under a vertical update pairs of adjacent $y$ undergo the pairwise update described in the main text. This allows occupied sites to `reproduce'. This is therefore a birth-death process of the directed percolation type \cite{cardy1996scaling}. When $p$ is large the death rate is small and the reproduction rate is large, and the process is in an `active' phase with $\<\tilde n\> > 0$, while when $p$ is small the population of occupied sites dies out.

\section{Shape of a spreading droplet for weakly varying $v(\phi)$}
\label{rthetaappendix}

Consider an asymptotic front shape described by the parameterized curve $(\theta, r_t(\theta))$ in polar coordinates, which grows simply by rescaling: $r_t(\theta) = t \times r(\theta)$. Let $\phi(\theta)$ be the angle of the front's normal (to the $x$ axis) at polar position $\theta$. The radial growth rate is ${\dot r_t(\theta) = v(\phi(\theta))/\cos[\phi(\theta)-\theta]}$. Since the curve grows by rescaling we have $\partial_\theta [ \dot r_t(\theta) / r_t(\theta)] = 0$. Note that
\be\label{eq:tan}
\partial_\theta\ln r(\theta) = - \tan[\phi(\theta)-\theta].
\ee
Combining these gives \cite{WolfWulff}
\be
\left( \tan[\phi(\theta) - \theta] + w(\phi(\theta)) \right) \phi'(\theta) = 0.
\ee
 Therefore at a location where $r(\theta)$ is smooth we either have $\phi'(\theta) = 0$, i.e. a straight segment, or 
 \be\label{eq:tanw}
 \tan[\phi(\theta) - \theta ] = - w(\phi(\theta)).
 \ee
If the solution is everywhere smooth then the above equation must be satisfied everywhere. (Such solutions exist for sufficiently weakly varying $v$.) It is straightforward to solve this equation  in  powers of $w$:
\be\label{wseries}
-  \tan[\phi(\theta) - \theta ] =
w(\theta) - \f{1}{2} \partial_\theta w(\theta)^2 + \f{1}{6} \partial_\theta^2 w(\theta)^3 + \ldots
\ee
We find that the RHS involves only total derivatives of periodic functions. (Just from looking at Eq.~\ref{eq:tanw} this is at first sight surprising since it emerges from various cancellations.) Therefore, integrating the right hand side according to (\ref{eq:tan}) gives a periodic $r(\theta)$. 

For a formal explanation for why the expansion of $\tan(\phi-\theta)$ contains only total derivatives of periodic functions, consider a flow in the space of functions $v(\phi)$ which interpolates between the function of interest and the trivial function $v(\phi)$=\text{const}. Let $v_1(\phi)$ and $v_2(\phi)$ be two functions that are infinitesimally close on this flow and let $\phi_1(\theta)$ and $\phi_2(\theta)$ be the corresponding solutions. Assuming that $\phi_1(\theta)$ is periodic and corresponds to a periodic $r(\theta)$ we show that this property is inherited by $\phi_2(\theta)$ to order $\phi_2-\phi_1$. 
Using (\ref{eq:tan}), (\ref{eq:tanw}) we obtain
\ba
& \tan [\phi_1(\theta)-\theta]-\tan [\phi_2(\theta)-\theta] \\ &= 
\partial_\theta \left[ 
\ln v_2(\phi_1(\theta))
-
 \ln v_1(\phi_1(\theta)) 
 \right]
\end{align}
As required, the RHS is indeed the total derivative of a periodic function (note that $\phi_2$ does not appear on the RHS). Integrating along the flow then establishes the property for general $v(\phi)$ at the formal level --- i.e. assuming that the solution evolves smoothly during the flow.

\section{Haar average formula}
\label{sec:HaarAverage}

Here we review a standard formula for the average of matrix elements of unitary matrix
with respect to the Haar probability measure $\mu$ on $U(N)$.
Let us abbreviate $\int_{U(N)} \rd \mu(U)$ as $\mathbb E_U$.
We are going to prove that
\begin{align}
& \mathbb E_U~ U \ket a \bra b U^\dagger \otimes U \ket c \bra d U^\dagger \\ &=
\sum_{s=\pm} \frac{I +s F}{2N(N +s 1)}  ( \delta_{ab} \delta_{cd} +s \delta_{cb} \delta_{ad} )
\label{eq:HaarAvg}
\end{align}
where $F$ is the swap operator on $(\mathbb C^N)^{\otimes 2}$.
Evaluating a particular matrix element, we have
\begin{align}
&\mathbb E_U ~ U_{a'a} U^*_{b'b} U_{c'c} U^*_{d'd} = \nonumber\\
&\quad \frac{1}{N^2-1} \Big[ 
\delta_{a'b'}\delta_{c'd'} \delta_{ab} \delta_{cd} 
+
\delta_{a'd'}\delta_{b'c'}\delta_{ad}\delta_{bc} \nonumber \\
&\qquad  \qquad
-\frac{1}{N} (
\delta_{ab}\delta_{cd}\delta_{a'd'}\delta_{b'c'}
+
\delta_{a'b'}\delta_{c'd'}\delta_{ad}\delta_{bc}
)
\Big].
\end{align}

\begin{proof}[Proof of Eq.~\eqref{eq:HaarAvg}]
The average is a matrix on $\mathcal H = (\mathbb C^N)^{\otimes 2} $
that commutes with every $U^{\otimes 2}$.
Hence, the average is block-diagonal in the basis 
where the representation of $U(d)$ is block-diagonal.
The irreps appearing in $\mathcal H$ are 
the symmetric subspace and the anti-symmetric subspace.
In each irrep, the average must be proportional to the identity $I_\pm$ by the Schur's lemma,
and we need to evaluate the trace in order to determine the constant of proportionality. 
The projection onto the (anti-)symmetric subspace is $(I \pm F)/2$ where $F$ is the swap operator: $F \ket{ac} = \ket{ca}$.
So the trace is
\ba
&\frac12 \Tr\Big[ U \ket a \bra b U^\dagger \otimes U \ket c \bra d U^\dagger 
 \pm 
 \notag\\
& \qquad\qquad\qquad \qquad  U \ket c  \bra b U^\dagger \otimes U \ket a \bra d U^\dagger \Big] \notag \\
&= \frac12 [ \delta_{ab} \delta_{cd} \pm \delta_{cb} \delta_{ad} ]
\end{align}
This must be equal to
$ C_{\pm} \Tr( I_\pm ) = C_{\pm} N (N \pm 1 ) /2
$
Therefore, the average is equal to
$
\sum_{s=\pm} C_s (I +s F)/2.
$
\end{proof}

\section{Proof of Eq.~\eqref{eq:fluctationDeepInLightcone}}
\label{app:fluctuationBound}

Let $N$ be the Hilbert space dimension of $n$ $q$-dimensional qudits; $N = q^n$.
For any $N \times N$ unitary $U$, denote by $U^{\otimes t,t}$ 
the tensor product $(U \otimes U^*)^{\otimes t}$,
where $U^*$ is the complex conjugate of $U$.
Let $\mu$ be the Haar probability distribution on $U(N)$, 
and define for any probability distribution $\nu$ on $U(N)$,
a real number
\begin{align}
g(\nu,t) = \opnorm{
\underbrace{\avg_{U \sim \nu} U^{\otimes t,t,}}_{\Pi_\nu}
-
\underbrace{\avg_{U \sim \mu} U^{\otimes t,t} }_{\Pi_\mu} 
}.
\end{align}
Here, $\opnorm{\cdot}$ denotes the maximum singular value.
Due to left and right invariance of $\mu$, it follows that $\Pi_\mu^2 = \Pi_\mu = \Pi_\nu \Pi_\mu = \Pi_\mu \Pi_\nu$.
($\Pi_\nu$ is not in general a projector.)
Therefore, 
\begin{align}
g(\nu^{* m},t) = \opnorm{\Pi_\nu^m - \Pi_\mu} = \opnorm{(\Pi_\nu - \Pi_\mu)^m} = g(\nu,t)^m,
\label{eq:m-fold-convolution}
\end{align}
where $\nu^{*m}$ is the $m$-fold convolution of $\nu$,
i.e., $\nu^{*m}$ is the distribution of the product $U_1 U_2 \cdots U_m$ when every $U_i$ obeys distribution $\nu$.

Now, let $\nu$ be the distribution on $U(N)$ obtained by applying one layer of 
even bond local Haar random unitaries ($U(q^2)$) and then one layer of odd bond local Haar random unitaries.
Brandao-Harrow-Horodecki's result~\cite{BHH2016} implies that
\begin{align}
&g(\nu,t) \le \exp( - 1/M_{t,q} ), \label{eq:design-gap}\\
&M_{t,q} = 4250 \lceil \log_q(4t) \rceil^2 q^2 t^5 t^{3.1/\log q} .
\end{align}
Their theorem does not directly cover this,
but they have lemmas that are good enough for our purpose; 
eq.~(48) of the CMP version is what we actually need.

Consider $f(U) = N^{-2} (\Tr UXU^\dagger Y UXU^\dagger Y )^2 \ge 0$ 
where all the matrices $U,X,Y$ are $N \times N$.
$f(U)$ can be thought of as $\bra{\tilde X} U^{\otimes 4,4} \ket{\tilde Y}$ for some vectors 
$\ket{\tilde X}$ and $\ket{\tilde Y}$.
Assume $\Tr(X) = \Tr(Y) = 0$, but $\Tr(X^2) = \Tr(Y^2) = N$.
Then, the Euclidean norms of $\ket{\tilde X}$ and $\ket{\tilde Y}$ are both $N^2$.
Normalizing so that $\ket X := \ket{\tilde X} / N^2$ and $\ket Y := \ket{\tilde Y}/N^2$,
we can write $f(U) = N^2 \bra X U^{\otimes 4,4} \ket Y$.

By Eqs.~\eqref{eq:m-fold-convolution} and \eqref{eq:design-gap}, we have
\begin{align}
\abs{
\avg_{U \sim \nu^{*m}} f(U) - \avg_{U \sim \mu} f(U)
}
\le
e^{-m/M_{4,q}} q^{2n}
\end{align}
If $\abs{\avg_{U \sim \mu} f(U)} \le q^{-c n}$,
then $\avg_{U \sim \nu^{*m}} f(U) \le 2q^{-cn}$ whenever $m/n \ge (c+2)M_{4,q} \log q$.

Hastings' Schwinger-Dyson trick~\cite{Hastings2007} gives
\begin{align}
\avg_{U\sim \mu} f(U) \le 10 N^{-2} = 10 q^{-2n}.
\end{align}
Therefore, whenever $m/n \ge 4 M_{4,q} \log q$,
we have
\begin{align}
\avg_{U \sim \nu^{*m}} f(U) \le 11 q^{-2n}.
\end{align}

\section{A mean field approximation}
\label{app:meanfield}

Ref.~\cite{Aleiner} argued, on the basis of Keldysh perturbation theory, that in various circumstances the out-of-time-order correlator  would satisfy a traveling wave equation such as the Fisher-KPP equation (the details of this equation depending on the physical system).  An example is the Fisher-KPP equation itself:
\be
\partial_t \otc = D \nabla^2 \otc + \lambda \otc (1-\otc).
\ee
The key feature is $\lambda$ term, which means that if $\otc$ is `seeded' with a small nonzero value, it will increase to a value close to one on a timescale of order $\lambda^{-1}$ (and then saturate). This equation has stable solutions describing a front propagating with a speed $v_B = 2\sqrt{D\lambda}$. This front does not broaden. 

This phenomenology is very different from the picture which we have obtained from the random circuit and the mapping to classical growth processes. Recall that in 1D we related $\otc$ to a homogeneous (linear) equation, and in higher dimensions we found that $\otc$ was not governed by a partial differential equation.

The purpose of this Appendix is to show that a traveling wave picture can emerge from our mappings if we  make a certain mean field approximation.  This mean field approximation is not valid in the systems we have studied --- it is an uncontrolled approximation which does not capture the true behavour either at short or at large times.  However in  variant models a small parameter could be present which justified the mean field approximation up to some finite but large timescale. In this situation we expect that mean field will nevertheless break down at asymptotically long times, with the  front  eventually roughening in the manner discussed in the text. 

Recall that for the random circuit we have 
\be\label{appendix_C_n_eq}
\overline \otc(x,t) = \f{q^2}{q^2-1} \< n(x,t) \>,
\ee
where $n(x,t)$ is the occupation number in the fictitious classical cluster growth problem.  Let us consider the \textit{joint} probability distribution $P(\{n\};t)$ for this occupation number. This distribution involves nontrivial correlations between sites which are crucial for capturing the correct asymptotic behaviour. Nevertheless let us explore the mean field approximation in which we pretend all sites are independent, $P(\{n\};t)=\prod_x P_x(n(x);t)$, with 
\be
P_x(n(x);t) =
  \big[ 1-\<n(x,t)\> \big] \delta_{n(x),0} + \<n(x,t)\> \delta_{n(x),1}.
\ee
For simplicity,  consider  a model on the hypercubic lattice in $d$ dimensions (with coordination number $z=2d$) in which unitaries (`updates') are applied to bonds in a Poissonian fashion at rate $\Gamma/2$ per bond. This continuous time protocol does not change the basic point but it simplifies the equations. Write $m=\<n\>$. Note that if we update  a bond which contains at least one fictitious particle, the subsequent (conditionally) averaged density on that bond is $1-p$. This implies
\ba
m(x,t+\Delta t) & = (1- z \Gamma \Delta t)  m(x,t) \\ 
&+ \Gamma \Delta t (1-p) \sum_{y\in x} \< \lf 1 - \delta_{n(x),0}\delta_{n(y),0} \ri \> \notag,
\end{align}
where the first term is the probability that site $x$ does not receive an update in the interval $\Delta t$. Making the mean field approximation, $\<  \delta_{n(x),0}\delta_{n(y),0}\>$ factorizes into $(1-m(x))(1-m(y))$, so that
\ba\label{lattice_travelling_wave_eq}
\partial_t m(x,t)  = \Gamma \sum_{y\in x} \big( &
- p m(x,t) + (1-p) m(y,t)
\\ &  - (1-p) m(x,t)m(y,t)
 \big).
\end{align}
The first term on the right is a `death rate'. The second term is spreading. The third term is a correction to overcounting in the second term. An analogous equation could be written down for the regular circuit considered in the main text, but we would have to use discrete time.

Eq.~\ref{lattice_travelling_wave_eq} is a lattice traveling wave equation. This is most apparent if we make a formal expansion in the lattice spacing $a$ to second order (valid, given the approximations already made,  if the solution is slowly varying). Recalling $p = 1/(q^2 + 1)$ and Eq.~\ref{appendix_C_n_eq},
\ba
\Gamma^{-1} \partial_t \overline{\otc}(x,t) = & a^2 \bigg( (1-p)- ({1-2p})\, \overline{\otc} \bigg) \nabla^2 \overline{\otc}  \\
& + 2d(1-2p) \, \overline{\otc}(1-\overline{\otc}).
\end{align}
This differs from the Fisher-KPP equation only in the $\otc$--dependence of the diffusion constant, and we expect similar properties.

Above, the mean field limit was an unjustified formal approximation. We could of course construct random circuit models in which the (lattice) mean field approximation was quantitatively accurate up to a large time, for example by using long range interactions or a large coordination number to reduce the effect of correlations. However at long times, in physical dimensionalities, we expect the front to roughen so that the mean field traveling wave picture breaks down. In (unphysically) high dimensions, mean field may be valid even at late times (recall that  the phase diagram of the KPZ equation allows for a non-roughening phase in high dimensions, as discussed in the text).

\bibliography{vb-refs}
\end{document}